{}
{}

\documentclass[11pt,a4paper]{article}

\usepackage{graphicx} 
\usepackage{float}
\usepackage{color}
\usepackage[toc,page]{appendix} 
\usepackage{enumerate}
\usepackage{latexsym,amsthm,amssymb,amsfonts,amsbsy,amsmath}
\usepackage{amsmath}
\usepackage{dsfont}
\usepackage{amsfonts}
\usepackage{amssymb}
\usepackage{mathrsfs}
\usepackage{bbm}
\usepackage{mathtools}
\usepackage{cite}

\usepackage{amsthm}
\usepackage{bm} 
\usepackage{bbm} 
\usepackage{slashed} 
\usepackage{cancel}
\usepackage{xcolor}
\usepackage{tikz}
\usetikzlibrary{arrows,matrix,snakes}
\usetikzlibrary{decorations.markings,shapes.geometric,shapes.misc} 
\usepackage{tikz-cd}

\usepackage{pdflscape}
\usepackage{scalefnt}

\definecolor{lightred}{RGB}{255,127,127}
\definecolor{lightgreen}{RGB}{127,255,127}
\definecolor{lightblue}{RGB}{127,127,255}
\definecolor{linkcolor}{rgb}{0,0,0.6}
\usepackage[ pdftex,colorlinks=true,
pdfstartview=FitV,
linkcolor= linkcolor,
citecolor= linkcolor,
urlcolor= linkcolor,
hyperindex=true,
hyperfigures=false,
linktocpage=true]
{hyperref}

\numberwithin{equation}{section}

\usepackage{tabularx} 

\usepackage{fancyhdr} 

\theoremstyle{plain}

\newcommand{\vp}{\varphi}

\newcommand{\p}{\partial}
\newcommand{\g}{\mathfrak{g}}

\newcommand{\dd}{\text{d}}

\newcommand{\Pc}{\mathcal{P}}
\newcommand{\Id}{\text{Id}}

\newcommand{\s}{\sigma}

\newcommand{\Lc}{\mathcal{L}}

\newcommand{\W}[1]{I_{\text{W}\hspace{-1pt}\text{Z}}\left[#1\right]}
\newcommand{\Ww}[1]{I_{\text{W}\hspace{-1pt}\text{Z}}\bigl[#1\bigr]}

\newcommand{\Ad}{\text{Ad}}

\newcommand{\rhom}{{\text{\large$\varrho$}}}
\newcommand{\yh}{\widehat{y}}
\newcommand{\A}{\mathcal{B}}
\newcommand{\B}{\mathcal{C}}
\newcommand{\Kop}{\mathcal{U}}
\newcommand{\Lop}{\mathcal{V}}
\newcommand{\gh}{\widehat{g}}

\DeclareFontEncoding{LS1}{}{}
\DeclareFontSubstitution{LS1}{stix}{m}{n}
\DeclareSymbolFont{stixsymbols}{LS1}{stixscr}{m}{n}
\SetSymbolFont{stixsymbols}{bold}{LS1}{stixscr}{b}{n}
\DeclareMathSymbol{\kay}{\mathalpha}{stixsymbols}{"6B}

\def\res{\mathop{\text{res}\,}}

\definecolor{myGreen}{rgb}{0.0,0.4,0.0}

\makeatletter
\let\@keywords\@empty
\let\@subject\@empty
\providecommand{\keywords}[1]{\gdef\@keywords{#1}}
\providecommand{\subject}[1]{\gdef\@subject{#1}}
\def\thetitle{\@title}
\def\theauthor{\@author}
\def\thesubject{\@subject}
\def\thedate{\@date}
\def\thekeywords{\@keywords}
\makeatother
\AtBeginDocument{
\hypersetup{pdftitle={\thetitle}}
\hypersetup{pdfauthor={\theauthor}}
\hypersetup{pdfsubject={\thesubject}}
\hypersetup{pdfkeywords={\thekeywords}}}

\title{Integrable deformations of coupled \texorpdfstring{$\bm\sigma$-models}{sigma-models}}

\author{C. Bassi and S. Lacroix}

\begin{document}

\begin{flushright}
[ZMP-HH/19-26]
\end{flushright}

\begin{center}

\vspace*{2cm}

\begingroup\Large\bfseries\thetitle\par\endgroup

\vspace{1.5cm}

\begingroup
Cristian Bassi$^{a,}$\footnote{E-mail:~cristian.bassi@desy.de} and Sylvain Lacroix$^{a,}$\footnote{E-mail:~sylvain.lacroix@desy.de}
\endgroup

\vspace{1cm}

\begingroup
$^a\,$\it II. Institut f\"ur Theoretische Physik, Universit\"at Hamburg,
\\
Luruper Chaussee 149, 22761 Hamburg, Germany
\\
Zentrum f\"ur Mathematische Physik, Universit\"at Hamburg, \\
Bundesstrasse 55, 20146 Hamburg, Germany

\endgroup

\end{center}

\vspace{2cm}

\begin{abstract}
We construct integrability-preserving deformations of the integrable $\s$-model coupling together $N$ copies of the Principal Chiral Model. These deformed theories are obtained using the formalism of affine Gaudin models, by applying various combinations of Yang-Baxter and $\lambda$-deformations to the different copies of the undeformed model. We describe these models both in the Hamiltonian and Lagrangian formulation and give explicit expressions of their action and Lax pair. In particular, we recover through this construction various integrable $\lambda$-deformed models previously introduced in the literature. Finally, we discuss the relation of the present work with the semi-homolomorphic four-dimensional Chern-Simons theory.
\end{abstract}

\setcounter{footnote}{0}

\newpage

\setcounter{tocdepth}{2}
\tableofcontents

\section{Introduction and summary of the results}

Integrable non-linear $\s$-models form an important class of two-dimensional classical integrable field theories. Their study was initiated more than 40 years ago and has found applications in various domains of physics, such as the AdS/CFT correspondence (see for instance the review~\cite{Beisert:2010jr}) and condensed matter theory~\cite{Haldane:1982rj}. A prototypical example of integrable $\s$-model is given by the Principal Chiral Model on a real semi-simple Lie group $G_0$, with or without Wess-Zumino term. It describes the dynamics of a $G_0$-valued field $g(x^+,x^-)$, where $x^\pm=(t \pm x)/2$ denote the two-dimensional light-cone coordinates. Let $\g_0$ be the Lie algebra of $G_0$ and $\kappa$ the opposite of its Killing form. The action of this model is then given by
\begin{equation}\label{Eq:IntroPCM}
S_{\text{PCM}}[g] = \rho \iint \dd t \, \dd x \; \kappa\bigl( g^{-1}\p_+ g, g^{-1}\p_- g \bigr) + \kay \, \W g,
\end{equation}
where $\rho$ and $\kay$ are constant parameters, $\p_\pm$ denote the derivatives with respect to $x^\pm$ and $\W g$ is the Wess-Zumino term of $g$. The integrability of this model relies on the fact that its equation of motion can be recast in the form of a zero curvature equation $\p_+ \Lc_-(z) - \p_- \Lc_+(z) + \bigl[\Lc_+(z),\Lc_-(z)\bigr] = 0$ on a Lax pair $\Lc_\pm(z)$. This Lax pair depends on an auxiliary complex parameter $z$, called the spectral parameter.

It was shown by Klim\v{c}\'ik in~\cite{Klimcik:2002zj,Klimcik:2008eq} that the Principal Chiral Model (without Wess-Zumino term) admits a continuous integrable deformation, called the Yang-Baxter model, which generalises to an arbitrary group $G_0$ a model constructed in~\cite{Cherednik:1981df} for the group $SU(2)$. This deformed model depends on the choice of a skew-symmetric $R$-matrix on $\g_0$, \textit{i.e.} a linear operator $R:\g_0\rightarrow\g_0$ satisfying the modified classical Yang-Baxter equation $[RX,RY]-R[RX,Y]-R[X,RY]=-c^2 [X,Y]$ for every $X,Y\in\g_0$, with $c$ equal to 1 or $i$. The action of the Yang-Baxter model is given by
\begin{equation*}
S_{\text{YB}}[g] = \rho \iint \dd t \, \dd x \; \kappa\left( g^{-1}\p_+ g, \frac{1}{1-\eta \,R_g} g^{-1}\p_- g \right),
\end{equation*}
where $\eta$ is the deformation parameter and $R_g = \Ad^{-1}_g \circ R \circ \Ad_g$. This construction was later extended in various directions. For instance, one can construct integrable Yang-Baxter deformations of symmetric space $\s$-models~\cite{Delduc:2013fga}, of superstrings on semi-symmetric spaces~\cite{Delduc:2013qra,Delduc:2014kha} and of the Principal Chiral Model with Wess-Zumino term~\cite{Delduc:2014uaa}. Alternatively, one can also consider deformations based on homogeneous $R$-matrices~\cite{Kawaguchi:2014qwa}, satisfying the non-modified ($c=0$) classical Yang-Baxter equation.

Another type of integrable deformed $\s$-model, called the $\lambda$-model, was constructed by Sfetsos in~\cite{Sfetsos:2013wia}\footnote{Let us note that this model can be reformulated as a theory on $G_0\!\times\!G_0\!\times\!G_0$, which is a special case of the model originally considered in~\cite{Tseytlin:1993hm} and whose classical integrability was first proven in~\cite{Bardakci:1996gs}. We thank A. Tseytlin for pointing out these references.}. It corresponds to a deformation of the non-abelian T-dual of the Principal Chiral Model (without Wess-Zumino term) and generalises a result obtained in~\cite{Balog:1993es} for the group $SU(2)$. Its action is defined as
\begin{equation*}
S_{\lambda}[g] = S_{\text{WZW},\kay}[g] + \kay \iint \dd t \, \dd x \; \kappa\left( \p_+ g g^{-1}, \frac{1}{\lambda^{-1}-\Ad_{g}^{-1}} g^{-1}\p_- g \right),
\end{equation*}
where $\kay$ and $\lambda$ are constant parameters and $S_{\text{WZW,}\,\kay}[g]$ is the action of the conformal Wess-Zumino-Witten model at level $\kay$ (\textit{i.e.} the action \eqref{Eq:IntroPCM} with $\rho=\kay/2$). Similarly to the Yang-Baxter deformation, the $\lambda$-deformation can be generalised to symmetric-space $\s$-models~\cite{Hollowood:2014rla} and superstrings on semi-symmetric spaces~\cite{Hollowood:2014qma}.\\

The existence of a Lax pair ensures that the model admits an infinite number of conserved charges, extracted from the monodromy of the Lax matrix $\Lc(z)=\frac{1}{2}\bigl(\Lc_+(z)-\Lc_-(z)\bigr)$. In order to show the integrability of this model, one has to prove that these conserved charges are in involution. For integrable $\s$-models, this is done by showing that the Poisson bracket of the Lax matrix takes the form of a non-ultralocal Maillet bracket~\cite{Maillet:1985fn,Maillet:1985ek}. This was proved for the Principal Chiral Model in~\cite{Maillet:1985ec}, for the Yang-Baxter model, with and without Wess-Zumino term, in~\cite{Delduc:2013fga} and~\cite{Delduc:2014uaa} and for the $\lambda$-model in~\cite{Itsios:2014vfa} (see also~\cite{Hollowood:2014rla} for first results on the Hamiltonian analysis of the $\lambda$-model). For all these cases, the Maillet bracket takes a particular form, which is encoded in a rational function $\vp(z)$ of the spectral parameter, called the twist function~\cite{Maillet:1985ec, Reyman:1988sf, Sevostyanov:1995hd, Vicedo:2010qd} (see also~\cite{Lacroix:2018njs}). These results shed light on the common algebraic structure underlying the integrability of this family of models and led to their reinterpretation as part of a larger class of integrable field theories, called (realisations of) affine Gaudin models~\cite{Vicedo:2017cge}. In this formalism, the twist function and the Lax matrix of the model arise naturally from representations of untwisted affine Kac-Moody algebras.

Recently, the formalism of affine Gaudin models has been applied to generate an infinite family of new integrable classical $\s$-models~\cite{Delduc:2018hty,Delduc:2019bcl,Lacroix:2019xeh}. More precisely, these models are obtained by coupling in a non-trivial way an arbitrary number of Principal Chiral Models with Wess-Zumino terms on the same Lie group $G_0$. The fact that these $\s$-models are constructed as realisations of affine Gaudin models ensures that they are integrable (more precisely, they possess a Lax pair, whose spatial component satisfies a Maillet bracket with twist function). Let us briefly describe the coupled model with $N$ copies. It is defined by the action
\begin{equation}\label{Eq:IntroCoupled}
S\bigl[g^{(1)},\cdots,g^{(N)}\bigr] = \iint \dd t\,\dd x\; \sum_{r,s=1}^N \rho_{rs}\, \kappa\bigl( g^{(r)\,-1}\p_+ g^{(r)},g^{(s)\,-1}\p_- g^{(s)} \bigr) + \sum_{r=1}^N \kay_r \, \Ww{g^{(r)}},
\end{equation}
depending on $N$ fields $g^{(1)},\cdots,g^{(N)}$ valued in $G_0$, where $\rho_{rs}$ and $\kay_r$ are constant parameters. For generic values of these coefficients, the model is not integrable. The particular model of~\cite{Delduc:2018hty,Delduc:2019bcl}, which is integrable since it is constructed as a realisation of affine Gaudin model, corresponds to a specific choice of these coefficients. More precisely, they are expressed in terms of $3N-1$ free parameters in a way which, for brevity, we will not describe in this introduction. The Lax connection of the model takes the form
\begin{equation*}
\Lc_\pm(z) = \sum_{r=1}^N \alpha_r(z)\, g^{(r)\,-1}\p_\pm g^{(r)},
\end{equation*}
where the $\alpha_r(z)$'s are rational functions of the spectral parameter, whose expressions in terms of the $3N-1$ defining parameters of the model are also known explicitly~\cite{Delduc:2018hty,Delduc:2019bcl}.\\

It is natural at this point to search for integrable deformations of this coupled $\s$-model. It was explained in~\cite{Delduc:2019bcl} that such deformations exist and that they can also be defined as realisations of affine Gaudin models. For instance, one can apply a Yang-Baxter deformation to any of the $N$ copies of the model. Moreover, if one of the copy has no Wess-Zumino term, it is possible to consider a corresponding $\lambda$-deformation, which would then be more precisely a deformation of the model where this copy of the Principal Chiral Model has been replaced by its non-abelian T-dual. In general, one can then consider any combinations of these deformations on the different copies, leading to a whole panorama of different models.

Although these integrable deformed coupled $\s$-models are known to exist, they have not been constructed explicitly so far and have yet to be fully understood. In particular, since they are defined as realisations of affine Gaudin models, they are inherently formulated in the Hamiltonian framework. It is then an important aspect in the understanding of these models and of their properties to formulate them in the Lagrangian framework and to find an explicit expression of their action. In particular, this would give us access to the geometry underlying these theories, \textit{i.e.} the deformed metric and $B$-field of the target space $G_0^N$ which define these $\s$-models. In addition to clarifying the structure of the models at the classical level, describing their Lagrangian formulation can also benefit the understanding of their quantum properties, as for example the one-loop renormalisation of $\s$-models is controlled by the curvature of their underlying geometry. It is also an important problem to express the Lax pair of the model in terms of the Lagrangian fields, in order to understand how the integrable structure of the model manifests itself in the Lagrangian formulation. The explicit construction of the action and Lagrangian Lax pair of these integrable deformed coupled $\s$-models is the main subject of this article.

Several examples of integrable $\s$-models coupling together $\lambda$-models were proposed by Georgiou and Sfetsos in~\cite{Georgiou:2016urf,Georgiou:2017jfi,Georgiou:2018hpd,Georgiou:2018gpe}, using a different approach than the one considered in this article. Moreover, it was shown very recently that these models satisfy a Maillet bracket and possess a twist function~\cite{Georgiou:2019plp}. As an application of the general construction developed in this article, we will show that these models can be obtained as limits of the ones obtained using affine Gaudin models.\\

Before sketching the methods used in this article to construct and study integrable deformations of coupled $\s$-models, let us illustrate briefly some of its main results. Let us first consider the model with $N$ copies of the Principal Chiral Model with Wess-Zumino term, each subject to a Yang-Baxter deformation. It is defined by $4N-1$ parameters, which can be thought of as the $3N-1$ parameters of the undeformed model and $N$ deformation parameters, and by the choice of $N$ $\!R$-matrices $R_r$ on $\g_0$\footnote{The $R$-matrix $R_r$ is assumed to satisfy the additional property $R_r^3=c_r^2R_r$, except if the $r$-th copy does not possess a Wess-Zumino term, \textit{i.e.} if $\kay_r=0$.}. The action of the model then takes the form
\begin{equation}\label{Eq:IntroYB}
S[\{g^{(r)}\}] = \frac{1}{2} \iint \dd t\,\dd x\; \sum_{r,s=1}^N \kappa \Bigl( g^{(r)\,-1}\p_+ g^{(r)},\, \bigl(\null^{t}\Kop_+^{-1}\;\null^{t\!}\rhom_+ + \rhom_- \, \Kop_-^{-1} \bigr)_{rs}\, \,g^{(s)\,-1}\p_- g^{(s)} \Bigr) + \sum_{r=1}^N \kay_r \, \Ww{g^{(r)}}.
\end{equation}
In this expression, $t$ denotes the transposition of operators and $\rhom_\pm$, $\Kop_\pm$ and hence $\null^{t}\Kop_+^{-1}\;\null^{t\!}\rhom_+ + \rhom_- \, \Kop_-^{-1}$ are operators on $\g_0^N$, which can be seen as $N \times N$ matrices whose entries are operators on $\g_0$. The entries of $\rhom_\pm$ are of the form $\rho^\pm_{rs}\,\Id$, with the coefficients $\rho^\pm_{rs}$ expressed explicitly in terms of the $4N-1$ defining parameters of the model. The entries $\Kop^\pm_{rs}$ in the $r$-th row of $\Kop_\pm$ are expressed as polynomials of degree two (or degree one if $\kay_r=0$) in $\Ad_{g^{(r)}}^{-1}\circ R_r \circ \Ad_{g^{(r)}}$, with coefficients also explicitly known as functions of the defining parameters. In the undeformed limit (where the $N$ deformation parameters are taken to 0), the coefficients $\rho_{rs}^+$ and $\rho_{rs}^-$ converge respectively to $\rho_{sr}-\delta_{rs}\kay_r/2$ and $\rho_{rs}+\delta_{rs}\kay_r/2$ and the operators $\Kop_\pm$ simply become the identity operator on $\g_0^N$. The action \eqref{Eq:IntroYB} then reduces to the action \eqref{Eq:IntroPCM} of the undeformed model. The operators $\Kop_\pm$ entering the action of the model also control the Lax pair of the model. Indeed, the latter is given by
\begin{equation*}
\Lc_\pm(z) = \sum_{r,s=1}^N \widetilde\alpha_r(z) \bigl( \Kop_\pm^{-1}\bigr)_{rs}\; g^{(s)\,-1}\p_\pm g^{(s)},
\end{equation*}
with $\widetilde\alpha_r(z)$ explicit deformations of the rational functions $\alpha_r(z)$ considered in the undeformed model.

Let us now consider the model coupling together $N$ copies of the $\lambda$-model. This is a deformation of the model coupling $N$ copies of the non-abelian T-dual of the Principal Chiral Model, which is equivalent to the model \eqref{Eq:IntroCoupled} with no Wess-Zumino terms. This undeformed model then possesses $2N-1$ free parameters and its $N$-fold $\lambda$-deformation is described by $3N-1$ parameters. The action of this model takes the form
\begin{equation}\label{Eq:IntroLambda}
S[\{g^{(r)}\}] = \sum_{r=1}^N S_{\text{WZW,}\,\kay_r}\bigl[g^{(r)}\bigr] + \iint \dd t\,\dd x\; \sum_{r,s=1}^N \kay_r\,\kappa\left( \p_+ g^{(r)} g^{(r)\,-1}, \left( \frac{1}{\mathcal{M} - \mathcal{D}^{-1}} \right)_{rs} \,g^{(s)\,-1}\p_- g^{(s)} \right),
\end{equation}
where $\mathcal{M}$ and $\mathcal{D}$ are operators on $\g_0^N$, with entries $\mathcal{M}_{rs}=\mu_{rs}\,\Id$ and $\mathcal{D}_{rs} = \Ad_{g^{(r)}}\,\delta_{rs}$. The model is then characterised by the coefficients $\kay_r$ and $\mu_{rs}$, which are expressed explicitly in terms of the $3N-1$ defining parameters of the models. Actions of this form were already considered in the article~\cite{Georgiou:2018gpe}. In particular, it was argued in this reference that the truncation of this model where all the coefficients $\mu_{rs}$ vanish except for the coefficients $\mu_{11},\cdots,\mu_{(N-1)1}$ and $\mu_{N 2},\cdots,\mu_{NN}$ defines an integrable model with $3N-2$ parameters. This truncation can be seen as a particular limit of the model constructed above, with one deformation parameter less. Although the model considered here extends this truncation by introducing only one additional parameter, this extension has a non-trivial effect on the structure of the model, as all the coefficients $\mu_{rs}$ become generically non-zero in this model.

In the main text of this article, we also construct explicitly the integrable model coupling together $N_1$ Yang-Baxter models and $N_2$ $\lambda$-models, whose action takes a form which mixes the structures of the above actions \eqref{Eq:IntroYB} and \eqref{Eq:IntroLambda}. For brevity, we will not describe this action in the introduction. All these deformed models involve the inverse of operators on $\g_0^N$. These operators can be seen as $N\times N$ matrices whose entries are operators on $\g_0$. In particular, the non-commutativity of these entries makes the explicit inversion of these operators a non-straightforward problem. In the case of models with two copies only, we show how to perform this inversion explicitly. More precisely, we find an expression of these inverse operators which involves inversions of operators on one copy of $\g_0$ only. Using this result, we give more explicit expressions of the models coupling together two Yang-Baxter models or two $\lambda$-models.\\

Let us now briefly sketch the methods used in this article, which are based on the formalism of affine Gaudin models. To illustrate these methods, it is useful to come back to the deformed models with only one copy and describe their structure as realisations of affine Gaudin models. An important object characterising affine Gaudin models is their twist function, which is the rational function of the spectral parameter controlling the Poisson bracket of their Lax matrix. For the Yang-Baxter model and the $\lambda$-model, this twist function possesses two simple poles $z^\pm$ in the complex plane. Each of these poles $z^\pm$ corresponds to a so-called site of the underlying affine Gaudin model and is associated with a Kac-Moody current $\mathcal{J}^\pm$, which is an observable on the phase space of the model. For both the Yang-Baxter model and the $\lambda$-model, this phase space consists of canonical fields in the cotangent bundle $T^*G_0$, corresponding to the field $g$ in $G_0$ and its associated conjugate momentum. The currents $\mathcal{J}^\pm$ satisfy the standard Poisson brackets of Kac-Moody currents and Poisson commute one with another.

It is a standard result in the literature that the Hamiltonian integrable structure of the Yang-Baxter and $\lambda$-models is characterised by two commuting Kac-Moody currents~\cite{Delduc:2013fga,Delduc:2014uaa,Hollowood:2014rla,Vicedo:2015pna}. This is what motivated their reinterpretation as affine Gaudin models in~\cite{Vicedo:2017cge}. An important remark to make here is that although the Yang-Baxter model and the $\lambda$-model both possess commuting Kac-Moody currents in the same phase space, the expression of these currents in terms of the canonical fields of this phase space is different. It is this expression which characterises the model one considers and in particular differentiates the Yang-Baxter model and the $\lambda$-model. In the terminology of~\cite{Delduc:2019bcl}, the datum of $N$ commuting Kac-Moody currents in a certain phase space is called a Kac-Moody realisation with $N$ sites. In particular, the Yang-Baxter and $\lambda$-models define two different Kac-Moody realisations with two sites, in the same phase space.

The integrable coupled deformed models considered in this article are constructed as realisations of affine Gaudin models with $2N$ sites. Their twist function is thus a rational function of the spectral parameter $z$ with $2N$ simple poles, that we gather in pairs $z_r^\pm$, $r\in\lbrace 1,\cdots,N\rbrace$. We attach to these pairs $N$ independent copies of either the Yang-Baxter realisation or the $\lambda$-realisation. The phase space of the models is then formed by $N$ copies of the canonical fields on $T^*G_0$. The models are defined in the Hamiltonian framework: in particular, their Hamiltonian is constructed as the spatial integral of a particular quadratic combination of the $2N$ Kac-Moody currents attached to the $2N$ sites, following the general formalism of affine Gaudin models~\cite{Vicedo:2017cge,Delduc:2019bcl}. This definition of the Hamiltonian ensures that these models are integrable: their equation of motion can be recast as a zero curvature equation on a Lax pair and the corresponding Lax matrix satisfies a Maillet bracket, controlled by the choice of twist function made above. As these models are defined in the Hamiltonian framework, one then has to perform an inverse Legendre transform to obtain their Lagrangian formulation, and in particular their action and Lagrangian Lax pair. In this article, we do this using interpolation methods, which generalise to the deformed case the techniques used in~\cite{Delduc:2019bcl} to treat the undeformed coupled model. The phase space of the models being formed by $N$ independent copies of canonical fields on $T^*G_0$, they are formulated in the Lagrangian framework in terms of $N$ $G_0$-valued fields $g^{(r)}$, which are the fields introduced earlier.

The most important building blocks for the construction of these models are the Yang-Baxter and $\lambda$-realisations, which are Kac-Moody realisations in $T^*G_0$. In this article, we treat these two realisations in a uniform way, by introducing a general ansatz for the form of the corresponding Kac-Moody currents in terms of the canonical fields in $T^*G_0$, which includes these two examples. Using the fact that this ansatz should describe Kac-Moody currents, we identify certain key properties that it should satisfy in general. These properties then allow us to obtain a general expression for the action and the Lax pair of the models based on the combinations of any number of Kac-Moody realisations obeying an ansatz of this form. We then apply these results to the case of a model constructed from Yang-Baxter realisations and/or $\lambda$-realisations. In this case, the particular form of these realisations allows for further simplifications of the action, which for instance lead to the action introduced above for the cases with $N$ Yang-Baxter realisations or $N$ $\lambda$-realisations.\\

As a side result, we comment in this article on the relation of the models constructed here with the 4d semi-holomorphic Chern-Simons theory. This theory was introduced in~\cite{Costello:2013zra} and was related to integrable systems and in particular integrable lattice models in~\cite{Costello:2013sla, Witten:2016spx, Costello:2017dso, Costello:2018gyb}. More recently, it was shown in~\cite{Costello:2019tri} how to generate integrable two-dimensional field theories from this four-dimensional theory (see also~\cite{Bittleston:2019gkq,Vicedo:2019dej,Delduc:2019whp} for further developments). The reference~\cite{Costello:2019tri} treated two different classes of models, corresponding to so-called order and disorder defects. In particular, the Principal Chiral Model with Wess-Zumino term \eqref{Eq:IntroPCM} and its coupled version \eqref{Eq:IntroCoupled} were obtained in this formalism as models with disorder defects. The canonical analysis of the general models with disorder defects was performed in a subsequent article~\cite{Vicedo:2019dej}, which showed in particular that all these integrable field theories are realisations of affine Gaudin models. Finally, it was shown in~\cite{Delduc:2019whp} how to obtain the Yang-Baxter model and the $\lambda$-model in this framework. It is thus natural to search for a construction of the deformed coupled $\s$-models considered here from the 4d semi-holomorphic Chern-Simons theory. In this article, we present this construction explicitly and relate it to the affine Gaudin model approach.\\

The plan of this article is the following. In Section \ref{Sec:Ham}, we explain the construction of the models in the Hamiltonian framework. More precisely, we first describe in details in Subsection \ref{Kac-Moody} the Kac-Moody realisations in $T^*G_0$ that serve as building blocks for this construction. We then proceed to construct the models as realisations of affine Gaudin models in Subsection \ref{The}. We go on to perform the inverse Legendre transform of these models in Section \ref{Sec:Lag}, constructing in particular their action and their Lagrangian Lax pair. The results of Sections \ref{Sec:Ham} and \ref{Sec:Lag} are obtained using the general ansatz for the Kac-Moody realisations mentioned above in this introduction. We then study the models obtained from combinations of Yang-Baxter realisations and $\lambda$-realisations in Section \ref{Sec:Examples}: in particular, we find a simple expression of the action of these field theories and show that the $\s$-models constructed in~\cite{Georgiou:2016urf,Georgiou:2017jfi,Georgiou:2018hpd,Georgiou:2018gpe} can be obtained as particular limits of the ones constructed in this section. Finally, in Section \ref{Sec:CS}, we explain the relation of this work with the 4d semi-holomorphic Chern-Simons theory. Some technical results are gathered in Appendices \ref{Identities1} and \ref{Lorentz}.

\section{Hamiltonian formulation}
\label{Sec:Ham}

In this section, we define the integrable field theories that we will consider in this article. These theories are constructed as realisations of affine Gaudin models (AGM), following the general terminology of \cite{Vicedo:2017cge,Delduc:2019bcl}, and as such are then naturally defined in the Hamiltonian formalism. As explained in \cite{Vicedo:2017cge,Delduc:2019bcl}, the basic building blocks for the construction of realisations of AGM are the so-called Takiff realisations. In this article, we will be interested in a particular class of such realisations, which are given concretely by a pair of Kac-Moody currents in a certain phase space. As these particular Kac-Moody realisations are the basic building blocks of the models we will consider, we will start by describing them in details in Subsection \ref{Kac-Moody}, before proceeding to the construction of the models themselves in Subsection \ref{The}. For conciseness, we will not reintroduce here the general formalism of AGM and Takiff realisations and refer to \cite{Vicedo:2017cge,Delduc:2019bcl} for the details.

\subsection[Kac-Moody realisations in $T^*G_0$]{Kac-Moody realisations in $\bm{T^*G_0}$}
\label{Kac-Moody}

\subsubsection[The phase space of canonical fields on $T^*G_0$]{The phase space of canonical fields on $\bm{T^*G_0}$}
\label{Thealgebra}

All the Kac-Moody realisations that we shall consider in this article are defined on the same phase space. Let us then begin by describing this phase space.

\paragraph{Conventions and notation.} Let us consider a finite-dimensional semi-simple real Lie algebra $\mathfrak{g}_0$. Let us also introduce the opposite of its Killing form $\kappa$, which is a non-degenerate bilinear form on $\mathfrak{g}_0$. We will denote a basis for $\mathfrak{g}_0$ by $(I_a)_{a \in \{1,\cdots,n\}}$ and its dual basis with respect to $\kappa$ by $(I^a)_{a \in \{1,\cdots,n\}}$. It is then possible to define the split quadratic Casimir of $\mathfrak{g}_0$ as the following element:
\begin{equation} \label{Casimir}
C_{\underline{\mathbf{12}}} = I_a \otimes I^a
\end{equation}
in $\mathfrak{g}_0 \otimes \mathfrak{g}_0$, which is independent of the choice of basis (here and in the following, we use the standard tensorial notations $\underline{\mathbf{i}}$).\\

The Lie algebra $\mathfrak{g}_0$ can be seen as the real form of a complex Lie algebra $\mathfrak{g}$, or, in other words, as the subalgebra of fixed points of an antilinear involutive automorphism $\tau$ of $\mathfrak{g}$. A basis for $\mathfrak{g}$ over $\mathbb{C}$ is then given by $(I^a)_{a \in \{1,\cdots,n\}}$. We note that the split quadratic Casimir \eqref{Casimir} of the algebra is real, in the sense that it satisfies
\begin{equation*}
\tau_{\underline{\mathbf{1}}}C_{\underline{\mathbf{12}}} = \tau_{\underline{\mathbf{2}}}C_{\underline{\mathbf{12}}} = C_{\underline{\mathbf{12}}}.
\end{equation*}

To conclude, let us also mention the fact that by choosing $\mathfrak{g}_0$ to be the compact form of $\mathfrak{g}$, the bilinear form $\kappa$ becomes a positive scalar product on $\mathfrak{g}_0$.

\paragraph{Canonical fields on $\bm{T^*G_0}$.} Let $G_0$ be a connected real Lie group with Lie algebra $\mathfrak{g}_0$. We will now consider fields taking values in the cotangent bundle $T^*G_0$ and depending on a space coordinate $x$ in a one dimensional space $\mathbb{D}$, which for us will be either the real line $\mathbb{R}$ or the circle $S^1$. These fields can be conveniently described in the following way.

Firstly, we note that multiplying by $p^{-1}$, it is always possible to send the cotangent space at a point $p \in G_0$ to the one at the identity $\text{Id} \in G_0$, which is just the dual $\mathfrak{g}^*_0$ of the Lie algebra $\mathfrak{g}_0$. As we supposed $\mathfrak{g}_0$ to be semi-simple, we then have a canonical isomorphism between $\mathfrak{g}^*_0$ and $\mathfrak{g}_0$ through the bilinear form $\kappa$. This further implies that also $T^*G_0$ and $G_0 \times \mathfrak{g}_0$ are isomorphic to each other. Hence, it is possible to describe a field on $T^*G_0$ by a pair of fields $g(x)$ in $G_0$ and $X(x)$ in $\mathfrak{g}_0$.\\

Now, as $T^*G_0$ is a cotangent bundle, it possesses a canonical symplectic structure. This means that the space of fields on $T^*G_0$ comes naturally equipped with a Poisson bracket, which makes it the phase space describing the physical observables of an Hamiltonian field theory\footnote{In mathematical terms, these observables correspond more precisely to the algebra of functionals on this phase space, which has the structure of a Poisson algebra. In particular, these observables include all the possible local combinations of the canonical fields and their derivatives and the integrals over $\mathbb{D}$ of such combinations.}. In terms of the fields $g$ and $X$, this Poisson bracket can be written as
\begin{subequations} \label{PoissongX}
\begin{align}
\{g_{\underline{\mathbf{1}}}(x),g_{\underline{\mathbf{2}}}(y)\} &= 0, \\
\{X_{\underline{\mathbf{1}}}(x),g_{\underline{\mathbf{2}}}(y)\} &= g_{\underline{\mathbf{2}}}(x) C_{\underline{\mathbf{12}}}\delta_{xy}, \\
\{X_{\underline{\mathbf{1}}}(x),X_{\underline{\mathbf{2}}}(y)\} &=  [C_{\underline{\mathbf{12}}},X_{\underline{\mathbf{1}}}(x)]\delta_{xy},
\end{align}
\end{subequations}
where $C_{\underline{\mathbf{12}}}$ is the split quadratic Casimir of $\mathfrak{g}_0$ and $\delta_{xy} = \delta(x-y)$ is the Dirac delta-distribution.

\paragraph{The current $\bm{j(x)}$ and the momentum.} Let us now define the following $\mathfrak{g}_0$-valued current:
\begin{equation*}
j(x) = g^{-1}(x)\partial_xg(x),
\end{equation*}
which, from \eqref{PoissongX}, satisfies the Poisson brackets
\begin{subequations}  \label{Poissonj}
\begin{align}
\{g_{\underline{\mathbf{1}}}(x),j_{\underline{\mathbf{2}}}(y)\} &= 0, \\
\{j_{\underline{\mathbf{1}}}(x),j_{\underline{\mathbf{2}}}(y)\} &= 0, \\
\{X_{\underline{\mathbf{1}}}(x),j_{\underline{\mathbf{2}}}(y)\} &=  [C_{\underline{\mathbf{12}}},j_{\underline{\mathbf{1}}}(x)]\delta_{xy} -C_{\underline{\mathbf{12}}}\delta'_{xy}.
\end{align}
\end{subequations}
Let us also consider the quantity
\begin{equation} \label{Momentum}
\mathcal{P}_{G_0} = \int_{\mathbb{D}} \text{d}x \ \kappa(j(x),X(x)).
\end{equation}
From \eqref{PoissongX} and \eqref{Poissonj}, one can check that its Hamiltonian flow generates the spatial derivatives on both $g(x)$ and $X(x)$:
\begin{equation*}
\{\mathcal{P}_{G_0},g(x)\} = \partial_x g(x) \hspace{30pt} \text{and} \hspace{30pt} \{\mathcal{P}_{G_0},X(x)\} = \partial_x X(x).
\end{equation*}
Hence, this is nothing but the momentum of the phase space of canonical fields on $T^*G_0$.

\paragraph{The current $\bm{W(x)}$ and the Wess-Zumino term.} As shown for example in \cite{Delduc:2019bcl}, it is also possible to define another $\mathfrak{g}_0$-valued current $W(x)$ with Poisson brackets
\begin{equation}\label{PoissonW1}
\{g_{\underline{\mathbf{1}}}(x),W_{\underline{\mathbf{2}}}(y)\} = 0, \hspace{40pt} \{j_{\underline{\mathbf{1}}}(x),W_{\underline{\mathbf{2}}}(y)\} = 0
\end{equation}
and
\begin{equation} \label{PoissonW}
\{X_{\underline{\mathbf{1}}}(x),W_{\underline{\mathbf{2}}}(y)\} + \{W_{\underline{\mathbf{1}}}(x),X_{\underline{\mathbf{2}}}(y)\} =  [C_{\underline{\mathbf{12}}},W_{\underline{\mathbf{1}}}(x) - j_{\underline{\mathbf{1}}}(x)]\delta_{xy}.
\end{equation}
However, in this article, we shall not need the precise definition of $W(x)$ and thus we refer to \cite{Delduc:2019bcl} for details. A further property of this current is that it satisfies the following orthogonality relation:
\begin{equation} \label{Orthogonality}
\kappa\bigl(j(x),W(x)\bigr) = 0.
\end{equation}

As a final remark, we note that through this current it is possible to define the Wess-Zumino term of $g$~\cite{Wess:1971yu,Novikov:1982ei,Witten:1983ar}. Indeed, briefly considering the field $g$ to be dependent on a time coordinate $t \in \mathbb{R}$ (in the Hamiltonian formulation, this time dependence is implicitly defined by the choice of a Hamiltonian), the Wess-Zumino term of $g$ is given by (see for instance \cite{Delduc:2019bcl})
\begin{equation} \label{Wesszumino}
I_{\text{WZ}}[g] = \iint_{\mathbb{R} \times \mathbb{D}} \text{d}t\,\text{d}x \ \kappa(W,g^{-1}\partial_tg).
\end{equation}

\subsubsection{Kac-Moody currents}
\label{Kac-Moodycurrents}

\paragraph{Commuting Kac-Moody currents.} We are now in a position to introduce the Kac-Moody realisations that will serve as basic building blocks for the construction of the integrable models of Subsection \ref{The}. Such realisations are characterised by two commuting Kac-Moody currents on the phase space of fields on $T^*G_0$, \textit{i.e.} two $\mathfrak{g}$-valued fields $\mathcal{J}_\pm(x)$ satisfying the Poisson brackets
\begin{subequations} \label{PoissonJ}
\begin{align}
\{\mathcal{J}_{\pm \underline{\mathbf{1}}}(x),\mathcal{J}_{\pm \underline{\mathbf{2}}}(y)\} &= [C_{\underline{\mathbf{12}}},\mathcal{J}_{\pm \underline{\mathbf{1}}}(x)]\delta_{xy} - \ell^\pm C_{\underline{\mathbf{12}}} \delta'_{xy}, \\
\{\mathcal{J}_{\pm \underline{\mathbf{1}}}(x),\mathcal{J}_{\mp \underline{\mathbf{2}}}(y)\} &= 0,
\end{align}
\end{subequations}
where $\ell^\pm$ are constant numbers called the levels. Currents of such kind have already been found to play an important role in the study of integrable deformations of $\sigma$-models~\cite{Delduc:2013fga,Delduc:2014uaa,Hollowood:2014rla,Vicedo:2015pna}, leading to examples of Kac-Moody realisations such as the Yang-Baxter realisation (with or without Wess-Zumino term) and the $\lambda$-realisation~\cite{Delduc:2019bcl}. These examples will be described more in detail in Subsection \ref{Examples}. For the time being, we focus on aspects which are common to all the realisations we shall describe, in order to keep the treatment as general and uniform as possible.

In particular, in all the examples we shall consider, the Kac-Moody currents $\mathcal{J}_\pm(x)$ are expressed as linear combinations of the $\mathfrak{g}_0$-valued currents $X(x)$, $j(x)$ and $W(x)$ introduced in Subsection \ref{Thealgebra}. Moreover, the currents $X(x)$ and $W(x)$ always appear through the unique combination
\begin{equation*}
Y(x) = X(x) - \kay\, W(x),
\end{equation*}
for some real constant $\kay$ which depends on the particular realisation. As one can see from \eqref{Wesszumino}, the current $W$ is related to the Wess-Zumino term of the corresponding field $g$. Because of this relation, and as we will see more precisely in Subsection \ref{Legendre}, the presence of the current $W$ in the realisation, \textit{i.e.} the non-vanishing of $\kay$, will lead to the presence of a corresponding Wess-Zumino term in the action of the model.

From now on, we will suppose that the Kac-Moody currents $\mathcal{J}_\pm(x)$ take the form
\begin{equation} \label{Currents}
\mathcal{J}_\pm(x) = \A_\pm Y(x) + \B_\pm j(x),
\end{equation}
where $\A_\pm, \B_\pm : \mathfrak{g} \to \mathfrak{g}$ are linear operators on the Lie algebra $\mathfrak{g}$. We will allow these operators to be dynamical (and thus have non-trivial Poisson brackets with other quantities in the phase space), but will suppose them to depend only on the field $g$ (that is, not on $X$ or derivatives of $g$). As we shall see in Subsection \ref{Examples}, both the Yang-Baxter realisation and the $\lambda$-realisation can be retrieved in this formalism by making some specific choices for the operators $\A_\pm$ and $\B_\pm$.

Let us note that, in general, these operators cannot be arbitrary. Indeed, they should be chosen such that the currents \eqref{Currents} satisfy the brackets \eqref{PoissonJ}. We will not try to write here the most general conditions on $\A_\pm$ and $\B_\pm$ for these brackets to hold. However, as explained in details in Appendix \ref{Identities1}, one can already obtain some useful constraints on these operators by focusing on the non-ultralocal terms in the brackets \eqref{PoissonJ}, \textit{i.e.} terms proportional to the derivative of the Dirac distribution.  More precisely, one finds that $\A_\pm$ and $\B_\pm$ should satisfy the following identities:
\begin{subequations} \label{Identities}
\begin{align}
\A_\pm {^t}\B_\pm + \B_\pm {^t}\A_\pm &= \ell^\pm \text{Id}, \\
\A_\pm {^t}\B_\mp + \B_\pm {^t}\A_\mp &= 0,
\end{align}
\end{subequations}
where we have introduced the transpose ${^t}\mathcal{O}$ with respect to the form $\kappa$ for an operator $\mathcal{O}$ on the Lie algebra $\mathfrak{g}$.

\paragraph{Reality conditions.} In order for the models that we will construct from these realisations to be real, one has to impose some reality conditions on both the currents $\mathcal{J}_\pm$ and the levels $\ell^\pm$. There are two possible types of conditions that we shall consider. In the first case, we suppose that the currents are invariant under the antilinear involutive automorphism $\tau$ (\textit{i.e.} they are $\mathfrak{g}_0$-valued) and the corresponding levels are real:
\begin{equation} \label{Realreal}
\tau(\mathcal{J}_\pm(x)) = \mathcal{J}_\pm(x)  \hspace{30pt} \text{and} \hspace{30pt}  \overline{\ell^\pm} = \ell^\pm.
\end{equation}
In the second case, one requires the currents to be conjugate with respect to $\tau$ and the levels to be complex conjugate to each other:
\begin{equation} \label{Realconjugate}
\tau(\mathcal{J}_\pm(x)) = \mathcal{J}_\mp(x) \hspace{30pt} \text{and} \hspace{30pt} \overline{\ell^\pm} = \ell^\mp.
\end{equation}

\paragraph{Momentum and suitability of the realisation.} To conclude this section, we will now prove that all the realisations that we are considering here are suitable (in the language of \cite{Delduc:2019bcl}). In particular, this will later allow a simple characterisation of Lorentz invariance for the integrable models that we will build from them. 

To start with, it is simple to check that from the relations \eqref{Identities} obeyed by the operators $\A_\pm$ and $\B_\pm$, one can derive the following additional identities:
\begin{subequations}\label{Eq:Identities2}
\begin{align}
^t\A_+\A_+/\ell^+ + {^t}\A_-\A_-/\ell^- &= 0, \\
^t\B_+\B_+/\ell^+ + {^t}\B_-\B_-/\ell^- &= 0, \\
^t\A_+\B_+/\ell^+ + {^t}\A_-\B_-/\ell^- &= \text{Id}.
\end{align}
\end{subequations}
These, together with the definition of the currents \eqref{Currents} above and the identity \eqref{Orthogonality}, allow one to prove that the momentum \eqref{Momentum} can be re-expressed as
\begin{equation*}
\mathcal{P}_{G_0} = \frac{1}{2\ell^+}\int_{\mathbb{D}} \text{d}x \ \kappa(\mathcal{J}_+(x),\mathcal{J}_+(x)) + \frac{1}{2\ell^-}\int_{\mathbb{D}} \text{d}x \ \kappa(\mathcal{J}_-(x),\mathcal{J}_-(x)).
\end{equation*}
From \cite{Vicedo:2017cge,Delduc:2019bcl}, one recognises on the right-hand side the Segal-Sugawara integrals of the Kac-Moody realisation. This implies that the realisations described above are indeed suitable.

\subsubsection{Examples of realisations}
\label{Examples}

\noindent We will now review some relevant examples of Kac-Moody realisations.

\paragraph{Inhomogeneous Yang-Baxter realisation without Wess-Zumino term.} Let us start by considering a solution $R: \mathfrak{g}_0 \to \mathfrak{g}_0$ of the modified classical Yang-Baxter equation (mCYBE):
\begin{equation}\label{Eq:mCYBE}
[RX,RY] -R([RX,Y]+[X,RY]) = -c^2[X,Y], \;\;\;\;\;\; \forall \, X,Y \in \mathfrak{g}_0,
\end{equation}
with $c=1$ (so-called split case) or $c=i$ (non-split case), which we suppose to be skew-symmetric with respect to the non-degenerate form $\kappa$:
\begin{equation*}
\kappa(RX,Y) = -\kappa(X,RY), \;\;\;\;\;\; \forall\, X,Y \in \mathfrak{g}_0.
\end{equation*}
The Kac-Moody currents for the inhomogeneous Yang-Baxter realisation without Wess-Zumino term then read \cite{Delduc:2013fga,Vicedo:2015pna,Delduc:2019bcl}
\begin{equation} \label{CurrentsYB}
\mathcal{J}_\pm = \left(\frac{1}{2}\text{Id} \mp \frac{1}{2c} R_{g}\right)X \pm \frac{1}{2c\gamma} j,
\end{equation}
where $\gamma$ is a real constant and
\begin{equation*}
R_g = \Ad_g^{-1} \circ R \circ \Ad_g.
\end{equation*}
The proof that these are Kac-Moody currents can be found in \cite{Delduc:2013fga}, where the levels are found to be
\begin{equation}\label{Eq:LevelsYB}
\ell^\pm = \pm \frac{1}{2c\gamma}.
\end{equation}
Note in particular that the levels $\ell^\pm$ are opposite to one another.

Moreover, the reality conditions discussed in Subsection \ref{Kac-Moodycurrents} are satisfied. In particular, in the split case ($c=1$) the currents $\mathcal{J}_\pm$ are $\mathfrak{g}_0$-valued and the levels $\ell_\pm$ are real, hence \eqref{Realreal} is satisfied. In the non-split case ($c=i$) instead, it is a simple check that the currents and the levels satisfy \eqref{Realconjugate}.

In the general language of Subsection \ref{Kac-Moodycurrents}, we see that the current $W$ does not appear in the expression \eqref{CurrentsYB}, which means that for this realisation we take the coefficient $\kay$ in \eqref{Currents} to be zero. According to what has been discussed in the previous subsection, this justifies the fact that the models constructed from this realisation will not contain the Wess-Zumino term of $g$. Finally, by comparing with \eqref{Currents} we read for the operators $\A_\pm$ and $\B_\pm$:
\begin{equation}\label{Eq:OperatorsYB}
\A_\pm = \frac{1}{2}\text{Id} \mp \frac{1}{2c}R_{g} \hspace{30pt} \text{and} \hspace{30pt} \B_\pm = \ell^\pm \text{Id}.
\end{equation}
One easily checks that these operators satisfy the identities \eqref{Identities}, as expected.

\paragraph{Inhomogeneous Yang-Baxter realisation with Wess-Zumino term.} The inhomogeneous Yang-Baxter realisation defined in the previous paragraph has no Wess-Zumino term, \textit{i.e.} does not contain the current $W(x)$ (or equivalently has $\kay=0$). Following~\cite{Delduc:2014uaa}, one can generalise this construction to include the current $W(x)$ and thus a non-zero coefficient $\kay$, at least when the $R$-matrix underlying the realisation satisfies the additional condition $R^3=c^2 R$, with $c$ as in the right-hand side of the mCYBE \eqref{Eq:mCYBE} (note in particular that the standard Drinfeld-Jimbo $R$-matrix satisfies this condition). The levels of this generalised realisation are given by
\begin{equation*}
\ell^\pm = \pm \frac{1}{2c\gamma} - \kay,
\end{equation*}
with $\gamma$ a real constant. Comparing to the levels \eqref{Eq:LevelsYB} of the realisation without Wess-Zumino term, one sees that turning on the coefficient $\kay$ corresponds to relaxing the fact that the levels $\ell^\pm$ are opposite one to another.

The Kac-Moody currents of the inhomogeneous Yang-Baxter realisation with Wess-Zumino term can be computed from the results of~\cite[Section 3]{Delduc:2014uaa}, up to a few differences in the conventions\footnote{For completeness, note that this reference only treats the non-split case $c=i$. The results generalise straightforwardly to the split case $c=1$.}. In the present notations, they read\vspace{-4pt}
\begin{equation*}
\mathcal{J}_\pm = \left(\frac{1}{2}\text{Id} \mp \frac{1}{2c}R_{g} \mp \frac{\delta}{2}\Pi_{g}\right) Y + \left(\left(\pm\frac{1}{2c\gamma} - \frac{\kay}{2}\right)\text{Id} \mp \frac{\kay}{2c} R_g \mp \frac{\kay\delta}{2}\Pi_{g}\right)j,
\end{equation*}
where we recall that $Y=X-\kay W$ and where we have defined the quantities
\begin{equation*}
\Pi_{g} = 1-\frac{R_g^2}{c^2} \;\;\;\;\;\;\;\; \text{ and } \;\;\;\;\;\;\;\; \delta = \frac{1-\sqrt{1-4c^2\kay^2\gamma^2}}{2c\kay\gamma}.
\end{equation*}
Similarly to the case without Wess-Zumino term, it is simple to check that the reality conditions are satisfied for both the choices $c=1$ and $c=i$. From the form of the currents, one can make the following identifications comparing to Equation \eqref{Currents}:\vspace{-4pt}
\begin{equation}\label{Eq:OperatorsYBWZ}
\A_\pm = \frac{1}{2}\text{Id} \mp \frac{1}{2c}R_{g} \mp \frac{\delta}{2}\Pi_{g} \hspace{30pt} \text{and} \hspace{30pt} \B_\pm = \left(\ell^\pm + \frac{\kay}{2}\right)\text{Id} \mp \frac{\kay}{2c} R_g \mp \frac{\kay\delta}{2}\Pi_{g}.
\end{equation}
Let us note that, as expected, the identities \eqref{Identities} are again satisfied by these operators $\A_\pm$ and $\B_\pm$ (using the fact that we restrict here to $R$-matrices satisfying $R^3=c^2R$).

\paragraph{$\bm\lambda$-realisation.} For the $\lambda$-realisation, the Kac-Moody currents are given by \cite{Hollowood:2014rla,Vicedo:2015pna,Delduc:2019bcl}:
\begin{align*}
\mathcal{J}_+ &= X - \kay W - \kay j = Y- \kay j, \\
\mathcal{J}_- &= -\Ad_g(X - \kay W + \kay j) = -\Ad_g(Y + \kay j),
\end{align*}
with levels 
\begin{equation*}
\ell^\pm = \mp 2\kay.
\end{equation*}
Note that, similarly to the inhomogeneous Yang-Baxter realisation without Wess-Zumino term, these levels $\ell^\pm$ are opposite one to another. In this case, the reality condition \eqref{Realreal} is satisfied, since the currents $\mathcal{J}_\pm$ are $\mathfrak{g}_0$-valued and the levels $\ell^\pm$ are real.

To conclude, comparing to Equation \eqref{Currents}, one sees that for the $\lambda$-realisation the operators $\A_\pm$ and $\B_\pm$ have the following form:
\begin{equation}\label{Eq:OperatorsLambda}
\A_+ = \text{Id}, \hspace{20pt} \A_- = -\Ad_g, \hspace{20pt} \B_+ = -\kay\,\text{Id}, \hspace{20pt} \B_- = -\kay \,\Ad_g,
\end{equation}
and again one can check that the identities \eqref{Identities} are satisfied.

\subsection{Construction of the models}
\label{The}

\subsubsection{Definition as realisations of affine Gaudin models}
\label{DefGaudin}

\paragraph{Sites, levels and twist function.} In this subsection, we proceed to constructing the integrable field theories that we will consider in this article as realisations of AGM, following the general formalism and terminology of \cite{Vicedo:2017cge,Delduc:2019bcl}. As AGM, the models that we will consider possess $2N$ sites of multiplicity one, which we gather in pairs $(r,+)$ and $(r,-)$ with $r \in \{1,\cdots,N\}$. The position of the site $(r,\pm)$ in the complex plane $\mathbb{C}$ will be denoted by $z_r^\pm$. Since each site $(r,\pm)$ is of multiplicity one, it is associated with one level, which is a non-zero constant number and which we will denote by $\ell_r^\pm$. Following \cite{Vicedo:2017cge,Delduc:2019bcl}, let us also fix a non-zero real number $\ell^\infty$. Altogether, this data specifies the so-called twist function of the AGM, which in the present case reads
\begin{equation}\label{Eq:Twist}
\varphi(z) = \sum_{r=1}^N\left(\frac{\ell_r^+}{z-z_r^+} + \frac{\ell_r^-}{z-z_r^-}\right) - \ell^\infty,
\end{equation}
where $z \in \mathbb{C}$ is an auxiliary complex parameter, called the spectral parameter.

\paragraph{Kac-Moody currents and phase space.} To each site $(r,\pm)$ we attach a $\mathfrak{g}$-valued field $\mathcal{J}^{(r)}_\pm(x)$ in the phase space of the model, which we now describe. As explained in \cite{Vicedo:2017cge,Delduc:2019bcl}, the Poisson brackets of these fields are specified by the choice of levels $\ell_r^\pm$ made above. More precisely, we have the following:
\begin{subequations}\label{Eq:PBKM}
\begin{align}
\bigl\{\mathcal{J}^{(r)}_{\pm \underline{\mathbf{1}}}(x),\mathcal{J}^{(s)}_{\pm \underline{\mathbf{2}}}(y)\bigr\} &= \delta_{rs}\left([C_{\underline{\mathbf{12}}},\mathcal{J}^{(r)}_{\pm \underline{\mathbf{1}}}(x)]\delta_{xy} - \ell_r^\pm C_{\underline{\mathbf{12}}} \delta'_{xy}\right), \\
\bigl\{\mathcal{J}^{(r)}_{\pm \underline{\mathbf{1}}}(x),\mathcal{J}^{(s)}_{\mp \underline{\mathbf{2}}}(y)\bigr\} &= 0.
\end{align}
\end{subequations}
Thus, the models that we consider are constructed from $N$ independent pairs of commuting Kac-Moody currents $(\mathcal{J}_+^{(r)},\mathcal{J}_-^{(r)})$, $r \in \{1,\cdots,N\}$. We have described in detail in Subsection \ref{Kac-Moody} how such a pair can be realised in the phase space of canonical fields on $T^*G_0$. A natural way to realise the $2N$ currents $\mathcal{J}_\pm^{(r)}$ is then to consider $N$ independent realisations in $T^*G_0$ of the type described in Subsection \ref{Kac-Moody}. Concretely, this means that we choose the phase space of the model to be the space of fields on the product $T^*G_0^N$, with the currents $\mathcal{J}_\pm^{(r)}$ depending on the fields in the $r^{\text{th}}$-factor of $T^*G_0^N$.

This $r^{\text{th}}$- factor is described by a pair of canonical fields $g^{(r)}(x)$ and $X^{(r)}(x)$, valued respectively in the group $G_0$ and the Lie algebra $\mathfrak{g}_0$, which are the equivalent of the fields $g(x)$ and $X(x)$ introduced in Subsection \ref{Thealgebra} in order to describe one copy of $T^*G_0$. Similarly, one can define from these canonical fields the equivalent of the currents $j(x)$ and $W(x)$, which we shall denote by $j^{(r)}(x)$ and $W^{(r)}(x)$. Following the discussion above, we then also define the currents $\mathcal{J}_\pm^{(r)}$ as the analogues for the $r^{\text{th}}$-factor of the Kac-Moody currents $\mathcal{J}_\pm$ described in Subsection \ref{Kac-Moodycurrents}. Therefore, they take the form
\begin{equation}\label{Currentsr}
\mathcal{J}^{(r)}_\pm(x) = \A^{(r)}_\pm Y^{(r)}(x) + \B^{(r)}_\pm j^{(r)}(x),
\end{equation}
where
\begin{equation}\label{Eq:Yr}
Y^{(r)}(x) = X^{(r)}(x) - \kay_r W^{(r)}(x)
\end{equation}
and $\kay_r$ is a real constant number depending on the choice of realisation in the $r^{\text{th}}$-copy. The $\A^{(r)}_\pm$'s and $\B^{(r)}_\pm$'s are linear operators on the Lie algebra $\mathfrak{g}$, which are the equivalent of the operators $\A_\pm$ and $\B_\pm$ introduced in Subsection \ref{Kac-Moodycurrents}. In particular, they depend only on $g^{(r)}$ and satisfy analogous identities to the ones of equation \eqref{Identities}.

\paragraph{Gaudin Lax matrix.} We are now in a position to define the remaining building block that, in the next section, will allow us to write down an Hamiltonian for the model. This is the so-called Gaudin Lax matrix of the model, which we define as the following $\mathfrak{g}$-valued field~\cite{Vicedo:2017cge,Delduc:2019bcl}:
\begin{equation} \label{GaudinL}
\Gamma(z,x) = \sum_{r=1}^N\left(\frac{\mathcal{J}^{(r)}_+(x)}{z-z_r^+} + \frac{\mathcal{J}^{(r)}_-(x)}{z-z_r^-}\right).
\end{equation}

\paragraph{Reality conditions.} As we discussed in Subsection \ref{Kac-Moodycurrents}, in order for the models which we construct in this article to be real, we have to impose some reality conditions. For each pair of sites $(r,\pm)$, there are two cases. In the first one, we suppose the positions of the two sites $z_r^\pm$ to be real and that the condition \eqref{Realreal} on the currents $\mathcal{J}_\pm^{(r)}$ and the levels $\ell_r^\pm$ holds. In the second case, we assume instead that the the positions of the sites are complex conjugate to each other and that the currents and levels satisfy the condition \eqref{Realconjugate}.

These conditions can be summarised in terms of the twist function and the Gaudin Lax matrix as the following equivariance relations:
\begin{equation*}
\tau(\Gamma(z,x)) = \Gamma(\bar{z},x) \hspace{30pt} \text{ and } \hspace{30pt} \overline{\varphi(z)} = \varphi(\bar{z}).
\end{equation*}

\subsubsection{Hamiltonian and momentum}
\label{SubSubSec:Dyn}

\paragraph{Hamiltonian.} In order to construct the Hamiltonian of the model, we start by rewriting the twist function in terms of its zeroes $\zeta_i$ ($i\in\lbrace 1,\cdots,2N\rbrace$), which, for future convenience, we will suppose to be real and distinct. As we assumed $\ell^\infty$ to be non-zero, we can thus rewrite the twist function as
\begin{equation}\label{Eq:TwistZeroes}
\varphi(z) = -\ell^{\infty}\frac{\prod_{i=1}^{2N}{(z-\zeta_i)}}{\prod_{s=1}^N{(z-z_s^+)(z-z_s^-)}}.
\end{equation}
Let us consider the spectral parameter dependent local charge
\begin{equation*}
\mathcal{Q}(z) = -\frac{1}{2\varphi(z)} \int_{\mathbb{D}} \text{d}x \ \kappa(\Gamma(z,x),\Gamma(z,x))
\end{equation*}
and define, for $i = 1,\cdots,2N$,
\begin{equation*}
\mathcal{Q}_i = \underset{z = \zeta_i}{\operatorname{res}} \mathcal{Q}(z) \text{d}z
\end{equation*}
or, more explicitly,
\begin{equation} \label{Charges}
\mathcal{Q}_i =-\frac{1}{2\varphi'(\zeta_i)} \int_{\mathbb{D}} \text{d}x \ \kappa(\Gamma(\zeta_i,x),\Gamma(\zeta_i,x)).
\end{equation}
These are local charges quadratic in the currents $\mathcal{J}^{(r)}_\pm$ which, as proven in \cite{Delduc:2019bcl}, satisfy $\{Q_i,Q_j\} = 0$ for all $i$ and $j$. We define the Hamiltonian of the model to be the linear combination
\begin{equation} \label{Hamiltonian}
\mathcal{H} = \sum_{i=1}^{2N} \epsilon_i \mathcal{Q}_i,
\end{equation}
for some real numbers $\epsilon_i$. This then generates the time evolution of the model through the Hamiltonian flow
\begin{equation*}
\partial_t = \{\mathcal{H}, \cdot \}.
\end{equation*}
Note that, as a consequence of the reality conditions we introduced, $\mathcal{H}$ is real \cite{Delduc:2019bcl}.

\paragraph{Momentum and relativistic invariance.} Recall that, in Subsection \ref{Kac-Moodycurrents}, we proved that the Kac-Moody realisations in $T^*G_0$ that we are considering are suitable. According to \cite{Delduc:2019bcl}, this gives some additional information on the space-time properties of the model. Firstly, its momentum (\textit{i.e.} the generator of spatial translations with respect to $x$) is given by the following expression:
\begin{equation}\label{Eq:Momentum}
\mathcal{P} = \sum_{i=1}^{2N} \mathcal{Q}_i.
\end{equation}
Secondly, requiring relativistic invariance of the model restricts the choice of the coefficients $\epsilon_i$ in the definition of $\mathcal{H}$ to
\begin{equation*}
\epsilon_i = + 1 \hspace{15pt} \text{ or } \hspace{15pt} \epsilon_i = -1,
\end{equation*}
for every $i\in\lbrace 1,\cdots,2N\rbrace$. We then see that there is a natural division of the indices $i\in\lbrace 1,\cdots,2N\rbrace$ labelling the zeroes $\zeta_i$ into the sets $\mathcal{I}_\pm = \{i \mid \epsilon_i = \pm 1\}$. In the rest of this article, we will suppose that there are as many $\epsilon_i$'s equal to $+1$ as $\epsilon_i$'s equal to $-1$ (\textit{i.e.} that the sets $\mathcal{I}_\pm$ are both of size $|\mathcal{I}_+|=|\mathcal{I}_-|=N$)\footnote{As first observed in~\cite{Delduc:2019bcl}, the models obtained when choosing $\mathcal{I}_+$ and $\mathcal{I}_-$ of different sizes would not possess a well-defined inverse Legendre transform.}.

\subsubsection{Integrability}
\label{SubSubSec:Int}

\paragraph{Lax pair and zero curvature equation.} We define the Lax matrix of the model to be the following $\mathfrak{g}$-valued field~\cite{Vicedo:2017cge,Delduc:2019bcl}:
\begin{equation}\label{Eq:DefLax}
\mathcal{L}(z,x) = \frac{\Gamma(z,x)}{\varphi(z)}.
\end{equation}
By construction, it has poles at the zeroes $\zeta_i$ of the twist function. More precisely, it can be rewritten as \cite{Delduc:2019bcl}
\begin{equation}\label{L}
\mathcal{L}(z,x) = \sum_{i=1}^{2N} \frac{1}{\varphi'(\zeta_i)} \frac{\Gamma(\zeta_i,x)}{z-\zeta_i}.
\end{equation}
From here, one can check that the time evolution of $\mathcal{L}(z,x)$ takes the form of a zero curvature equation,
\begin{equation}\label{Zerocurvature}
\partial_t\mathcal{L}(z,x) - \partial_x\mathcal{M}(z,x) + [\mathcal{M}(z,x),\mathcal{L}(z,x)] = 0,
\end{equation}
where $\mathcal{M}(z,x)$ is defined as
\begin{equation}\label{M}
\mathcal{M}(z,x) = \sum_{i=1}^{2N} \frac{\epsilon_i}{\varphi'(\zeta_i)} \frac{\Gamma(\zeta_i,x)}{z-\zeta_i}.
\end{equation}
Therefore, the model admits a Lax pair representation with Lax pair $(\mathcal{L},\mathcal{M})$.

\paragraph{Maillet bracket and integrability.} The fact that the equations of motion of the model take the form of a zero curvature equation allows one to extract an infinite number of charges from the monodromy of the Lax matrix $\mathcal{L}(z,x)$. The integrability of the model then follows from the fact that these are in involution, which is a consequence of the Lax matrix satisfying the following Poisson bracket:
\begin{multline}
\label{PoissonL}
 \ \ \ \ \ \ \ \ \{\mathcal{L}_{\underline{\mathbf{1}}}(z,x),\mathcal{L}_{\underline{\mathbf{2}}}(w,y)\} = [\mathcal{R}_{\underline{\mathbf{12}}}(z,w),\mathcal{L}_{\underline{\mathbf{1}}}(z,x)] \delta_{xy} - [\mathcal{R}_{\underline{\mathbf{21}}}(w,z),\mathcal{L}_{\underline{\mathbf{2}}}(w,x)] \delta_{xy} \\ 
- (\mathcal{R}_{\underline{\mathbf{12}}}(z,w) + \mathcal{R}_{\underline{\mathbf{21}}}(w,z)) \delta'_{xy}, \ \ \ \ \ \ \ \ \  \ \ \ \ \ \ \ \  \ \ \ \ \ \ \ \  \ \ \ \ \ \ \ \  
\end{multline}
where the $\mathcal{R}$-matrix is defined to be
\begin{equation*}
\mathcal{R}_{\underline{\mathbf{12}}}(z,w) = \frac{C_{\underline{\mathbf{12}}}}{w-z}\varphi(w)^{-1}.
\end{equation*}
The bracket \eqref{PoissonL} is an example of a Maillet non-ultralocal bracket \cite{Maillet:1985fn,Maillet:1985ek}. One can check that it satisfies the Jacobi identity due to the fact that the $\mathcal{R}$-matrix is a solution of the classical Yang-Baxter equation
\begin{equation*}
[\mathcal{R}_{\underline{\mathbf{12}}}(z_1,z_2),\mathcal{R}_{\underline{\mathbf{13}}}(z_1,z_3)] + [\mathcal{R}_{\underline{\mathbf{12}}}(z_1,z_2),\mathcal{R}_{\underline{\mathbf{23}}}(z_2,z_3)] + [\mathcal{R}_{\underline{\mathbf{32}}}(z_3,z_2),\mathcal{R}_{\underline{\mathbf{13}}}(z_1,z_3)] = 0.
\end{equation*}

\paragraph{Lax pair in light-cone coordinates.} As we will need this in Section \ref{Sec:Lag}, let us briefly discuss the reparametrisation of the Lax pair in light-cone components. Let us firstly introduce the light-cone coordinates $x^\pm = (t \pm x)/2$ and the corresponding derivatives $\partial_\pm = \partial_t \pm \partial_x$. The zero curvature equation \eqref{Zerocurvature} can then be rewritten as
\begin{equation*}
\partial_+\mathcal{L}_-(z,x) - \partial_-\mathcal{L}_+(z,x) + [\mathcal{L}_+(z,x),\mathcal{L}_-(z,x)] = 0,
\end{equation*}
where we have introduced the light-cone Lax pair
\begin{equation*}
\mathcal{L}_\pm(z,x) = \mathcal{L}(z,x) \pm \mathcal{M}(z,x).
\end{equation*}
Finally, from Equations \eqref{L} and \eqref{M}, one finds the following expression for $\Lc_\pm(z,x)$: 
\begin{equation} \label{Laxlight}
\mathcal{L}_\pm(z,x) = \pm2\sum_{i\in\mathcal{I}_\pm} \frac{1}{\varphi'(\zeta_i)} \frac{\Gamma(\zeta_i,x)}{z-\zeta_i},
\end{equation}
in terms of the split of the zeroes into the two sets $\mathcal{I}_\pm$ introduced in Subsection \ref{SubSubSec:Dyn}.

\subsubsection{Exploring the ``space of models''}
\label{SubSubSec:Space}

\paragraph{Gaudin parameters.} Let us describe the ``space of models'' that we are considering in this article by summarising what are the defining parameters of the integrable field theories that we have constructed so far.  As affine Gaudin models, these theories are characterised by the following quantities, that we shall refer to as Gaudin parameters:
\vspace{-4pt}
\begin{itemize}\setlength\itemsep{0.05em}
\item the positions $z^\pm_r$ ;
\item the levels $\ell^\pm_r$ ;
\item the constant term $\ell^\infty$ in the twist function ;
\item the Kac-Moody realisations with levels $\ell^\pm_r$ attached to each pair of sites $(r,\pm)$.\vspace{-4pt}
\end{itemize}

As explained in~\cite[Subsection 1.4.2]{Delduc:2019bcl}, there exists a redundancy between the Gaudin parameters of the model, corresponding to the freedom of translating and dilating the spectral parameter. Indeed, the model with parameters $z^\pm_r$, $\ell^\pm_r$ and $\ell^\infty$ as above is invariant under the transformation
\begin{equation}\label{Eq:Redundancy}
z^\pm_r \longmapsto a z^\pm_r + b \;\;\;\;\;\;\;\; \text{and} \;\;\;\;\;\;\;\;  \ell^\infty \longmapsto a^{-1} \ell^\infty,
\end{equation} 
where $a$ and $b$ are real numbers with $a\neq 0$ and where we keep the levels $\ell^\pm_r$ and the Kac-Moody realisations fixed. Note that one can fix the dilation redundancy (corresponding to the parameter $a$ in the transformation above) by setting the constant term $\ell^\infty$ to a specific value. Similarly, one can fix the translation redundancy (corresponding to the parameter $b$) by setting one of the positions $z^\pm_r$ to a specific point.\\

Note that the Gaudin parameters introduced above are in general not all real but should satisfy the reality conditions described in Subsections \ref{Kac-Moodycurrents} and \ref{DefGaudin}. Let us then discuss what are the real parameters of the models. Note first that the constant term $\ell^\infty$ is always assumed to be real. Moreover, recall that for each pair of sites $(r,\pm)$, there are two possible reality conditions: either the positions $z^\pm_r$ and the levels $\ell^\pm_r$ are real or they form pairs of complex conjugate numbers. We will encode the choice of reality condition for the sites $(r,\pm)$ by introducing a number $c_r$, which is defined to be $1$ in the first case and $i$ in the second one. In particular, $z^\pm_r$ and $\ell^\pm_r$ can then be written using the following parametrisation:
\begin{equation}\label{Eq:ParamReal}
z^\pm_r = z_r \pm c_r \eta_r \;\;\;\;\;\; \text{ and } \;\;\;\;\;\; \ell^\pm_r = \frac{\ell^{[0]}_r}{2} \pm \frac{\ell^{[1]}_r}{2c_r\eta_r},
\end{equation}
where the parameters $z_r$, $\eta_r$, $\ell^{[0]}_r$ and $\ell^{[1]}_r$ are real. As we shall see, this particular choice of parametrisation will also be convenient for the interpretation of the models as deformations in the next subsection. Note that it is equivalent to defining
\begin{equation}\label{Eq:ParamDef}
z_r = \frac{z^+_r+z^-_r}{2}, \;\;\;\;\;\;\; \eta_r = \frac{z^+_r-z_r^-}{2c_r}, \;\;\;\;\;\;\; \ell^{[0]}_r = \ell^+_r + \ell^-_r \;\;\;\;\; \text{ and } \;\;\;\;\; \ell^{[1]}_r = \frac{z^+_r - z^-_r}{2}(\ell^+_r - \ell^-_r).
\end{equation}

\paragraph{Choice of realisations.} As explained in Subsection \ref{DefGaudin}, the choice of the Kac-Moody realisation attached to the sites $(r,\pm)$ corresponds to specifying the explicit form of the operators $\A^{(r)}_\pm$ and $\B^{(r)}_\pm$ and the value of the coefficient $\kay_r$ appearing in Equations \eqref{Currentsr} and \eqref{Eq:Yr}. In particular, one can choose this realisation among the examples described in Subsection \ref{Examples}.

For instance, if one takes the inhomogeneous Yang-Baxter realisation (with Wess-Zumino term), $\kay_r$ is set to $-(\ell^+_r+\ell^-_r)/2=-\ell^{[0]}_r/2$ and the operators $\A^{(r)}_\pm$ and $\B^{(r)}_\pm$ are given by Equation \eqref{Eq:OperatorsYBWZ} (replacing $g$ by $g^{(r)}$ and $c$ by the number $c_r\in\lbrace 1,i\rbrace$ defined in the previous paragraph, which encodes the choice of reality conditions for the sites $(r,\pm)$).

Similarly, if one chooses the $\lambda$-realisation, the operators $\A^{(r)}_\pm$ and $\B^{(r)}_\pm$ are given by Equation \eqref{Eq:OperatorsLambda} (with $g$ replaced by $g^{(r)}$), while $\kay_r$ is given by $-\ell^+_r/2$. Note however that one can choose the $\lambda$-realisation only if the levels $\ell^\pm_r$ are real (\textit{i.e.} $c_r=1$ in the notations of the previous paragraph) and are such that
\begin{equation}\label{Eq:ConstraintLevel}
\ell_r^{[0]}=\ell^+_r + \ell^-_r = 0. 
\end{equation}
This is in contrast with the case of the inhomogeneous Yang-Baxter realisation with Wess-Zumino term considered above, where the levels $\ell^\pm_r$ are not subject to any constraints (other than the reality conditions).

Note that the choice of a Yang-Baxter realisation at the sites $(r,\pm)$ comes with the additional freedom of choosing a skew-symmetric $R$-matrix $R_r$, solution of the mCYBE \eqref{Eq:mCYBE}. As explained in Subsection \ref{Examples}, this operator should in general satisfy the additional property $R_r^3=c_r^2 R_r$. However, if the levels $\ell^\pm_r$ satisfy the constraint \eqref{Eq:ConstraintLevel}, \textit{i.e.} if one considers a Yang-Baxter realisation without Wess-Zumino term, one does not need to require this additional condition on $R_r$.

\paragraph{The space of models.} The discussion above concerns the choice of realisation for one pair of sites $(r,\pm)$. One can then construct different models by considering different combinations of realisations for the $N$ pairs $(1,\pm),\cdots,(N,\pm)$ describing the models. In particular, one can consider a model with $N_1$ copies of the Yang-Baxter realisation and $N_2$ copies of the $\lambda$-realisation, where $N_1+N_2=N$. Let us discuss what are the free parameters of this theory. As explained in the previous paragraphs, the model is described by the $4N+1$ Gaudin parameters $z^\pm_r$, $\ell^\pm_r$ and $\ell^\infty$, or equivalently by the $4N+1$ real parameters $z_r,\eta_r,\ell^{[0]}_r$, $\ell^{[1]}_r$ and $\ell^\infty$. Taking into account the translation and dilation redundancy \eqref{Eq:Redundancy} and the fact that the levels corresponding to the $\lambda$-realisations should satisfy the constraints \eqref{Eq:ConstraintLevel}, we arrive at the conclusion that this model is described by $3N+N_1-1$ free parameters. Note that in addition to these parameters, which specify its structure as an AGM, the model is also determined by the choice of $N_1$ $R$-matrices for the Yang-Baxter realisations (which do not need to be identical).

As was explained in~\cite{Vicedo:2017cge}, see also~\cite{Delduc:2019bcl}, the models with only one realisation, \textit{i.e.} with $N=1$, correspond to well-known integrable $\sigma$-models, which served as basis for defining the Yang-Baxter and $\lambda$-realisations. Indeed, the inhomogeneous Yang-Baxter realisation (without or with Wess-Zumino term) is defined in such a way that the AGM with one copy of this realisation, corresponding in the above paragraph to $N_1=1$ and $N_2=0$, coincides with the so-called Yang-Baxter $\sigma$-model, without~\cite{Klimcik:2002zj,Klimcik:2008eq} or with~\cite{Delduc:2014uaa} Wess-Zumino term. Similarly, the AGM with one copy of the $\lambda$-realisation, \textit{i.e.} with $N_1=0$ and $N_2=1$, yields the so-called $\lambda$-model~\cite{Sfetsos:2013wia}. The model defined above with arbitrary numbers $N_1$ and $N_2$ is thus a generalisation of these models. According to the general coupling procedure described in~\cite[Subsection 2.3.3]{Delduc:2019bcl}, it corresponds to coupling together $N_1$ copies of the Yang-Baxter model and $N_2$ copies of the $\lambda$-model in a non-trivial way which however ensures the integrability of this interacting model (as, by construction, it is a realisation of AGM).

\paragraph{Zeroes versus levels.} Let us end this subsection with some remarks about a possible more convenient reparametrisation of the models that we are considering. Recall from Subsections \ref{SubSubSec:Dyn} and \ref{SubSubSec:Int} that in order to define the Hamiltonian and express the Lax pair of the models, one uses the zeroes $\zeta_i$, $i\in\lbrace 1,\cdots,2N\rbrace$, of the twist function. These zeroes are related implicitly to the Gaudin parameters $z^\pm_r$, $\ell^\pm_r$ and $\ell^\infty$ through the equation $\vp(\zeta_i)=0$, with the twist function $\vp(z)$ defined in terms of the Gaudin parameters as in \eqref{Eq:Twist}. This equation is equivalent to a polynomial equation of degree $2N$ in $\zeta_i$. Thus, it is in general impossible to give an explicit expression of the zeroes $\zeta_i$ in terms of the Gaudin parameters.

One way of bypassing this difficulty is to consider as defining parameters of the models the positions $z^\pm_r$, the zeroes $\zeta_i$ and the constant term $\ell^\infty$. One then defines the twist function of the model by Equation \eqref{Eq:TwistZeroes} instead of Equation \eqref{Eq:Twist} and the levels $\ell^\pm_r$ as the corresponding residues:
\begin{equation*}
\ell^\pm_r = \res_{z=z_r^\pm} \vp(z)\,\text{d}z = \mp \frac{\ell^\infty}{z_r^+-z_r^-} \frac{\prod_{i=1}^{2N}(z^\pm_r-\zeta_i)}{\prod_{s=1, s\neq r}^{N} (z_r^\pm-z_s^\pm)(z_r^\pm-z_s^\mp)}.
\end{equation*}
The main advantage of this re-parametrisation is that all the relevant quantities that are used to describe the models, in particular the levels $\ell^\pm_r$ and the Hamiltonian $\mathcal{H}$, can be written as rational expressions of the parameters $z_r^\pm$, $\zeta_i$ and $\ell^\infty$. Note however that this parametrisation has a disadvantage when one wants to consider $\lambda$-realisations and/or Yang-Baxter realisations without Wess-Zumino terms. Indeed, for these realisations, the levels should satisfy the additional constraint \eqref{Eq:ConstraintLevel}, which translates in a rather complicated algebraic condition on the parameters $z_r^\pm$ and $\zeta_i$, using the above expressions for the levels. Finally, let us note that the translation and dilation redundancy \eqref{Eq:Redundancy} among the Gaudin parameters can be re-expressed in terms of this new parametrisation as the invariance of the model under the transformation
\begin{equation*}
z^\pm_r \longmapsto a z^\pm_r + b, \;\;\;\;\;\;\;\;\; \zeta_i \longmapsto a\zeta_i+b \;\;\;\;\;\;\;\;\; \text{and} \;\;\;\;\;\;\;\;\; \ell^\infty \longmapsto a^{-1} \ell^\infty.
\end{equation*}

\subsubsection{Recovering undeformed models}
\label{SubSubSec:UndefHam}

In this subsection, following the results of~\cite{Delduc:2019bcl}, we discuss how the model defined above by taking $N_1$ Yang-Baxter realisations and $N_2$ $\lambda$-realisations can be interpreted as a deformation of a simpler model. This result generalises the well known facts that the Yang-Baxter model (with or without Wess-Zumino term) is a deformation of the Principal Chiral Model (PCM, with or without Wess-Zumino term) and the $\lambda$-model is a deformation of the non-abelian T-dual of the PCM. In the present language, these correspond respectively to the cases $(N_1=1,N_2=0)$ and $(N_1=0,N_2=1)$. The undeformed limit of the model with arbitrary $N_1$ and $N_2$ corresponds to a theory coupling together $N_1$ copies of the PCM (with Wess-Zumino terms) and $N_2$ copies of its non-abelian T-dual.  

This undeformed model is also defined as a realisation of AGM but possesses a sligthly different sites structure. Indeed, in the language of \cite{Vicedo:2017cge,Delduc:2019bcl}, instead of the $2N$ sites $(r,\pm)$ of multiplicity one, it possesses $N$ sites $(r)$ of multiplicity two. These sites correspond to double poles in the twist function and the Gaudin Lax matrix of the model and are associated with so-called Takiff realisations of multiplicity two, which generalise the notion of Kac-Moody realisations for sites of multiplicity greater than one. As we shall now explain, the site $(r)$ of multiplicity two is obtained from the pair of sites $(r,\pm)$ in the deformed model by making their positions $z_r^+$ and $z_r^-$ collide, while controlling the behaviour of the corresponding levels $\ell^\pm_r$.

\paragraph{Colliding two simple poles into a double pole.} Let us focus here on one pair of sites $(r,\pm)$. In order to isolate the parts of the twist function and the Gaudin Lax matrix of the model corresponding to this pair, let us rewrite them as
\begin{align*}
\varphi(z) &= \frac{\ell_r^+}{z-z_r^+} + \frac{\ell_r^-}{z-z_r^-} + \widetilde{\varphi}(z), \\
\Gamma(z) &= \frac{\mathcal{J}^{(r)}_+}{z-z_r^+} + \frac{\mathcal{J}^{(r)}_-}{z-z_r^-} + \widetilde{\Gamma}(z),
\end{align*}
where $\widetilde{\varphi}$ and $\widetilde{\Gamma}$ contain all the information related to the other sites. Using the parameters $c_r$, $z_r$, $\eta_r$, $\ell^{[0]}_r$ and $\ell^{[1]}_r$ introduced in the previous Subsection (see Equation \eqref{Eq:ParamDef}), one can rewrite the twist function as
\begin{equation}\label{Eq:TwistDef}
\varphi(z) = \frac{\ell^{[1]}_r}{(z-z_r)^2 - c_r^2 \eta_r^2 } + \frac{\ell^{[0]}_r(z-z_r)}{(z-z_r)^2-c_r^2 \eta_r^2} + \widetilde{\varphi}(z).
\end{equation}

As mentioned above, the undeformed limit corresponds to making the two positions $z_r^+$ and $z_r^-$ collide at the point $z_r$ and thus to taking $\eta_r \to 0$. In particular, this leads us to interpret $\eta_r$ as a deformation parameter. We aim here to recover, in the limit $\eta_r\to0$, a model with a site of multiplicity two, \textit{i.e.} with a double pole in its twist function. It is then clear from Equation \eqref{Eq:TwistDef} that this is the case if one supposes that the quantities $\ell^{[0]}_r$ and $\ell^{[1]}_r$ stay finite when $\eta_r$ goes to 0. From now on, we will thus define the undeformed limit as taking $\eta_r \to 0$ while keeping $\ell^{[0]}_r$ and $\ell^{[1]}_r$ finite (let us note that the levels $\ell^\pm_r$ of the sites $(r,\pm)$ then diverge, as one can see from Equation \eqref{Eq:ParamReal}). In this limit, the twist function becomes
\begin{equation*}
\vp(z) \xrightarrow{\eta_r\to 0} \frac{\ell^{[1]}_r}{(z-z_r)^2} + \frac{\ell^{[0]}_r}{z-z_r} + \widetilde{\varphi}(z).
\end{equation*}
Following the terminology of~\cite{Vicedo:2017cge,Delduc:2019bcl}, this corresponds to the twist function of an AGM with a site $(r)$ of multiplicity two, with position $z_r$ and Takiff levels $\ell_r^{[0]}$ and $\ell^{[1]}_r$ (and with the other sites, contained in $\widetilde{\varphi}(z)$, as in the deformed model).\\

A similar argument applies to the Gaudin Lax matrix of the model. Let us suppose that the Kac-Moody currents $\mathcal{J}^{(r)}_\pm$ are such that the limits
\begin{equation}\label{Eq:LimTakiff}
\mathcal{J}^{(r)}_{[0]} = \lim_{\eta_r \to 0}\left(\mathcal{J}^{(r)}_+ + \mathcal{J}^{(r)}_-\right) \;\;\;\;\;\;\;  \text{ and }  \;\;\;\;\;\;\;\;\; 
\mathcal{J}^{(r)}_{[1]} = \lim_{\eta_r \to 0}c_r \eta_r\left(\mathcal{J}^{(r)}_+ - \mathcal{J}^{(r)}_-\right)
\end{equation}
are finite. Then the Gaudin Lax matrix becomes in the undeformed limit:
\begin{equation*}
\Gamma(z) \xrightarrow{\eta_r\to 0} \frac{\mathcal{J}^{(r)}_{[1]}}{(z-z_r)^2} + \frac{\mathcal{J}^{(r)}_{[0]}}{z-z_r} + \widetilde{\Gamma}(z).
\end{equation*}
Thus, $\mathcal{J}^{(r)}_{[0]}$ and $\mathcal{J}^{(r)}_{[1]}$ are the Takiff currents attached to the site $(r)$ of the undeformed model\footnote{Starting from the Kac-Moody Poisson brackets \eqref{Eq:PBKM} of the currents $\mathcal{J}^{(r)}_\pm$, one can indeed show that in the undeformed limit, the currents $\mathcal{J}^{(r)}_{[0]}$ and $\mathcal{J}^{(r)}_{[1]}$ satisfy the brackets of Takiff currents with levels $\ell_r^{[0]}$ and $\ell^{[1]}_r$.}. Let us now discuss this undeformed limit for the Yang-Baxter realisation and the $\lambda$-realisation.

\paragraph{From the Yang-Baxter to the PCM realisation.} Let us suppose that the sites $(r,\pm)$ are associated with a Yang-Baxter realisation with Wess-Zumino term, as described in Subsection \ref{Examples}. Let us first note that for this realisation, the Wess-Zumino coefficient is given by $\kay_r=-\ell^{[0]}_r/2$. In particular, the undeformed limit defined in the previous paragraph can then be seen as taking $\eta_r$ to $0$ while keeping $\kay_r$ and $\ell^{[1]}_r$ finite. Let us denote by $R_r$ the $R$-matrix associated with this Yang-Baxter limit and introduce $R^{(r)}=\Ad_{g^{(r)}}^{-1}\circ R_r\circ\Ad_{g^{(r)}}$ and $\Pi^{(r)}=\Id-R^{(r)\,2}/c_r^2$. The Kac-Moody currents of the realisation are then given by
{\begin{equation}\label{Eq:YBDef}
\mathcal{J}^{(r)}_\pm = \frac{1}{2}\left(\text{Id} \mp \frac{R^{(r)}}{c_r} \mp \delta_r\Pi^{(r)}\right) Y^{(r)} + \left(\left(\pm\frac{\ell^{[1]}_r}{2c_r\eta_r} - \frac{\kay_r}{2}\right)\text{Id} \pm \frac{\kay_r}{2c_r} R^{(r)} \pm \frac{\kay_r\delta_r}{2}\Pi^{(r)}\right)j^{(r)},
\end{equation}}
with
\begin{equation}\label{Eq:YDelta}
Y^{(r)} = X^{(r)}-\kay_r\, W^{(r)} \;\;\;\;\; \text{ and } \;\;\;\;\; \delta_r = \ell^{[1]}_r \frac{1-\sqrt{1-\left(2c_r  \eta_r \kay_r/\ell^{[1]}_r\right)^2}}{2c_r  \eta_r \kay_r}.
\end{equation}
Let us now consider the undeformed limit, \textit{i.e.} taking $\eta_r$ to 0 while keeping $\kay_r$ and $\ell^{[1]}_r$ finite. One first observes that in this limit, the coefficient $\delta_r$ tends to 0. Using this, one finds that the limits $\mathcal{J}^{(r)}_{[0]}$ and $\mathcal{J}^{(r)}_{[1]}$ defined in Equation \eqref{Eq:LimTakiff} are indeed finite and simply read
\begin{equation*}
\mathcal{J}^{(r)}_{[0]} = X^{(r)} - \kay_r\, W^{(r)} - \kay_r\, j^{(r)} \;\;\;\;\; \text{ and } \;\;\;\;\; \mathcal{J}^{(r)}_{[1]} = \ell^{[1]}_r j^{(r)}.
\end{equation*}
Thus, the undeformed limit described in the previous paragraph is well defined. Moreover, one recognises in the above equation the Takiff currents of the PCM+WZ realisation (with levels $\ell^{[0]}_r=-2\kay_r$ and $\ell^{[1]}_r$), as defined in~\cite[Subsection 3.1.3]{Delduc:2019bcl}.

\paragraph{From the $\bm\lambda$-realisation to the non-abelian T-dual realisation.} A similar mechanism to the one described above for the Yang-Baxter realisation provides the underformed limit of the $\lambda$-realisation, yielding the so-called non-abelian T-dual realisation, as defined in~\cite[Subsection 4.3.1]{Delduc:2019bcl}. This limit requires however a more subtle treatment. Indeed, if one were to consider the currents $\mathcal{J}^{(r)}_\pm$ of the $\lambda$-realisation in terms of the fields $g^{(r)}$ and $X^{(r)}$ and take the limits \eqref{Eq:LimTakiff} ``naively'', one would encounter divergent expressions, making the undeformed limit procedure ill-defined. In order to obtain a well defined limit, one has to consider the fields $g^{(r)}$ and $X^{(r)}$ as depending on the deformation parameter $\eta_r$ and suppose that they obey a well-chosen asymptotic expansion when $\eta_r$ goes to 0. In particular, one of the consequences of this more subtle limit is that it changes the phase space of the realisation: from the space of canonical fields on $T^*G_0$ (described by $g^{(r)}$ and $X^{(r)}$), one goes in the limit to the space of canonical fields on $T^*\g_0$, which is the phase space of the non-abelian T-dual realisation. For brevity, we will not re-explain this procedure in the present article and refer to~\cite[Subsection 4.4.3]{Delduc:2019bcl} for details.

\paragraph{Undeformed limits of the coupled models.} Let us consider the model defined in the previous subsection by coupling together $N_1$ copies of the Yang-Baxter model and $N_2$ copies of the $\lambda$-model. For each pair of sites $(r,\pm)$, one can consider the corresponding undeformed limit $\eta_r \to 0$. One would then obtain a model where the $r$-th copy reduces to either an undeformed PCM with Wess-Zumino term or a non-abelian T-dual of the PCM (depending on whether we started with a Yang-Baxter realisation or a $\lambda$-realisation at the sites $(r,\pm)$), still interacting non-trivially with the other $N-1$ copies. One can then consider different combinations of these undeformed limits on any number of copies, yielding various limits of the model. All these limits can be seen as deformations of a completely undeformed model, obtained by taking the limit where all the deformation parameters $\eta_1,\cdots,\eta_N$ are sent to 0. This undeformed model is the coupling  of $N_1$ copies of the PCM with Wess-Zumino terms and $N_2$ copies of the non-abelian T-dual of the PCM. In particular, if one considers $N_2=0$, one obtains the model coupling together $N$ copies of the PCM with Wess-Zumino term: this is the integrable coupled $\sigma$-model first introduced in~\cite{Delduc:2018hty} and whose detailed construction was presented in~\cite[Subsection 3.3]{Delduc:2019bcl}.

Although it is defined in a different way, let us note also that the undeformed model with $N_2 \neq 0$ copies of the non-abelian T-dual of the PCM is in fact canonically equivalent to the model with $N=N_1+N_2$ copies of the PCM, where $N_2$ of these copies have no Wess-Zumino term. This is because the non-abelian T-dual realisation is related to the PCM realisation without Wess-Zumino term by a canonical transformation~\cite{Lozano:1995jx}. Thus, the general model with $N_1$ Yang-Baxter realisations and $N_2$ $\lambda$-realisations can be seen as a deformation of the model coupling $N_1$ PCM with Wess-Zumino term and $N_2$ PCM without Wess-Zumino term (which is a particular case of the model introduced in~\cite{Delduc:2018hty}) after having first T-dualised the $N_2$ copies without Wess-Zumino term.

\paragraph{Homogeneous Yang-Baxter limit.} For completeness, let us end this subsection by mentioning briefly another possible limit of the models considered here, which corresponds to going from an inhomogeneous Yang-Baxter realisation to a homogeneous Yang-Baxter realisation\footnote{This idea was first applied in the article~\cite{Kawaguchi:2014qwa} in the context of the deformed superstring on $AdS_5 \times S^5$.}. Let us consider an inhomogeneous Yang-Baxter realisation without Wess-Zumino term and with $R$-matrix $R$, which satisfies the mCYBE \eqref{Eq:mCYBE}. So far, we considered the coefficient $c$ appearing in the mCYBE as being either $1$ or $i$, depending on the type of reality conditions imposed on the realisation. However, one easily checks that the construction of the Yang-Baxter realisation as recalled in Subsection \ref{Examples} holds without changes for any $c\neq 0$ (the realisation is then equivalent to the one with $c=1$ or $c=i$ by rescaling the matrix $R$). The homogeneous limit consists in taking the limit $c\to 0$ of this realisation while also making the corresponding simple poles in the twist function collide (see for example~\cite{Lacroix:2019xeh}). Similarly to what happens for the undeformed limit described in this subsection, this yields a model with a site of multiplicity two, to which is attached the so-called homogeneous Yang-Baxter realisation, as defined in~\cite[Subsection 4.1.1]{Delduc:2019bcl}. This realisation corresponds to a deformation of the PCM realisation without Wess-Zumino term by a homogeneous $R$-matrix, \textit{i.e.} a solution of the (non-modified) CYBE:
\begin{equation*}
[RX,RY] -R([RX,Y]+[X,RY]) = 0, \;\;\;\;\;\;\;\; \forall \, X,Y \in \mathfrak{g}_0,
\end{equation*}
which corresponds to the limit $c\to 0$ of the mCYBE.

\paragraph{Summary.} Although we introduced them as limits, the PCM, non-abelian T-dual and homogeneous Yang-Baxter realisations can be constructed independently, as was done for example in~\cite{Delduc:2019bcl}. One can then consider AGM containing these realisations. In general, one can construct a model coupling together any combination of PCMs, non-abelian T-dual models, homogeneous and inhomogeneous Yang-Baxter models and $\lambda$-models. Up to taking appropriate limits, the present article then covers all these possibilities. In particular, one can obtain a model with $N-1$ copies of the PCM and one homogeneous Yang-Baxter realisation: one then recovers the model studied in~\cite[Appendix D]{Delduc:2019bcl} as the simplest illustration of the various possible integrable deformations of coupled integrable $\sigma$-models.

\section{Lagrangian formulation}
\label{Sec:Lag}

In this section, our aim will be to describe the models introduced in Section 2.2 in the Lagrangian formulation. Recall that in the Hamiltonian formulation, the degrees of freedom of these models are the fields $g^{(r)}(x)$ and $X^{(r)}(x)$, describing canonical fields valued in $N$ independent copies of the cotangent bundle $T^*G_0$. The fields $g^{(r)}(x)$ are the ``coordinate fields'' valued in the space $G_0$. The momentum fields conjugate to these coordinates are then encoded in the fields $X^{(r)}(x)$ (see for instance \cite[Subsection 3.1.1]{Delduc:2019bcl} for details). In order to pass to the Lagrangian formulation, one has to consider the coordinate fields $g^{(r)}(x,t)$ as depending explicitly on the time variable $t\in \mathbb{R}$, defined by the Hamiltonian of the model, and express the momentum fields of the theory, encoded in $X^{(r)}$, in terms of these Lagrangian fields $g^{(r)}(x,t)$ and their derivatives $\partial_t g^{(r)}(x,t)$ and $\partial_x g^{(r)}(x,t)$. Finally, one obtains the action of the model as a functional of $g^{(r)}(x,t)$ by performing an inverse Legendre transform on their Hamiltonian.

In the present case, we will obtain the Lagrangian expression of the fields $X^{(r)}$ in a rather indirect way. Indeed, as we shall see, these fields can be expressed naturally in terms of the Lax pair of the model. For this reason, we will start by determining the Lagrangian expression of the latter.

\subsection{Lax pair in the Lagrangian formulation}
\label{Lax}

\paragraph{Maurer-Cartan currents in terms of the Lax pair.} Let us begin by considering the time evolution of the fields $g^{(r)}$. In the Hamiltonian formulation, this is given by their Poisson bracket with the Hamiltonian. More explicitly, recalling the definition \eqref{Hamiltonian} of the latter, one expresses the temporal Maurer-Cartan current $j_t^{(r)} = g^{(r)}(x)^{-1} \partial_t g^{(r)}(x)$ as
\begin{equation*}
j_t^{(r)}(x) = g^{(r)}(x)^{-1}\{\mathcal{H},g^{(r)}(x)\} = \sum_{i=1}^{2N} \epsilon_i \, g^{(r)}(x)^{-1}\{\mathcal{Q}_i,g^{(r)}(x)\}.
\end{equation*}
From the expression \eqref{Charges} of the charges $\mathcal{Q}_i$, we then have
\begin{equation*}
j_t^{(r)}(x) = \sum_{i=1}^{2N} \frac{\epsilon_i}{\varphi'(\zeta_i)} \int_{\mathbb{D}} dy \ \kappa_{\underline{\mathbf{2}}} \left(g^{(r)}_{\underline{\mathbf{1}}}(x)^{-1}\{g^{(r)}_{\underline{\mathbf{1}}}(x),\Gamma_{\underline{\mathbf{2}}}(\zeta_i,y)\},\Gamma_{\underline{\mathbf{2}}}(\zeta_i,y)\right). \nonumber
\end{equation*}
The Poisson bracket in the integrand is calculated by inserting the definition \eqref{GaudinL} of $\Gamma(z,x)$, yielding:
\begin{equation*}
\{g^{(r)}_{\underline{\mathbf{1}}}(x), \Gamma_{\underline{\mathbf{2}}}(\zeta_i,y)\} = \frac{1}{\zeta_i - z_r^+}\{g^{(r)}_{\underline{\mathbf{1}}}(x), \mathcal{J}^{(r)}_{+\underline{\mathbf{2}}}(y)\} + \frac{1}{\zeta_i - z_r^-}\{g^{(r)}_{\underline{\mathbf{1}}}(x), \mathcal{J}^{(r)}_{-\underline{\mathbf{2}}}(y)\},
\end{equation*}
where we have also used the fact that $\mathcal{J}_\pm^{(s)}$ depends on the fields in the $s^{\text{th}}$-factor of $T^*G_0^N$ and thus Poisson commutes with $g^{(r)}$ if $r \ne s$. In order to calculate the Poisson brackets on the right hand side we then use the definition \eqref{Currentsr} of the currents $\mathcal{J}^{(r)}_\pm$ in terms of $Y^{(r)}$ and $j^{(r)}$. Note that, firstly, the Poisson brackets of $g^{(r)}$ with $j^{(r)}$ vanish. Moreover, the brackets of $g^{(r)}$ with the operators $\A^{(r)}_\pm$ and $\B^{(r)}_\pm$ also give no contribution as we assumed that these operators depend only on $g^{(r)}$. Thus, we have to take into account only the terms coming from the Poisson bracket of $g^{(r)}$ with $Y^{(r)}$, so that
\begin{equation*}
g^{(r)}_{\underline{\mathbf{1}}}(x)^{-1} \{g^{(r)}_{\underline{\mathbf{1}}}(x), \mathcal{J}^{(r)}_{\pm\underline{\mathbf{2}}}(y)\} = -\A^{(r)}_{\pm\underline{\mathbf{2}}}C_{\underline{\mathbf{12}}}\delta_{xy} = - {^t}\A^{(r)}_{\pm\underline{\mathbf{1}}}C_{\underline{\mathbf{12}}}\delta_{xy},
\end{equation*}
where we have used the fact that for any operator $\mathcal{O}$ on $\g$, one has $\mathcal{O}_{\underline{\mathbf{2}}} C_{\underline{\mathbf{12}}}= \null^t\mathcal{O}_{\underline{\mathbf{1}}} C_{\underline{\mathbf{12}}}$. Putting everything together, we conclude that
\begin{equation*}
j_t^{(r)}(x) = \sum_{i=1}^{2N} \frac{\epsilon_i}{\varphi'(\zeta_i)}\left(\frac{{^t}\A^{(r)}_+}{z_r^+ - \zeta_i} + \frac{{^t}\A^{(r)}_-}{z_r^- - \zeta_i}\right)\Gamma(\zeta_i,x).
\end{equation*}
Using the expression of the temporal component of the Lax pair \eqref{M}, this can be re-expressed in the following way:
\begin{equation*}
j_t^{(r)} = {^t}\A^{(r)}_+ \mathcal{M}(z_r^+) + {^t}\A^{(r)}_- \mathcal{M}(z_r^-).
\end{equation*}
Moreover, by repeating this argument replacing the Hamiltonian by the momentum $\mathcal{P}$, expressed in terms of the charges $\mathcal{Q}_i$ as in \eqref{Eq:Momentum}, and using the expression \eqref{L} for the spatial component of the Lax pair, one finds a similar relation for the currents $j^{(r)}$:
\begin{equation*}
j^{(r)} = {^t}\A^{(r)}_+ \mathcal{L}(z_r^+) + {^t}\A^{(r)}_- \mathcal{L}(z_r^-).
\end{equation*}
Therefore, using light-cone coordinates, we find that the Maurer-Cartan currents
\begin{equation*}
j_\pm^{(r)} = g^{(r)\,-1} \partial_\pm g^{(r)}
\end{equation*}
take the following rather simple form in terms of the Lax pair:
\begin{equation}\label{Eq:jL}
j_\pm^{(r)} = {^t}\A^{(r)}_+ \mathcal{L}_\pm(z_r^+) + {^t}\A^{(r)}_- \mathcal{L}_\pm(z_r^-).
\end{equation}

\paragraph{Lagrangian Lax pair from interpolation.} Our goal in this subsection is to find a Lagrangian expression of the Lax pair, \textit{i.e.} an expression of $\mathcal{L}_\pm(z)$ in terms of the Maurer-Cartan currents $j_\pm^{(r)}$. We note that Equation \eqref{Eq:jL} relates these currents to the evaluations $\mathcal{L}_\pm(z_r^+)$ and $\mathcal{L}_\pm(z_r^-)$ of the Lax pair at the positions $z_r^+$ and $z_r^-$. As we shall now explain, this relation is enough to reconstruct the expression of $\mathcal{L}_\pm(z)$ in terms of  $j_\pm^{(r)}$ for all values of the spectral parameter $z$. Let us define
\begin{equation} \label{Eq:J}
J_\pm^{(r)} = \mathcal{L}_\pm(z_r^\pm),
\end{equation}
for $r = 1,\cdots,N$. From Equation \eqref{Laxlight}, one sees that $\mathcal{L}_\pm(z)$ is a rational function of $z$ with $N$ simple poles, situated at the zeroes of the twist function $\zeta_i$, for $i \in \mathcal{I}_\pm$ (recall that we have supposed that the subsets $\mathcal{I}_\pm$ are both of size $N$). It is a standard result that such a function is completely determined by its evaluation at $N$ pairwise distinct points. In particular, $\mathcal{L}_\pm(z)$ can be expressed in terms of its evaluations at the positions $z_r^\pm$, \textit{i.e.} the currents $J_\pm^{(r)}$ introduced above. More precisely, one has the following interpolation formula (see also Lemma B.2 of \cite{Delduc:2019bcl})
\begin{equation}\label{Eq:Interpolation}
\mathcal{L}_\pm(z) = \sum_{r=1}^N \frac{\varphi_{\pm,r}(z_r^\pm)}{\varphi_{\pm,r}(z)}J_\pm^{(r)},
\end{equation}
where
\begin{equation}\label{Eq:PhiPmR}
\varphi_{\pm,r}(z) = \frac{\prod_{i\in \mathcal{I}_\pm}{(z-\zeta_i)}}{\prod_{\substack{s=1 \\ s\ne r}}^N{(z-z_s^{\pm})}}.
\end{equation}
We are now in a position to rewrite the Lax pair in terms of the currents $j_\pm^{(r)}$. Indeed, the above equation \eqref{Eq:jL} can now be rewritten as a system of linear equations between the currents $j_\pm^{(r)}$ and $J_\pm^{(r)}$, which (at least formally) can be inverted. More precisely, reinserting \eqref{Eq:Interpolation} in \eqref{Eq:jL}, we have that
\begin{equation}\label{Eq:J2j}
j_\pm^{(r)} = \sum_{s=1}^N \Kop^\pm_{rs} J_\pm^{(s)},
\end{equation}
where we have defined
\begin{equation}\label{Eq:DefK}
\Kop^\pm_{rs} = \delta_{rs}\, {^t}\A_\pm^{(r)} + \frac{\vp_{\pm,s}(z_s^\pm)}{\vp_{\pm,s}(z_r^\mp)}\: {^t}\A_\mp^{(r)}.
\end{equation}
In the following, we will see the operators $\Kop^\pm_{rs}$ as the entries of some matrix operators $\Kop_\pm$, so that $\Kop^\pm_{rs} = (\Kop_\pm)_{rs}$. Note that $\Kop_\pm$ are then $N \times N$ matrices with non-commutative entries. To conclude, we rewrite the currents $J_\pm^{(r)}$ in terms of the $j_\pm^{(r)}$'s by means of the inversion
\vspace{-4pt}\begin{equation}\label{Eq:j2J}
J_\pm^{(r)} = \sum_{s=1}^N (\Kop_\pm^{-1})_{rs} j_\pm^{(s)},\vspace{-4pt}
\end{equation}
where $(\Kop_\pm^{-1})_{rs}$ denote the entries of the inverse of the matrix operators $\Kop_\pm$. Reinserting now \eqref{Eq:j2J} in \eqref{Eq:Interpolation} then gives an expression of $\mathcal{L}_\pm(z)$ in terms of the currents $j_\pm^{(r)}$:
\begin{equation}\label{Eq:j2L}
\Lc_\pm(z) = \sum_{s=1}^N \left( \sum_{r=1}^N \frac{\varphi_{\pm,r}(z_r^\pm)}{\varphi_{\pm,r}(z)} (\Kop^{-1}_\pm)_{rs} \right) j^{(s)}_\pm.
\end{equation}
Note that this is a formal relation, as it involves the inverse of the matrix operators $\Kop_\pm$. Performing explicitly this inversion is in general not straightforward because of the non-commutativity of the entries of $\Kop_\pm$ (for example, one cannot use the general expression for the inverse of a matrix in terms of its comatrix). We will explain in Subsection \ref{2copies} how this is done explicitly in the case of two copies.

\paragraph{Different interpolations and factorisations of the twist function.} We conclude this section by making an important remark about Equations \eqref{Eq:J} and \eqref{Eq:Interpolation}. In these equations, we decided to express the component $\Lc_+(z)$, resp. $\Lc_-(z)$, of the Lax pair in terms of its evaluations at the positions $z_r^+$, resp. $z_r^-$. Let us stress here that this choice is arbitrary, as one could have chosen for example to interpolate $\Lc_+(z)$ and $\Lc_-(z)$ through their evaluations at the positions $z_r^-$ and $z_r^+$ respectively\footnote{Note in particular that the indices $\pm$ of $\Lc_\pm(z)$ are conceptually totally unrelated to the labels $\pm$ of the positions $z_r^\pm$. Indeed, the former are space-time indices corresponding to the light-cone directions in $\mathbb{R}\times\mathbb{D}$ while the latter are abstract labels distinguishing the two sites $(r,+)$ and $(r,-)$.}. More generally, one could have considered the evaluations
\begin{equation*}
\widetilde{J}_\pm^{\,(r)} = \mathcal{L}_\pm(z_r^{\pm\sigma_r}),
\end{equation*}
where the $\sigma_r$'s take values in the set $\{+1,-1\}$ for every $r$. The interpolation equation \eqref{Eq:Interpolation} would then become
\begin{equation}\label{Eq:Interpolation2}
\mathcal{L}_\pm(z) = \sum_{r=1}^N \frac{\widetilde{\varphi}_{\pm,r}(z_r^{\pm\sigma_r})}{\widetilde{\varphi}_{\pm,r}(z)}\widetilde{J}_\pm^{\,(r)}, \;\;\;\;\; \text{ where now } \;\;\;\;\; \widetilde{\varphi}_{\pm,r}(z) = \frac{\prod_{i\in \mathcal{I}_\pm}{(z-\zeta_i)}}{\prod_{\substack{s=1 \\ s\ne r}}^N{(z-z_s^{\pm\sigma_s})}}.
\end{equation}
Following the method developed in the previous paragraph, one would then express the currents $\widetilde{J}^{\,(r)}_\pm$ in terms of the Maurer-Cartan currents $j^{(r)}_\pm$ by a relation similar to Equation \eqref{Eq:j2J}, with the operators $\Kop_\pm$ replaced by some different operators $\widetilde{\Kop}_\pm$. Re-inserting this expression in Equation \eqref{Eq:Interpolation2} would then give $\Lc_\pm(z)$ in terms of $j^{(r)}_\pm$, similarly to Equation \eqref{Eq:j2L}. This expression can be shown to coincide with Equation \eqref{Eq:j2L} as one should expect, considering that they correspond to two ways of expressing the same object $\Lc_\pm(z)$. Similarly, all the methods and computations developed in the rest of this subsection can be applied starting from an arbitrary choice of interpolation, \textit{i.e.} from an arbitrary choice of $\sigma_r$'s: the end results (in particular the expression of the action of the model in terms of the Maurer-Cartan currents that will be obtained in the next subsection) can then be shown to be independent of this choice. For this reason, and to avoid unnecessary cumbersome notations, we will use in the rest of this article a particular choice of $\sigma_r$'s, namely $\sigma_r=+1$ for every $r$, corresponding to the choice made originally in the previous paragraph.\\

To conclude this paragraph, let us discuss a reinterpretation of the functions $\widetilde\varphi_{\pm,r}(z)$ appearing in the interpolation formula \eqref{Eq:Interpolation2} and of the freedom encoded in the choice of $\sigma_r$'s in terms of the twist function \eqref{Eq:TwistZeroes} of the model. Let us rewrite the latter in the following factorised form:
\begin{equation} \label{Twistlight}
\varphi(z) = -\ell^\infty\widetilde{\varphi}_+(z)\widetilde{\varphi}_-(z), \;\;\;\;\; \text{ where } \;\;\;\;\; \widetilde{\varphi}_\pm(z) = \frac{\prod_{i\in \mathcal{I}_\pm}{(z-\zeta_i)}}{\prod_{s=1}^N{(z-z_s^{\pm\sigma_s})}}.
\end{equation}
The functions $\widetilde{\varphi}_{\pm,r}(z)$ can then be re-expressed as $\widetilde{\varphi}_{\pm,r}(z) = (z-z_r^{\pm\sigma_r})\widetilde{\varphi}_\pm(z)$. Moreover, we observe that the freedom in the choice of the $\sigma_r$'s gets now reinterpreted as the existence of different ways of factorising the twist function. Indeed, redistributing the pairs of factors $(z-z_r^+)$ and $(z-z_r^-)$ associated to the paired sites $(r,\pm)$ into the definition \eqref{Twistlight} of $\widetilde{\varphi}_\pm(z)$ amounts to changing the values of the $\sigma_r$'s\footnote{Note that contrarily to the poles $z_r^\pm$, the zeroes $\zeta_i$ of the twist function cannot be redistributed differently between the functions $\widetilde\vp_+(z)$ and $\widetilde{\vp}_-(z)$, as they are naturally associated with one or the other depending on whether the index $i$ belongs to the set $\mathcal{I}_+$ or $\mathcal{I}_-$.}. In the rest of this article and in agreement with the notations of the previous paragraph, we will denote by $\vp_\pm(z)$ the functions $\widetilde\vp_\pm(z)$ corresponding to the choice $\sigma_r=+1$ for every $r\in\lbrace 1,\cdots,N\rbrace$.

\subsection{Inverse Legendre transform and action of the models}
\label{Legendre}

\paragraph{Lagrangian expression of the momentum.} We are now in a position to perform the first step towards writing down the inverse Legendre transform of the model, \textit{i.e.} re-expressing the fields $X^{(r)}$, which encode the momentum fields of the theory, in terms of Lagrangian fields. Let us first note from Equation \eqref{Eq:Yr} that the fields $Y^{(r)}$ and $X^{(r)}$ are related through the current $W^{(r)}$. As explained in Subsection \ref{Thealgebra}, this current $W^{(r)}$ is expressed in terms of the field $g^{(r)}$ and its spatial derivative  (and not the momentum fields) and has thus a direct Lagrangian expression. Thus, finding the Lagrangian expression of $X^{(r)}$ is equivalent to finding the Lagrangian expression of $Y^{(r)}$. As we shall now see, the latter is easier to find, using the Lagrangian expression of the Lax pair obtained in the previous paragraph. From the definition \eqref{Eq:DefLax} of the Lax matrix $\Lc(z)$, one can prove that (see also \cite[Equation (2.22)]{Delduc:2019bcl})
\begin{equation*}
\mathcal{L}(z_r^\pm) = \frac{\mathcal{J}^{(r)}_\pm}{\ell_r^\pm} = \frac{\A^{(r)}_\pm}{\ell_r^\pm} Y^{(r)} + \frac{\B^{(r)}_\pm}{\ell_r^\pm} j^{(r)},
\end{equation*}
where to obtain the second equality we have used the definition \eqref{Currentsr} of the currents $\mathcal{J}_\pm^{(r)}$. Then, using the identities \eqref{Eq:Identities2} satisfied by the operators $\A_\pm^{(r)}$ and $\B_\pm^{(r)}$, we find the following expression for $Y^{(r)}$:
\begin{equation}\label{Eq:YfromL}
Y^{(r)} = {^t}\B_+^{(r)} \mathcal{L}(z_r^+) + {^t}\B_-^{(r)} \mathcal{L}(z_r^-).
\end{equation}
Using the light-cone components of the Lax pair, this can be rewritten as
\begin{equation*}
Y^{(r)} = \frac{{^t}\B^{(r)}_+ \mathcal{L}_+(z_r^+) + {^t}\B^{(r)}_- \mathcal{L}_+(z_r^-)}{2} - \frac{{^t}\B^{(r)}_+ \mathcal{L}_-(z_r^+) + {^t}\B^{(r)}_- \mathcal{L}_-(z_r^-)}{2}.
\end{equation*}
From the Lagrangian expression \eqref{Eq:Interpolation} of $\mathcal{L}_\pm(z)$, one then finds that
\begin{equation}\label{Eq:J2Y}
Y^{(r)} = \sum_{s=1}^N \left[ \Lop_{rs}^+ J_+^{(s)} + \Lop_{rs}^- J_-^{(s)} \right],
\end{equation}
where we have defined
\begin{equation}\label{Eq:DefL}
\Lop^\pm_{rs} = \pm\frac{1}{2}\left(\delta_{rs}\, {^t}\B_\pm^{(r)} + \frac{\vp_{\pm,s}(z_s^\pm)}{\vp_{\pm,s}(z_r^\mp)}\; {^t}\B_\mp^{(r)}\right).
\end{equation}
Similarly to the operators $\Kop^\pm_{rs}$ in the previous subsection, we will see the operators $\Lop^\pm_{rs}$ as the entries of some $N\times N$ matrix of operators $\Lop_\pm$, so that $\Lop^\pm_{rs} = (\Lop_\pm)_{rs}$.

\paragraph{Action in terms of $\bm{j_\pm^{(r)}}$'s and $\bm{J_\pm^{(r)}}$'s.} The action of the models is obtained as the following inverse Legendre transform of the Hamiltonian (see for instance \cite{Delduc:2019bcl}):
\begin{equation*}
S[g^{(1)},\cdots,g^{(N)}] = \sum_{r=1}^N\iint\text{d}t \,\text{d}x \ \kappa\left(X^{(r)},j_t^{(r)}\right) - \int \text{d}t \ \mathcal{H},
\end{equation*}
where both $X^{(r)}$ and $\mathcal{H}$ should be replaced by their expressions in terms of Lagrangian fields. Recalling the definitions \eqref{Eq:Yr} and \eqref{Wesszumino}, one can rewrite the action in terms of the fields $Y^{(r)}$ making the Wess-Zumino terms of $g^{(r)}$ appear:
\begin{equation*}
S[g^{(1)},\cdots,g^{(N)}] = \sum_{r=1}^N\iint\text{d}t \,\text{d}x \ \kappa\left(Y^{(r)},j_t^{(r)}\right) - \int \text{d}t \ \mathcal{H} + \sum_{r=1}^N \kay_r\, \Ww{g^{(r)}}.
\end{equation*}
From here, reinserting the expression \eqref{Eq:J2Y} of $Y^{(r)}$ in terms of the currents $J_\pm^{(r)}$, we find:
\begin{multline}\label{Eq:ActionNonLorentz}
S[g^{(1)},\cdots,g^{(N)}] = \frac{1}{2} \sum_{r,s}^N \iint\text{d}t \,\text{d}x \ \left[\kappa\left(\Lop_{sr}^+ J_+^{(r)},j_-^{(s)}\right) + \kappa\left(j_+^{(r)},\Lop_{rs}^- J_-^{(s)}\right) \right] - \int \text{d}t \ \mathcal{H} + \sum_{r=1}^N \kay_r\, \Ww{g^{(r)}} \\
 + \frac{1}{2} \sum_{r,s}^N \iint\text{d}t \,\text{d}x \ \left[\kappa\left(\Lop_{sr}^+ J_+^{(r)},j_+^{(s)}\right) + \kappa\left(j_-^{(r)},\Lop_{rs}^- J_-^{(s)}\right) \right]. \ \ \ \ \ \ \ \ \ \ \ \ \ \ \ \ \ \ \ \ \
\end{multline}
We note that the terms in the second line are not Lorentz invariant. However, one shows that these are cancelled by the term containing the Hamiltonian (for brevity, we give the proof of this result in Appendix \ref{Lorentz}), so that we eventually get
\begin{equation}\label{Eq:ActionJj}
S[g^{(1)},\cdots,g^{(N)}] = \frac{1}{2} \sum_{r,s}^N \iint\text{d}t \,\text{d}x \ \left[\kappa\left(\Lop_{sr}^+ J_+^{(r)},j_-^{(s)}\right) + \kappa\left(j_+^{(r)},\Lop_{rs}^- J_-^{(s)}\right) \right] + \sum_{r=1}^N \kay_r\, \Ww{g^{(r)}}.
\end{equation}

\paragraph{Action in terms of Maurer-Cartan currents.} To conclude this subsection, we proceed to compute the expression of the action in terms of the $j_\pm^{(r)}$'s only. This is done through the formal inversion relation \eqref{Eq:j2J}. As a final result we obtain
\begin{equation}\label{Eq:Action}
S[g^{(1)},\cdots,g^{(N)}] = \iint\text{d}t \,\text{d}x \ \sum_{r,s=1}^N \kappa\left(j_+^{(r)},\mathcal{O}_{rs}j_-^{(s)}\right) + \sum_{r=1}^N \kay_r \,\Ww{g^{(r)}},
\end{equation}
where we have defined $\mathcal{O}_{rs}$ as the entries of the following matrix operator:
\begin{equation} \label{O}
\mathcal{O} = \frac{1}{2}\left({^t}\Kop_+^{-1}\,{^t}\Lop_+ + \Lop_-\Kop_-^{-1}\right).
\end{equation}
Finally, using the identities \eqref{Identities}, one proves that the second term in this definition is equal to the first one, so that we get:
\begin{equation} \label{Opm}
\mathcal{O} = {^t}\Kop_+^{-1}\,{^t}\Lop_+ = \Lop_-\Kop_-^{-1}.
\end{equation}

\paragraph{Model with two copies.} \label{2copies} In this paragraph, we give an explicit expression for the inversion of the operator matrices $\Kop_\pm$ and consequently for the coupling operator $\mathcal{O}$ in the case of a model with two copies only, \text{i.e.} with $N = 2$. In order to do so, one has to make a further assumption about the operators $\A_\pm^{(r)}$ appearing in the ansatz \eqref{Currentsr} made for the Kac-Moody currents $\mathcal{J}^{(r)}_\pm$. More precisely, we will suppose that they satisfy the following commutation relation
\begin{equation}\label{Eq:BCom}
\left[ \A_+^{(r)}, \A_-^{(r)} \right] = 0, \;\;\;\;\;\; \forall \, r\in\lbrace 1,2 \rbrace.
\end{equation}
Let us note that, crucially, this additional condition is satisfied by the Yang-Baxter realisation (with or without Wess-Zumino term) and the $\lambda$-realisation, as can be checked easily from Equations \eqref{Eq:OperatorsYB}, \eqref{Eq:OperatorsYBWZ} and \eqref{Eq:OperatorsLambda}\footnote{It is not obvious whether this condition is an accidental property of these particular realisations or if it can be derived more generally as a consequence of the fact that $\mathcal{J}^{(r)}_\pm$ are Kac-moody currents, as was for example the case for the identities \eqref{Identities} (see Appendix \ref{Identities1}).}.

As we have noted in Subsection \ref{Lax}, the fact that it is not straightforward to invert the operator matrices $\Kop_\pm$ is due to the non-commutativity of their entries. However, using the additional assumption \eqref{Eq:BCom} made on the operators $\A^{(r)}_\pm$, one shows that:
\begin{equation}\label{Eq:ComK}
\bigl[\Kop^\pm_{rs},\Kop^\pm_{rt}\bigr] = 0, \ \ \ \ \ \ \forall \, r,s,t \in \lbrace 1,2 \rbrace.
\end{equation}
Thus, even if the entries of $\Kop_\pm$ are not all commutative, this shows that the ones on a same line commute with one another. This fact will allow us to find an explicit expression of the inverse of $\Kop_\pm$.\\

Let us introduce the operators
\begin{equation}\label{Eq:OpDet}
\Delta^\pm_1 = (\Kop^\pm_{11}\;\Kop^\pm_{22} - \Kop^\pm_{12}\;\Kop^\pm_{21})^{-1} \;\;\;\;\;\; \text{ and } \;\;\;\;\;\; \Delta^\pm_2 = (\Kop^\pm_{22}\;\Kop^\pm_{11} - \Kop^\pm_{21}\;\Kop^\pm_{12})^{-1}.
\end{equation}
If the entries $\Kop^\pm_{rs}$ of $\Kop_\pm$ were commutative, the objects $\Delta^\pm_1$ and $\Delta^\pm_2$ would be equal and would correspond to the inverse of the determinant of the $2\times2$ matrix $\Kop_\pm$. In the present case, these operators $\Delta^\pm_r$ are the inverse of non-commutative versions of the determinant. In terms of these, the inverse of the operator $\Kop_\pm$ is then given by
\begin{equation}\label{Eq:KInv}
\Kop^{-1}_\pm = \begin{pmatrix}
\Kop^\pm_{22} \,\Delta^\pm_1 & -\Kop^\pm_{12}\, \Delta^\pm_2 \vspace{4pt} \\
-\Kop^\pm_{21} \,\Delta^\pm_1 & \Kop^\pm_{11}\, \Delta^\pm_2
\end{pmatrix}.
\end{equation}
Indeed, one checks explicitly that
\begin{equation*}
\begin{pmatrix}
\Kop^\pm_{11}  & \Kop^\pm_{12}  \vspace{4pt} \\
\Kop^\pm_{21} & \Kop^\pm_{22} 
\end{pmatrix} \begin{pmatrix}
\Kop^\pm_{22} \,\Delta^\pm_1 & -\Kop^\pm_{12}\, \Delta^\pm_2 \vspace{4pt} \\
- \Kop^\pm_{21}\, \Delta^\pm_1 & \Kop^\pm_{11}\, \Delta^\pm_2
\end{pmatrix}  = \begin{pmatrix}
(\Kop^\pm_{11}\;\Kop^\pm_{22} - \Kop^\pm_{12}\;\Kop^\pm_{21})\Delta^\pm_1 & (\Kop^\pm_{12}\;\Kop^\pm_{11}-\Kop^\pm_{11}\;\Kop^\pm_{12}) \Delta^\pm_2 \vspace{4pt} \\
(\Kop^\pm_{21}\;\Kop^\pm_{22} - \Kop^\pm_{22}\;\Kop^\pm_{21}) \Delta^\pm_1 & (\Kop^\pm_{22}\;\Kop^\pm_{11} - \Kop^\pm_{21}\;\Kop^\pm_{12}) \Delta^\pm_2
\end{pmatrix}.
\end{equation*}
The property \eqref{Eq:ComK} then ensures that the off-diagonal terms vanish, while the definition \eqref{Eq:OpDet} of the operators $\Delta_r^\pm$ is such that the diagonal terms are the identity operator, thus proving that the matrix \eqref{Eq:KInv} is the inverse of $\Kop_\pm$\footnote{More precisely, this proves that it is the right inverse of $\Kop_\pm$. However, recalling that the entries of $\Kop_\pm$ are operators on $\g$, one can see $\Kop_\pm$ as a $2\dim\g \times 2\dim\g$ matrix, for which the left and right inverses coincide.}. The expression \eqref{Eq:KInv} is a non-commutative generalisation of the standard comatrix formula for the inverse of a $2\times 2$ matrix, where in particular one takes into account the non-commutativity of the entries by considering different ``inverse determinants'' $\Delta_r^\pm$ in the different columns.

To give a more compact expression of the entries of $\Kop_\pm^{-1}$, let us introduce the notation $\bar r$, defined for every $r\in\{1,2\}$ by $\bar{r} \in \{1,2\} \setminus r$ (\textit{i.e.} $\bar 1=2$ and $\bar 2=1$). Then, one has
\begin{equation*}
(\Kop^{-1}_\pm)_{rs} = (-1)^{r+s} \, \Kop^\pm_{\bar s \bar r} \,\Delta_s^\pm.
\end{equation*}

Reinserting the above results into the expression \eqref{Opm} of the operator $\mathcal{O}$, one can compute its entries $\mathcal{O}_{rs}$, which appear in the action \eqref{Eq:Action} of the model, yielding
\begin{equation} \label{O2}
\mathcal{O}_{rs} = {^t\!}\Delta_{r}^+ \left({^t}\Kop_{\bar{r}\bar{r}}^+ \;{^t}\Lop_{sr}^+ - {^t}\Kop_{\bar{r}r}^+ \; {^t}\Lop_{s\bar{r}}^+\right) = \left(\Lop_{rs}^-\; \Kop_{\bar{s}\bar{s}}^- - \Lop_{r\bar{s}}^- \;\Kop_{\bar{s}s}^- \right)\Delta_{s}^-.
\end{equation}

\subsection{Parameters of the models}
\label{SubSec:Param}

In Subsection \ref{SubSubSec:Space}, we have discussed what are the defining parameters of the models, from their construction as realisations of affine Gaudin models. Let us briefly give some additional comments on the subject in the light of the Lagrangian formulation of the models. 

\paragraph{Functions $\bm{\vp_\pm(z)}$.} Recall the functions $\vp_\pm(z)$ and $\vp_{\pm,r}(z)=(z-z_r^\pm)\vp_\pm(z)$ introduced in Subsection \ref{Lax}. It is clear from the results of this section that these functions play an important role in describing the Lagrangian formulation of the models. For example, they are used to obtain the Lagrangian expression \eqref{Eq:Interpolation} of the Lax pair. Similarly, they enter the definitions \eqref{Eq:DefK} and \eqref{Eq:DefL} of the operators $\Kop_\pm$ and $\Lop_\pm$, which are then used to express the operator $\mathcal{O}$ appearing in the action \eqref{Eq:Action} of the model. Note that the definition of the operators $\Kop_\pm$ and $\Lop_\pm$ also involves the operators $\A_\pm^{(r)}$ and $\B_\pm^{(r)}$, which characterise the choice of Kac-Moody realisations of the model. In particular, these realisations depend on the levels $\ell^\pm_r$. For completeness, let us thus note that the latter can also be expressed quite easily using the functions $\vp_\pm(z)$ and $\vp_{\pm,r}(z)$: indeed, we have
\begin{equation} \label{Eq:LevelPhiPm}
\ell^\pm_r = -\ell^{\infty} \vp_{\pm,r}(z_r^\pm) \vp_{\mp}(z_r^\pm) = \mp \frac{\ell^\infty}{z_r^+-z_r^-} \vp_{+,r}(z_r^\pm)\vp_{-,r}(z_r^\pm). 
\end{equation}
Finally, let us note that these levels also determine the coefficients $\kay_r$ of the Wess-Zumino terms in the action \eqref{Eq:Action} of the model. Thus, the datum of the functions $\vp_\pm(z)$ is enough to describe completely the model in its Lagrangian formulation.

\paragraph{Parameters $\bm{(z_r^\pm,\nu_r^\pm,\ell^\infty)}$.} Recall that in Subsection \ref{SubSubSec:Space}, we discussed two possible sets of parameters for the model: the ``Gaudin parameters'' $(z_r^\pm,\ell_r^\pm,\ell^\infty)$ and the parameters $(z_r^\pm,\zeta_i,\ell^\infty)$, where the datum of the levels $\ell^\pm_r$ has been replaced by the datum of the zeroes $\zeta_i$ of the twist function. In particular, recall that the second parametrisation is in general more convenient as the zeroes play an important role in the description of the model and as they cannot be expressed explicitly in terms of the levels, whereas the levels can be expressed rationally in terms of the zeroes. Recall also that choosing the parametrisation using the zeroes is however less convenient to describe models with $\lambda$-realisations and/or Yang-Baxter realisations without Wess-Zumino term. Indeed, these realisations require that the levels $\ell^\pm_r$ satisfy the additional constraint \eqref{Eq:ConstraintLevel}, which translates into a complicated algebraic condition on the parameters $(z^\pm_r,\zeta_i)$.

These observations motivate the introduction of a third possible set of parameters $(z_r^\pm,\nu_r^\pm,\ell^\infty)$, which is in some sense intermediate between the two sets described above and which circumvents the various issues related to solving algebraic equations. In this parametrisation, the datum of the levels $\ell^\pm_r$ or of the zeroes $\zeta_i$ is replaced by the datum of the coefficients 
\begin{equation*}
\nu_r^\pm = \vp_{\pm,r}(z_r^\pm).
\end{equation*}
Note that these coefficients characterise the partial fraction decomposition of the functions $\vp_\pm(z)$:
\begin{equation*}
\vp_\pm(z) = 1 + \sum_{r=1}^N \frac{\nu_r^\pm}{z-z_r^\pm}.
\end{equation*}
In particular, the levels $\ell^\pm_r$ can then be expressed in terms of these parameters as
\begin{equation*}
\ell^\pm_r = - \ell^\infty \nu_r^\pm \left( 1 + \sum_{s=1}^N \frac{\nu_s^\mp}{z_r^\pm-z_s^\mp} \right).
\end{equation*}
Thus, the condition \eqref{Eq:ConstraintLevel}, which the levels $\ell^\pm_r$ should satisfy in order to attach a $\lambda$-realisation or a Yang-Baxter realisation without Wess-Zumino term to the sites $(r,\pm)$, becomes
\begin{equation}\label{Eq:ConstraintNu}
\nu_r^+ \left( 1 + \sum_{s=1}^N \frac{\nu_s^-}{z_r^+-z_s^-} \right) + \nu_r^- \left( 1 + \sum_{s=1}^N \frac{\nu_s^+}{z_r^--z_s^+} \right) = 0.
\end{equation}
If one considers a model with $N_2$ $\lambda$-realisations, one has to impose $N_2$ relations as the one above, which form a linear system on the corresponding set of coefficients $\nu_r^+$ (or equivalently on the corresponding $\nu_r^-$). This is the advantage of this parametrisation, as one then has to solve linear constraints on the parameters instead of algebraic ones when using the zeroes. In particular, the solutions of these constraints are rational expressions of the remaining free parameters (however potentially quite complicated). This will be useful later in Subsection \ref{SubSubSec:Nlambda} when we will study the model with $N$ coupled $\lambda$-models.

\paragraph{Coefficients $\bm{\rho_{rs}^\pm}$.} Let us end this subsection by introducing some coefficients which will be useful to study the undeformed limit of the models in Subsection \ref{SubSec:Undef} and specific examples of models in Section \ref{Sec:Examples}. We define
\begin{subequations}\label{Eq:Rho}
\begin{align}
\rho_{rr}^\pm &= \mp\frac{\ell^{\infty}}{2}\varphi_{\pm,r}(z_r^\pm)\frac{\varphi_{\mp,r}(z_r^\mp)-\varphi_{\mp,r}(z_r^\pm)}{z_r^\mp - z_r^\pm},\\
\rho_{rs}^\pm &= \mp\frac{\ell^{\infty}}{2} \frac{\varphi_{\mp,r}(z_r^\mp)\varphi_{\pm,s}(z_s^\pm)}{z_r^\mp-z_s^\pm}, \ \ \ \text{for} \ r \ne s.
\end{align}
\end{subequations}
Using the expression \eqref{Eq:LevelPhiPm} of the levels $\ell^\pm_r$, one shows that these coefficients can be rewritten as
\begin{equation}\label{Eq:Rho2}
\rho^\pm_{rs} =\pm \frac{1}{2} \left( \ell^\pm_r \, \delta_{rs} + \ell_r^\mp\,\frac{\vp_{\pm,s}(z_s^\pm)}{\vp_{\pm,s}(z_r^\mp)} \right).
\end{equation}
Using this expression, the operators $\Kop^\pm_{rs}$ and $\Lop^\pm_{rs}$ introduced in \eqref{Eq:DefK} and \eqref{Eq:DefL} can be re-expressed as
\begin{equation}\label{Eq:KandLfromRho}
\Kop^\pm_{rs} = \left({^t}\A_\pm^{(r)} - \frac{\ell_r^\pm}{\ell_r^\mp}\, {^t}\A_\mp^{(r)} \right)\delta_{rs} \pm \frac{2\rho_{rs}^\pm}{\ell_r^\mp}\; {^t}\A_\mp^{(r)}, \;\;\;\;\;\;\; \Lop^\pm_{rs} = \pm\frac{1}{2}\left({^t}\B_\pm^{(r)} - \frac{\ell_r^\pm}{\ell_r^\mp}\, {^t}\B_\mp^{(r)} \right)\delta_{rs} + \frac{\rho_{rs}^\pm}{\ell_r^\mp}\; {^t}\B_\mp^{(r)}.
\end{equation}

\subsection{Undeformed limit}
\label{SubSec:Undef}

As explained in Subsection \ref{SubSubSec:UndefHam}, one can see the model constructed from $N_1$ inhomogeneous Yang-Baxter realisations and $N_2$ $\lambda$-realisations as a deformation of a simpler model, coupling together $N_1$ copies of the PCM with Wess-Zumino term and $N_2$ copies of the non-abelian T-dual of the PCM. This was understood by means of the so-called undeformed limit, in which the positions $z_r^+$ and $z_r^-$ collide for every pair of sites $(r,\pm)$, or equivalently by letting the parameters $\eta_r$ go to $0$, while keeping $\ell^{[0]}_r$ and $\ell^{[1]}_r$ finite (see Subsection \ref{SubSubSec:UndefHam} for details). The goal of this subsection will be to complete this discussion by studying this limit in the Lagrangian formulation of the model, focusing mostly on Yang-Baxter realisations (as explained in Subsection \ref{SubSubSec:UndefHam}, the undeformed limit of $\lambda$-realisations requires a more subtle treatment which we will not detail here for conciseness). In particular, this will allow us to compare the methods and results presented in the previous subsections for deformed models to the ones presented in~\cite{Delduc:2018hty,Delduc:2019bcl} for undeformed ones.

\paragraph{Interpolation formula.} Let us focus for the moment on a pair of sites $(r,\pm)$, which we suppose to be associated with a Yang-Baxter realisation with Wess-Zumino term. The corresponding operators $\A^{(r)}_\pm$ are then given by
\begin{equation*}
\A^{(r)}_\pm = \frac{1}{2} \left( \text{Id} \mp \frac{1}{c_r} R^{(r)} \mp \delta_r \Pi^{(r)} \right),
\end{equation*}
where $\delta_r$ is defined in Equation \eqref{Eq:YDelta}, $R^{(r)}=\Ad_{g^{(r)}}^{-1}\circ R_r\circ\Ad_{g^{(r)}}$ and $\Pi^{(r)}=\Id-R^{(r)\,2}/c_r^2$. Recall that in Subsection \ref{Lax}, we found the Lagrangian expression of the Lax pair by interpolation methods, using the fact that one can express the Maurer-Cartan currents $j^{(r)}_\pm$ in terms of the evaluation of the Lax pair at the positions $z_r^\pm$ by Equation \eqref{Eq:jL}. In the present case, this equation can be rewritten as
\begin{equation}\label{Eq:InterpolationYB}
j^{(r)}_\pm = \frac{\Lc_\pm(z_r^+) + \Lc_\pm(z_r^-)}{2} + \left( R^{(r)} - c_r\delta_r \Pi^{(r)} \right) \frac{\Lc_\pm(z_r^+) - \Lc_\pm(z_r^-)}{2c_r}.
\end{equation}
As recalled above, the undeformed limit corresponds to making the positions $z_r^\pm$ collide to the same point $z_r$. It is then clear that in the undeformed limit, the above formula simply becomes
\begin{equation}\label{Eq:InterpolationPCM}
j_\pm^{(r)} =  \mathcal{L}_\pm(z_r).
\end{equation}
This is precisely the interpolation formula obtained in~\cite[Equation (3.33)]{Delduc:2019bcl} for the model coupling $N$ PCM with Wess-Zumino terms. In this reference, this formula plays a key role in obtaining the Lagrangian expression of the Lax pair of this model. The method developed in Subsection \ref{Lax} of this article is thus a generalisation of the one of~\cite{Delduc:2019bcl} to include deformed realisations.

Recall from Equation \eqref{Eq:J} that in the deformed model, the currents $J_\pm^{(r)}$ are defined as the evaluations $\Lc_\pm(z^\pm_r)$. It is then clear from the above equation that in the undeformed limit, these currents $J_\pm^{(r)}$ coincide with the Maurer-Cartan currents $j^{(r)}_\pm$. The expression \eqref{Eq:Interpolation} of the Lax pair in the present article is thus a natural deformation of the one (3.34) of~\cite{Delduc:2019bcl} for the undeformed model. Moreover, this implies that the operator $\Kop_\pm$, which relates the currents $j^{(r)}_\pm$ and $J^{(r)}_\pm$ (see Equation \eqref{Eq:J2j}), becomes the identity in the undeformed limit, or equivalently, in components:
\begin{equation}\label{Eq:LimitK}
\Kop^\pm_{rs} \xrightarrow{\eta_1,\cdots,\eta_N\to 0} \delta_{rs} \text{Id}.
\end{equation}

For completeness, let us comment briefly on the homogeneous Yang-Baxter limit considered at the end of Subsection \ref{SubSubSec:UndefHam} (note that we considered the homogeneous limit only for realisations without Wess-Zumino term, in which case $\kay_r=\delta_r=0$). Recall that this limit corresponds to taking the coefficient $c_r$ to $0$. Recall also that the positions $z_r^\pm$ are given by $z_r\pm c_r\eta_r$. Thus, in the limit $c_r \to 0$, the equation \eqref{Eq:InterpolationYB} becomes
\begin{equation}\label{Eq:InterpolationhYB}
j_\pm^{(r)} =  \mathcal{L}_\pm(z_r) + \eta_r R^{(r)} \Lc_\pm'(z_r),
\end{equation}
where $\Lc_\pm'(z)$ denotes the derivative of $\Lc_\pm(z)$ with respect to the spectral parameter $z$. This is the equivalent of the equation (D.7) of~\cite{Delduc:2019bcl}, which was obtained when studying a model with $N-1$ PCM realisations and one homogeneous Yang-Baxter realisation. It is interesting to compare the equations \eqref{Eq:InterpolationYB}, \eqref{Eq:InterpolationPCM} and \eqref{Eq:InterpolationhYB}: the undeformed interpolation formula \eqref{Eq:InterpolationPCM} is corrected by a derivative term $\Lc_\pm'(z_r)$ for an homogeneous Yang-Baxter deformation and by a finite difference term $\bigl(\Lc_\pm(z_r+c_r\eta_r)-\Lc_\pm(z_r-c_r\eta_r)\bigr)/2c_r$ for an inhomogeneous Yang-Baxter realisation.

\paragraph{Lagrangian expression of $\bm{Y^{(r)}}$.} Recall from Subsection \ref{Legendre} that after the derivation of the Lagrangian expression of the Lax pair, the next step for performing the inverse Legendre transform of the model is to find the Lagrangian expression of the field $Y^{(r)}$, which encodes the momentum fields of the model. This was done using Equation \eqref{Eq:YfromL}, which expresses $Y^{(r)}$ in terms of the Lax pair, through the operators $\B^{(r)}_\pm$. For a Yang-Baxter realisation, it can be re-written, after a few manipulations, as
\begin{equation*}
Y^{(r)} = \left( \ell^{[1]}_r \text{Id} - \eta_r\kay_r R^{(r)} + c_r \eta_r \kay_r\delta_r \Pi^{(r)} \right) \frac{\Lc(z^+_r)-\Lc(z_r^-)}{2c_r\eta_r} - \kay_r  \frac{\Lc(z^+_r)+\Lc(z_r^-)}{2}.
\end{equation*}
The undeformed limit correspond to taking $\eta_r$ to 0 while keeping $\ell^{[1]}_r$ and $\kay_r=-\ell^{[0]}_r/2$ finite. Recalling that $z^\pm_r=z_r\pm c_r\eta_r$, the above equation then becomes in this limit
\begin{equation*}
Y^{(r)} = \ell^{[1]}_r \Lc'(z_r) - \kay_r \Lc(z_r).
\end{equation*}
This then coincides with the equation (3.36) of~\cite{Delduc:2019bcl}.\\

Recall from Subsection \ref{Legendre} that Equation \eqref{Eq:YfromL} allows us to rewrite $Y^{(r)}$ in terms of the currents $J^{(r)}_\pm$ and the operators $\Lop^\pm_{rs}$, in Equation \eqref{Eq:J2Y}. In the undeformed limit, the currents $J_\pm^{(r)}$ are identified with the Maurer-Cartan currents $j_\pm^{(r)}$. Moreover, one can study the behaviour of the undeformed limit of the operators $\Lop^\pm_{rs}$ using their expression \eqref{Eq:KandLfromRho}. In particular, the coefficients $\rho^\pm_{rs}$ in this expression, defined by Equation \eqref{Eq:Rho}, can be shown to converge in the undeformed limit to:
\begin{equation}\label{Eq:LimRho}
\rho^+_{rs} \xrightarrow{\eta_1,\cdots,\eta_N\to 0} \rho_{sr} - \frac{\kay_r}{2}\delta_{rs} \;\;\;\;\;\; \text{ and } \;\;\;\;\; \rho^-_{rs} \xrightarrow{\eta_1,\cdots,\eta_N\to 0} \rho_{rs} + \frac{\kay_r}{2}\delta_{rs},
\end{equation}
with the coefficients $\rho_{rs}$ as defined in~\cite[Equation (3.40)]{Delduc:2019bcl}. Note that in this limit, the expression of the coefficient $\kay_r$ also coincides with its expression in~\cite[Equation (3.38)]{Delduc:2019bcl}. Using the above limit of the coefficients $\rho_{rs}^\pm$, as well as the expression \eqref{Eq:ParamReal} of the levels $\ell^\pm_r$ in terms of the coefficients $\ell^{[0]}_r=-2\kay_r$ and $\ell^{[1]}_r$ which stay finite in the undeformed limit, one can compute the limit of the operators $\Lop^\pm_{rs}$ starting from their expression \eqref{Eq:KandLfromRho}:
\begin{equation}\label{Eq:LimitL}
\Lop^+_{rs} \xrightarrow{\eta_1,\cdots,\eta_N\to 0} \rho_{sr}\, \text{Id} \;\;\;\;\;\; \text{ and } \;\;\;\;\;\; \Lop^-_{rs} \xrightarrow{\eta_1,\cdots,\eta_N\to 0} \rho_{rs} \,\text{Id}.
\end{equation}
In particular, Equation \eqref{Eq:J2Y} agrees with~\cite[Equation (3.39)]{Delduc:2019bcl} in the undeformed limit:
\begin{equation*}
Y^{(r)} \xrightarrow{\eta_1,\cdots,\eta_N\to 0} \sum_{s=1}^N \left(\rho_{sr} j_+^{(s)} + \rho_{rs} j_-^{(s)} \right).
\end{equation*}

\paragraph{Action.} Finally, we are now in a position to calculate the undeformed limit of the action of the model with $N$ copies of the Yang-Baxter realisation. By reinserting the limits \eqref{Eq:LimitK} and \eqref{Eq:LimitL} in the expression \eqref{O} for the operator $\mathcal{O}$, we find:
\begin{equation*}
\mathcal{O}_{rs} \xrightarrow{\eta_1,\cdots,\eta_N\to 0} \rho_{rs} \,\text{Id}.
\end{equation*}
Comparing to Equation (3.49) of \cite{Delduc:2019bcl}, one sees that the action \eqref{Eq:Action} then reduces to the one of $N$ coupled copies of the PCM with Wess-Zumino term:
\begin{equation}\label{Eq:ActionUndef}
S\bigl[g^{(1)},\cdots,g^{(N)}\bigr] = \iint\text{d}t\, \text{d}x \ \sum_{r,s=1}^N \rho_{rs}\, \kappa\left(j_+^{(r)},j_-^{(s)}\right) + \sum_{r=1}^N \kay_r \Ww{g^{(r)}}.
\end{equation}

\paragraph{Undeformed and $\bm q$-deformed symmetries.} The undeformed model \eqref{Eq:ActionUndef} possesses $N$ global symmetries acting by left translation on the fields $g^{(r)}$:
\begin{equation}\label{Eq:LeftSym}
g^{(1)}(t,x) \longmapsto h_1 g^{(1)}(t,x), \;\;\; \cdots \;\; , \;\;\; g^{(N)}(t,x) \longmapsto h_N g^{(N)}(t,x),
\end{equation}
where $h_1,\cdots,h_N$ are constant elements of $G_0$. Indeed, these transformations leave the Maurer-Cartan currents $j^{(r)}_\pm=g^{(r)\,-1}\p_\pm g^{(r)}$ and the Wess-Zumino terms $\Ww{g^{(r)}}$ invariant.

These global symmetries are broken by the introduction of deformations. Indeed, let us consider the model with $N$ copies of the Yang-Baxter model studied in this subsection. The entries of the operators $\Kop_\pm$ and $\Lop_\pm$ are expressed in terms of the operators $R^{(r)}=\Ad_{g^{(r)}}^{-1}\circ R_r \circ \Ad_{g^{(r)}}$ and $\Pi^{(r)}=\Ad_{g^{(r)}}^{-1}\circ \Pi_r \circ \Ad_{g^{(r)}}$. Because of their dependence on the fields $g^{(r)}$, these operators are not invariant under the left translations \eqref{Eq:LeftSym}, making the operator $\mathcal{O}$ appearing in the deformed action \eqref{Eq:Action} not invariant. Thus, the transformations \eqref{Eq:LeftSym} are not symmetries of the action \eqref{Eq:Action}.

It is a well-known result~\cite{Delduc:2013fga} (see also
\cite{Kawaguchi:2011pf,Kawaguchi:2012ve,Kawaguchi:2012gp}) that in the Yang-Baxter model (with one copy and without Wess-Zumino term), this broken symmetry is in fact deformed into a $q$-deformed Poisson-Lie symmetry. Based on this result, it was explained in~\cite{Delduc:2019bcl} that this is in general the case for every affine Gaudin model with a Yang-Baxter realisation (without Wess-Zumino term). In particular, the model coupling $N$ copies of the Yang-Baxter models without Wess-Zumino term then possesses $N$ $q$-deformed symmetries, which replace the translation symmetries \eqref{Eq:LeftSym}. Their action on the fields of the model can be computed using the results of~\cite{Delduc:2016ihq}: in particular, let us note that this action is non-local.

As the bilinear form $\kappa$ is invariant under conjugacy, the undeformed model also possesses a global symmetry acting by simultaneous right translation on all the fields $g^{(r)}$:
\begin{equation}\label{Eq:RightSym}
g^{(1)}(t,x) \longmapsto g^{(1)}(t,x)\,h, \;\;\; \cdots\;\;, \;\;\; g^{(1)}(t,x) \longmapsto g^{(N)}(t,x)\,h,
\end{equation}
with $h$ a constant element of $G_0$. As explained in~\cite{Delduc:2019bcl}, it corresponds to the diagonal symmetry of the underlying affine Gaudin model. As such, it is not broken by applying Yang-Baxter deformations to the various copies of the model. Indeed, one checks that under the transformation \eqref{Eq:RightSym}, the operators $\mathcal{O}_{rs}$ entering the action of the model with $N$ Yang-Baxter realisations become $\Ad_h^{-1} \circ \mathcal{O}_{rs} \circ \Ad_h$. Since the Maurer-Cartan currents $j^{(r)}_\pm$ become $\Ad_h^{-1} j^{(r)}_\pm$ and the Wess-Zumino terms $\Ww{g^{(r)}}$ are invariant under this transformation, it is thus a symmetry of the deformed action \eqref{Eq:Action}. Note that a similar result holds for models involving $\lambda$-realisations: in this case, the corresponding fields $g^{(r)}$ should not transform by right multiplication but by conjugacy $g^{(r)}\mapsto h^{-1} g^{(r)} h$, while the fields corresponding to Yang-Baxter realisations still transform by right multiplication by $h$.

\section[Yang-Baxter and \texorpdfstring{$\bm\lambda$}{$\lambda$}-deformed coupled models]{Yang-Baxter and $\bm\lambda$-deformed coupled models}
\label{Sec:Examples}

The action \eqref{Eq:Action} presented in the previous section was obtained using the general ansatz introduced in Subsection \ref{Kac-Moodycurrents} for the form of the Kac-Moody realisations defining the model. In this section, we specialise these results to the model constructed from $N_1$ copies of the Yang-Baxter realisation and $N_2$ copies of the $\lambda$-realisation. As we shall see, the particular form of these realisations will allow us to rewrite the action of this model in a simpler form. In particular, we will show that the integrable $\s$-model introduced in~\cite{Georgiou:2018gpe} corresponds to a particular limit of the model constructed from $N$ copies of the $\lambda$-realisation. We will then focus on models with two copies and will rewrite their action in a more explicit form, using the expressions~\eqref{Eq:KInv} and~\eqref{O2} of the inverse of $\Kop_\pm$ and of the operators $\mathcal{O}_{rs}$ obtained in this case.

\subsection[Deformed model with $N_1$ Yang-Baxter realisations and $N_2$ $\lambda$-realisations]{Deformed model with $\bm{N_1}$ Yang-Baxter realisations and $\bm{N_2}$ $\bm\lambda$-realisations}

Let us consider a model made up of $N_1$ copies of the Yang-Baxter realisation with Wess-Zumino term and $N_2$ copies of the $\lambda$-realisation. Let us now associate the former to the first $N_1$ pairs of sites $(r,\pm)$ and the latter to the last $N_2$ pairs. Then, from \eqref{Eq:OperatorsYBWZ} and \eqref{Eq:OperatorsLambda}, one obtains, for the operators $\A^{(r)}_\pm$ and $\B^{(r)}_\pm$, the following expression
\begin{align*}
\A^{(r)}_\pm &= \frac{1}{2}\text{Id} \mp \frac{1}{2c_r}R^{(r)} \mp \frac{\delta_r}{2}\Pi^{(r)}, \;\;\;\; \B^{(r)}_\pm = \left(\ell_r^\pm + \frac{\kay_r}{2}\right)\text{Id} \mp \frac{\kay_r}{2c_r} R^{(r)} \mp \frac{\kay_r\delta_r}{2}\Pi^{(r)}, & \;\;\;\;\;\;\;\;\;\; 1 \le r \le N_1, \\
\A^{(r)}_+ &= \text{Id}, \;\; \A^{(r)}_- = -\Ad_{g^{(r)}}, \;\; \B^{(r)}_+ = -\kay_r\,\text{Id}, \;\; \B^{(r)}_- = -\kay_r\, \Ad_{g^{(r)}}, & N_1 < r \le N_2,
\end{align*}
where $R^{(r)}=\Ad_{g^{(r)}}^{-1}\circ R_r \circ \Ad_{g^{(r)}}$ and $\Pi^{(r)}=\Ad_{g^{(r)}}^{-1}\circ \Pi_r \circ \Ad_{g^{(r)}}$. We observe that the relations $\B^{(r)}_\pm = \ell_r^\pm + \kay_r \A^{(r)}_\pm$ and $\B^{(r)}_\pm = \mp \kay_r \A^{(r)}_\pm$ respectively hold in the first and in the second case. Thus, from \eqref{Eq:DefK} and \eqref{Eq:DefL} and after a few manipulations, we obtain for the entries of the operator $\Lop_\pm$:
\begin{subequations} \label{N1N2L}
\begin{align}
\Lop_{rs}^\pm &= \rho_{rs}^\pm\text{Id} \pm \frac{\kay_r}{2} \Kop_{rs}^\pm, & \hspace{-75pt} 1 \le r \le N_1, \label{Eq:KLYB} \\
\Lop_{rs}^\pm &= -\kay_r\delta_{rs}{^t}\A^{(r)}_\pm + \frac{\kay_r}{2} \Kop_{rs}^\pm, & \hspace{-55pt}  N_1 < r \le N_2.
\end{align}
\end{subequations}
where the coefficients $\rho_{rs}^\pm$ have been defined in \eqref{Eq:Rho}. 

From the expressions \eqref{O} and \eqref{Opm} of the operator $\mathcal{O}$ found in the previous section, we are now in a position to write the action of the model. We choose to express the entries $\mathcal{O}_{rs}$ of this operator as in \eqref{O} for $1 \le r \le N_1$ and as in the second equality in \eqref{Opm} for $N_1 < r \le N_2$\footnote{This choice makes the discussion of the cases $(N_1 = N,N_2 = 0)$ and $(N_1 = 0,N_2 = N)$ in the next subsections simpler. However, we note that due to the relation \eqref{Opm}, different choices are possible in general (see for example Subsection \ref{SubSec:Models2}).}. Reinserting \eqref{N1N2L} in the form of the action \eqref{Eq:Action}, we obtain
\begin{align}\label{N1N2Action}
S\bigl[g^{(1)},\cdots,g^{(N)}\bigr] &= \frac{1}{2}\iint\text{d}t\, \text{d}x \ \sum_{r=1}^{N_1}\sum_{s,t=1}^{N} \kappa\left(j_+^{(r)},\left(\alpha^+_{st}\,{^t}(\Kop_+^{-1})_{tr} + \alpha^-_{rt} \, (\Kop_-^{-1})_{ts}\right)j_-^{(s)}\right) + \sum_{r=1}^{N_1} \kay_r\, \Ww{g^{(r)}}\nonumber \\
& \hspace{20pt} + \iint\text{d}t\, \text{d}x\sum_{r=N_1+1}^{N}\sum_{s=1}^{N} \kay_r \, \kappa\left(\Ad_{g^{(r)}} j_+^{(r)},(\Kop_-^{-1})_{rs}j_-^{(s)}\right)  + \sum_{r=N_1+1}^{N} S_{\text{WZW,}\,\kay_r}[g^{(r)}],
\end{align}
with
\begin{align*}
\alpha^\pm_{rs} &= \rho_{rs}^\pm, &  \hspace{-105pt} 1 \le r \le N_1, \\
\alpha^+_{rs} &= -\kay_{r}\delta_{rs}, &  \hspace{-105pt} N_1 < r \le N
\end{align*}
and where $S_{\text{WZW,}\,\kay_r}[g^{(r)}]$ denotes the Wess-Zumino-Witten action of $g^{(r)}$ with level $\kay_r$:
\begin{equation*}
S_{\text{WZW,}\,\kay_r}[g^{(r)}] = \frac{\kay_r}{2} \iint\text{d}t \,\text{d}x \; \kappa\bigl(j^{(r)}_+, j^{(r)}_-\bigr) + \kay_r\,\Ww{g^{(r)}}.
\end{equation*}

\subsubsection[Model with $N$ Yang-Baxter realisations]{Model with $\bm N$ Yang-Baxter realisations}

Let us now briefly discuss the model with copies of the Yang-Baxter realisation only. In this case, the action \eqref{N1N2Action} gets simplified to
\begin{equation}\label{N1Action}
S\bigl[\bigl\lbrace g^{(r)}\bigr\rbrace\bigr] = \frac{1}{2} \iint \dd t\,\dd x\, \sum_{r,s = 1}^N \kappa\Bigl(g^{(r)\,-1}\p_+ g^{(r)},\, \bigl(\null^{t}\Kop_+^{-1}\;\null^{t\!}\rhom_+ + \rhom_- \, \Kop_-^{-1} \bigr)_{rs}\, \,g^{(s)\,-1}\p_- g^{(s)} \Bigr)  + \sum_{r=1}^N \kay_r \, \Ww{g^{(r)}},
\end{equation}
where $\rhom_\pm$ are operators on $\g_0^N$ which can be seen as $N\times N$ matrices with entries $(\rhom_\pm)_{rs}=\rho^\pm_{rs}\,\Id$.

Let us describe more explicitly the operators $\Kop_\pm$ appearing in the action \eqref{N1Action}. From the expressions of the operators $\A_\pm^{(r)}$ and $\B_\pm^{(r)}$ for a Yang-Baxter realisation, one finds that
\begin{equation}\label{Eq:UYB}
\mathcal{U}_\pm = \Id \pm \widetilde{\mathcal{R}}^\pm \vartheta^\pm,
\end{equation}
where we defined
\begin{equation*}
\widetilde{\mathcal{R}}^\pm_{rs} = \left(R^{(r)}  \mp c_r \Id - c_r\delta_r \Pi^{(r)} \right)\delta_{rs},
\end{equation*}
with $R^{(r)}=\Ad_{g^{(r)}}^{-1}\circ R_r\circ\Ad_{g^{(r)}}$ and $\Pi^{(r)}=\Id-R^{(r)\,2}/c_r^2$, and
\begin{equation}\label{Eq:Theta}
\vartheta^{\pm}_{rs} = \theta^\pm_{rs}\,\Id, \hspace{50pt} \theta_{rs}^\pm = \frac{\mp\rho_{rs}^\pm - \kay_r\delta_{rs}}{c_r\ell_r^\mp}.
\end{equation}

Let us end this subsection by presenting an alternative form of the action of the model. Let us introduce the operator $c$, with entries $c_{rs}=c_r\delta_{rs}\,\Id$. Then, one can further rewrite the operator $\Kop_\pm$ as
\begin{equation*}
\Kop_\pm = \bigl(\Id \pm \widetilde{\mathcal{R}}\,\widetilde{\vartheta}^\pm\bigr)\bigl(\Id-c\vartheta^\pm\bigr),
\end{equation*}
where
\begin{equation*}
\widetilde{\mathcal{R}}_{rs} = \left( R^{(r)} - c_r \delta_r \Pi^{(r)} \right)\delta_{rs} \;\;\;\;\;\;\; \text{ and } \;\;\;\;\;\;\; \widetilde{\vartheta}^\pm = \frac{\vartheta^\pm}{\Id - c \vartheta^\pm}.
\end{equation*}
Finally, introducing $\widetilde{\rhom}_\pm=\rhom_\pm(\Id-c\vartheta^\pm)^{-1}$, one can rewrite the action of the model in the form
\begin{align*}
S[\{g^{(r)}\}] &= \frac{1}{2} \iint \dd t\,\dd x\, \sum_{r,s = 1}^N \kappa\left(g^{(r)\,-1}\p_+ g^{(r)}, \left(\frac{1}{\Id+\null^{t}\widetilde{\vartheta}^+\,\null^{t}\widetilde{\mathcal{R}}}\;\null^{t}\widetilde{\rhom}_+ + \widetilde{\rhom}_- \, \frac{1}{\Id-\widetilde{\mathcal{R}}\;\widetilde{\vartheta}^-} \right)_{rs}\, g^{(s)\,-1}\p_- g^{(s)} \right) \\ & \hspace{20pt} + \sum_{r=1}^N \kay_r \, \Ww{g^{(r)}}.
\end{align*}
This way of writing the action of the model is quite similar to the way the action of the Yang-Baxter model with one copy is expressed and thus seems rather natural. Let us note however that it has some downsides compared to the expression \eqref{N1Action}. Indeed, the entries $\widetilde{\rho}^{\,\pm}_{rs}$ and $\widetilde{\theta}^{\,\pm}_{rs}$ of the operators $\widetilde{\rhom}_\pm$ and $\widetilde{\vartheta}^\pm$ appearing in the expression above are not straightforwardly expressed in terms of the parameters of the models (contrarily to the coefficients $\rho^\pm_{rs}$ and $\theta^\pm_{rs}$ which were used in the previous formulation) as their definition involves the inversion of the operator $\Id-c\vartheta^\pm$.

From the expression of the action above, one can simply check that its undeformed limit yields the action of $N$ coupled PCMs with Wess-Zumino terms presented in \cite{Delduc:2019bcl}. Indeed, in this limit, the parameters $\theta^\pm_{rs}$ and thus also the operators $\widetilde{\theta}^\pm$, go to zero. In particular, the coefficients $\widetilde{\rho}_{rs}^{\,\pm}$ and $\rho^\pm_{rs}$ have the same limit. From Equation \eqref{Eq:LimRho}, we then see that in this limit, $\widetilde{\rho}_{sr}^{\,+}+\widetilde{\rho}_{rs}^{\,-}\to 2\,\rho_{rs}$, with $\rho_{rs}$ as defined in \cite{Delduc:2019bcl}.

\subsubsection[Model with $N$ $\lambda$-realisations]{Model with $\bm N$ $\bm\lambda$-realisations}
\label{SubSubSec:Nlambda}

\paragraph{Action.} Let us now discuss the case where we take $\lambda$-realisations only. For this model, the action reads
\begin{equation} \label{N2Action}
S[\{g^{(r)}\}] = \sum_{r=1}^N S_{\text{WZW,}\,\kay_r}\bigl[g^{(r)}\bigr] + \iint \dd t\,\dd x\; \sum_{r,s=1}^N \kay_r\,\kappa\left( \p_+ g^{(r)} g^{(r)\,-1}, \left(\Kop_-^{-1}\right)_{rs} \,g^{(s)\,-1}\p_- g^{(s)} \right).
\end{equation}
From the expression of the operators $\A_\pm^{(r)}$ of the $\lambda$-realisation, one can rewrite the operator $\Kop_-$ as
\begin{equation}\label{Eq:ULambda}
\Kop_- = \mathcal{M} - \mathcal{D}^{-1}, \;\;\;\;\;\;\;\; \text{ with } \;\;\;\;\;\;\;\;\; \mathcal{M}_{rs} = \mu_{rs}\,\Id \;\;\;\;\;\; \text{ and } \;\;\;\;\;\; \mathcal{D}_{rs} = \Ad_{g^{(r)}}\delta_{rs}, 
\end{equation}
where the coefficients $\mu_{rs}$ are defined as
\begin{equation}\label{Eq:Mu1}
\mu_{rs} = \frac{\vp_{-,s}(z_s^-)}{\vp_{-,s}(z_r^+)}.
\end{equation}
The action of the model then takes the simple form
\begin{equation}\label{Eq:ActionNLambda}
S[\{g^{(r)}\}] = \sum_{r=1}^N S_{\text{WZW,}\,\kay_r}\bigl[g^{(r)}\bigr] + \iint \dd t\,\dd x\; \sum_{r,s=1}^N \kay_r\,\kappa\left( \p_+ g^{(r)} g^{(r)\,-1}, \left(\frac{1}{\mathcal{M}-\mathcal{D}^{-1}}\right)_{rs} \,g^{(s)\,-1}\p_- g^{(s)} \right).
\end{equation}

\paragraph{Parameters.} Let us discuss what are the defining parameters of the model. We will use the parameterisation $(z_r^\pm,\nu_r^\pm,\ell^\infty)$ introduced in Subsection \ref{SubSec:Param}. As explained in this subsection, these parameters are convenient to take into account the fact that the levels $\ell^\pm_r$ of the models should satisfy the constraints $\ell^+_r+\ell^-_r=0$, that one has to impose to consider $\lambda$-realisations. Indeed, these constraints translate into the conditions \eqref{Eq:ConstraintNu} on the parameters $z_r^\pm$ and $\nu_r^\pm$. One can solve this condition by expressing the parameters $\nu_r^+$ in terms of $z_r^\pm$ and $\nu_r^-$:
\begin{equation}\label{Eq:NuLambda}
\nu_r^+ = \sum_{s=1}^N (\beta^{-1})_{rs} \nu_s^-, \;\;\;\;\;\;\; \text{ where } \;\;\;\;\;\; \beta = \left( \left( 1 + \sum_{t=1}^N \frac{\nu_t^-}{z_r^+-z_t^-} \right) \delta_{rs} + \frac{\nu_r^-}{z_r^--z_s^+} \right)_{r,s=1,\cdots,n}.
\end{equation}
The remaining $3N+1$ parameters $(z_r^\pm,\nu_r^-,\ell^\infty)$ are unconstrained: taking into account the translation and dilation redundancy among these parameters (see Subsection \ref{SubSubSec:Space}), the model is thus defined by $3N-1$ free parameters (for concreteness, one can for example fix this redundancy by fixing the values of $\ell^\infty$ and of one of the positions $z_r^\pm$). The coefficients $\mu_{rs}$ defined in Equation \eqref{Eq:Mu1} can be expressed in terms of this parametrisation as
\begin{equation}\label{Eq:Mu}
\mu_{rs} = \frac{\nu_s^-}{z_r^+-z_s^-} \left( 1 + \sum_{t=1}^N \frac{\nu_t^-}{z_r^+-z_t^-} \right)^{-1}.
\end{equation}
Similarly, the coefficient $\kay_r$ appearing in the action \eqref{Eq:ActionNLambda} is given by
\begin{equation*}
\kay_r = \frac{1}{2} \ell^\infty \nu_r^+ \left( 1 + \sum_{s=1}^N \frac{\nu_s^-}{z_r^+-z_s^-} \right),
\end{equation*}
where $\nu_r^+$ is replaced by its expression \eqref{Eq:NuLambda}.

\paragraph{Comparison with~\cite{Georgiou:2018gpe}.} Actions of the form~\eqref{Eq:ActionNLambda} have been considered in~\cite{Georgiou:2018gpe} (and in~\cite{Georgiou:2016urf,Georgiou:2017jfi,Georgiou:2018hpd} for the case $N=2$, see Subsection \ref{SubSubSec:TwoLambdas}). More precisely, the action \eqref{Eq:ActionNLambda} is identical to the action (2.13) of~\cite{Georgiou:2018gpe}, with the matrix $\lambda^{-1}$ in this reference identified in the present language with the matrix whose components are $\lambda^{-1}_{rs}=\sqrt{\kay_r/\kay_s}\,\mu_{rs}$.

It was shown in~\cite{Georgiou:2018gpe} that the model defined by taking all entries of $\lambda^{-1}$ to be zero except for $\lambda^{-1}_{11},\cdots,\lambda^{-1}_{(N-1)1}$ and $\lambda^{-1}_{N2},\cdots,\lambda^{-1}_{NN}$ is integrable. Let us now explain how this model can be obtained as a limit of the one constructed above by coupling together $N$ $\lambda$-realisations. We introduce the following reparametrisation of the positions $z_r^\pm$ of the model:
\begin{equation}\label{Eq:y}
z^+_r = y_r \; \text{ for } \; r\in\lbrace 1,\cdots,N-1 \rbrace, \;\;\;\;\;\;\; z_N^+ = \frac{1}{\gamma}, \;\;\;\;\;\;\; z_1^- = 0, \;\;\;\;\;\;\; z_r^- = \yh_r + \frac{1}{\gamma} \; \text{ for } \; \, r\in\lbrace 2,\cdots,N \rbrace,
\end{equation}
in terms of new parameters $y_1,\cdots,y_{N-1},\yh_2,\cdots,\yh_N$ and $\gamma$. We used here the translation redundancy on the parameters $z_r^\pm$ to fix the value of $z_1^-$ to 0. Recall that one can also use the dilation redundancy to fix the value of $\ell^\infty$: for future convenience, we choose here to fix it to
\begin{equation*}
\ell^\infty = 2\left(1 + \sum_{r=1}^N \frac{\nu_r^-}{z_N^+-z_r^-} \right)^{-1} = \frac{2}{\vp_-\bigl(z_N^+\bigr)}.
\end{equation*}
Using this parametrisation, the model is then described by the $3N-1$ free parameters $y_1,\cdots,y_{N-1}$, $\yh_2,\cdots,\yh_N$, $\nu^-_1,\cdots,\nu^-_N$ and $\gamma$. The limit we shall consider in this paragraph is $\gamma \to 0$, while keeping the remaining parameters fixed.

Using the expression \eqref{Eq:Mu} of the coefficients $\mu_{rs}$, one checks that in the limit $\gamma\to 0$, these coefficients all vanish except $\mu_{11},\cdots,\mu_{(N-1)1}$ and $\mu_{N2},\cdots,\mu_{NN}$. The matrix $\lambda^{-1}$, which has components $\lambda^{-1}_{rs}=\sqrt{\kay_r/\kay_s}\,\mu_{rs}$, then takes the same form as in the integrable truncation considered in~\cite{Georgiou:2018gpe}. As one considered the limit $\gamma\to0$, the model is now described by the $3N-2$ parameters $y_1,\cdots,y_{N-1}$, $\yh_2,\cdots,\yh_N$ and $\nu^-_1,\cdots,\nu^-_N$. This coincides with the number of free parameters considered for the integrable model of~\cite{Georgiou:2018gpe}. More precisely, the parameters used in this reference are the Wess-Zumino levels $\kay_1,\cdots,\kay_N$ and the coefficients $\mu_{11},\cdots,\mu_{(N-1)1}$ and $\mu_{N2},\cdots,\mu_{NN}$ (or equivalently the corresponding coefficients $\lambda^{-1}_{rs}$). Using the expressions of these coefficients obtained in the previous paragraph and considering the limit $\gamma\to 0$, one can relate explicitly the parametrisation used here with the parametrisation used in~\cite{Georgiou:2018gpe}. More precisely, one finds after several computational steps:
\begin{equation*}
y_r = \left(\frac{1}{\mu_{r1}}-1\right)\frac{\kay_1-a}{b}, \;\;\;\;\;\;\; \yh_s = \frac{\kay_s}{\mu_{Ns}} - \kay_N, \;\;\;\;\;\;\; \nu_1^- = \frac{\kay_1-a}{b} \;\;\;\;\;\;\; \text{and} \;\;\;\;\;\;\; \nu_s^-=\frac{\kay_s-\kay_N\mu_{Ns}}{b},
\end{equation*}
for $r\in\lbrace 1,\cdots,N-1\rbrace$ and $s\in\lbrace 2,\cdots,N\rbrace$, where we define $a = \sum_{r=1}^{N-1} \kay_r \mu_{r1}$ and $b = \sum_{s=2}^N \mu_{Ns} - 1$.\\

Let us comment on the limit $\gamma\to0$ considered above. This limit consists in singling out two sets of positions $\mathcal{Z}_1=\lbrace z_1^-, z_1^+,\cdots,z_{N-1}^+ \rbrace$ and $\mathcal{Z}_2=\lbrace z_2^-,\cdots,z_N^-,z_N^+\rbrace$ and sending the distance between these two sets to infinity. It is thus quite similar to the decoupling procedure considered in~\cite[Subsection 2.3.3]{Delduc:2019bcl}\footnote{The main difference with this procedure comes from the realisations attached to the sites. Indeed, in~\cite{Delduc:2019bcl}, the two sets $\mathcal{Z}_1$ and $\mathcal{Z}_2$ are associated with independent realisations, in the sense that the phase space of the model takes the form $\Pc_1 \times \Pc_2$ and the positions in $\mathcal{Z}_1$ are associated with Kac-Moody (or Takiff) currents in the first factor $\Pc_1$ and the positions in $\mathcal{Z}_2$ are associated with currents in $\Pc_2$. In the decoupling limit, where the sites $\mathcal{Z}_1$ cease to interact with the sites $\mathcal{Z}_2$, one then obtains two independent models on $\Pc_1$ and $\Pc_2$ respectively. In the present case, the currents associated with the sets $\mathcal{Z}_1$ and $\mathcal{Z}_2$ do not belong to independent parts of the phase space.}. According to this procedure, the sites $(1,-),(1,+),\cdots,(N-1,+)$ corresponding to the positions $\mathcal{Z}_1$ thus cease to interact with the sites $(2,-),\cdots,(N,-),(N,+)$ corresponding to the positions $\mathcal{Z}_2$ in the limit $\gamma\to 0$. This explains the structure of the model considered in~\cite{Georgiou:2018gpe}, where the fields $g^{(2)},\cdots,g^{(N-1)}$ have no interactions one with another. The theory before taking the limit $\gamma\to 0$ then defines a non-trivial integrable generalisation of this model: indeed, although it corresponds to adding only one parameter, this introduces non-trivial interactions between all the different fields $g^{(r)}$, as the coefficients $\mu_{rs}$ then become generically all non-zero.

Following the decoupling procedure of~\cite{Delduc:2019bcl}, one describes the integrability of the model in the limit $\gamma\to 0$ using two independent Lax pairs, which are obtained as two different limits of the initial Lax pair $\Lc_\pm(z)$. More precisely, let us consider:
\begin{equation}\label{Eq:LaxLimit}
\Lc_\pm^{(1)}(z) = \lim_{\gamma\to 0} \, \Lc_\pm(z) \;\;\;\;\;\; \text{ and } \;\;\;\;\;\; \Lc_\pm^{(2)}(z) = \lim_{\gamma\to0} \, \Lc_\pm\left(z+\frac{1}{\gamma}\right).
\end{equation}
It is clear that, before taking the limit $\gamma\to0$, both $\Lc_\pm(z)$ and $\Lc_\pm(z+\gamma^{-1})$ satisfy a zero curvature equation (as $\Lc_\pm(z)$ does) and thus still do after taking the limit. The reason behind the necessity of considering these two Lax pairs is that, loosely speaking, the Lax pair $\Lc_\pm(z)$ loses the information about the positions $\mathcal{Z}_2$ in the limit $\gamma\to 0$: the Lax pair $\Lc_\pm^{(1)}(z)$ then only ``corresponds to'' the positions $\mathcal{Z}_1$ (see~\cite{Delduc:2019bcl} for a more precise treatment). Considering the shift of the spectral parameter by $\gamma^{-1}$, as done in the definition of $\Lc_\pm^{(2)}(z)$, exchanges the roles of the sets $\mathcal{Z}_1$ and $\mathcal{Z}_2$, so that the second Lax pair $\Lc_\pm^{(2)}(z)$ contains the information about the positions $\mathcal{Z}_2$. This is coherent with~\cite{Georgiou:2018gpe}, where the integrable truncation was described using two Lax pairs.

The Hamiltonian analysis of the corresponding Lax matrices was performed recently in~\cite{Georgiou:2019plp}, where it was shown that their Poisson brackets are described by twist functions. In the language of affine Gaudin models used above, these twist functions are obtained from the twist function $\vp(z)$ of the original model by a limit similar to the one of Equation \eqref{Eq:LaxLimit} (see~\cite{Delduc:2019bcl}):
\begin{equation*}
\vp^{(1)}(z) = \lim_{\gamma\to0}\,\vp(z) \;\;\;\;\;\; \text{ and } \;\;\;\;\;\; \vp^{(2)}(z) = \lim_{\gamma\to0} \, \vp\left(z+\frac{1}{\gamma}\right).
\end{equation*}
One then finds that the twist function $\vp^{(1)}$ has poles at the points $\lbrace y_1,\cdots, y_{N-1}, 0\rbrace$ while the twist function $\vp^{(2)}(z)$ has poles at the points $\lbrace 0,\yh_2,\cdots,\yh_N\rbrace$. Up to dilation and translation, these poles coincide with the ones obtained in~\cite{Georgiou:2019plp}.

\subsection{Deformed models with two copies}
\label{SubSec:Models2}

Recall from Subsection \ref{2copies} that in the case of a model with two copies, one can rewrite the operators $\Kop_\pm^{-1}$ and $\mathcal{O}_{rs}$ more explicitly as in Equations \eqref{Eq:KInv} and \eqref{O2}. Using these results, we study in this subsection the models with two Yang-Baxter realisations and two $\lambda$-realisations.

\subsubsection{Model with two Yang-Baxter realisations}

Let us consider first the model with two Yang-Baxter realisations. In this case, we will use the first expression of the operators $\mathcal{O}_{rs}$ in Equation \eqref{O2}. The entries of the operator $\Kop_+$ can be read from \eqref{Eq:UYB} while the entries of $\Lop_+$ are related to the ones of $\Kop_+$ by equation \eqref{Eq:KLYB}. Using the notation $\bar r$ introduced in Subsection \ref{2copies}, one then obtains the following expression for the operators $\mathcal{O}_{rs}$:
\begin{equation*}
\mathcal{O}_{rs} = \frac{1}{1+ \theta^+_{rr} \widehat{R}^{(r)} + \theta^+_{\bar{r}\bar{r}} \widehat{R}^{(\bar r)} + \text{det}(\theta^+) \widehat{R}^{(\bar r)} \widehat{R}^{(r)}} \left(\rho_{sr}^+ + \bigl( \rho_{sr}^+ \theta^+_{\bar{r}\bar{r}} - \rho_{s\bar{r}}^+ \theta^+_{\bar{r}r}\bigr)\widehat{R}^{(\bar r)}\right) +\delta_{rs}\frac{\kay_r}{2},
\end{equation*}
where $\widehat{R}^{(r)} = c_r \Id + R^{(r)} + c_r\delta_r \Pi^{(r)}$, $\det(\theta^+)=\theta^+_{11}\theta^+_{22}-\theta^+_{12}\theta^+_{21}$ and $\theta^+_{rs}$ is given by Equation \eqref{Eq:Theta}.

\subsubsection[Model with two $\lambda$-realisations]{Model with two $\bm\lambda$-realisations}
\label{SubSubSec:TwoLambdas}

Let us now consider the model with two $\lambda$-realisations. Its action is given by Equation \eqref{N2Action} with $N=2$. Reinserting the explicit form \eqref{Eq:ULambda} of the operator $\Kop_-$ and calculating its inverse through \eqref{Eq:KInv}, we find that in this case the operator $\Kop_-^{-1}$ appearing in the action is explicitly given by
\begin{equation*}
(\Kop_-^{-1})_{rs} = (-1)^{r+s} \, \left(\mu_{\bar{s}\bar{r}}-\delta_{rs}\Ad_{g^{(\bar{s})}}^{-1}\right)\frac{1}{\text{det}(\mu) -\mu_{\bar{s}\bar{s}} \Ad_{g^{(s)}}^{-1} -\mu_{ss} \Ad_{g^{(\bar{s})}}^{-1} + \Ad_{g^{(s)}}^{-1}\Ad_{g^{(\bar{s})}}^{-1}},
\end{equation*}
with $\det(\mu)=\mu_{11}\mu_{22}-\mu_{12}\mu_{21}$ and $\mu_{rs}$ given by Equation \eqref{Eq:Mu1}.\\

Let us end this subsection by comparing this result with the ones of~\cite{Georgiou:2016urf,Georgiou:2017jfi,Georgiou:2018hpd}. Indeed, the integrable $\s$-models introduced in these references can be obtained from the model above by taking limits similar to the one considered in Subsection \ref{SubSubSec:Nlambda} (which allowed us to compare the model with $N$ copies of the $\lambda$-realisation with the integrable model of~\cite{Georgiou:2018gpe}).

Let us first consider the following reparametrisation of the positions $z_r^\pm$: $z_1^+=y$, $z_2^+=\gamma^{-1}$, $z_1^-=\yh+\gamma^{-1}$ and $z_2^-=0$, similar to the parametrisation \eqref{Eq:y} used in the model with $N$ copies. We then take the limit $\gamma\to 0$. One checks that in this limit, $\mu_{11}=\mu_{22}=0$. The model is then identical to the model (2.12) of~\cite{Georgiou:2017jfi} (see also~\cite{Georgiou:2016urf} for the case with equal Wess-Zumino levels $\kay_1=\kay_2$), where the remaining coefficients $\mu_{12}$ and $\mu_{21}$ are identified with $\mu_{12} = \lambda_0^{-1}\lambda_2^{-1}$ and $\mu_{21} = \lambda_0 \lambda_1^{-1}$, in terms of the parameters $\lambda_i$ of~\cite{Georgiou:2017jfi}.

Let us now consider another reparametrisation $z_1^+=0$, $z_2^+=y_2$, $z_1^-=y_1$ and $z_2^-=\gamma^{-1}$ and then take the limit $\gamma\to 0$. In this case, $\mu_{12}$ and $\mu_{22}$ vanish. The model is then identical to the model (3.1) of~\cite{Georgiou:2018hpd}, where the remaining coefficients $\mu_{11}$ and $\mu_{21}$ are identified with $\mu_{11} = \lambda_0^{-1}\lambda_4^{-1}$ and $\mu_{21} = \lambda_0 \lambda_1^{-1}$, in terms of the parameters $\lambda_i$ of~\cite{Georgiou:2018hpd}.

\section{Relation with 4d semi-holomorphic Chern-Simons theory}
\label{Sec:CS}

In this section, we explain how the models considered in this article can be obtained using the approach proposed recently by Costello and Yamazaki to generate integrable 2d field theories from 4d semi-holomorphic Chern-Simons theory ~\cite{Costello:2019tri} (see~\cite{Costello:2013zra, Costello:2013sla, Witten:2016spx, Costello:2017dso, Costello:2018gyb, Bittleston:2019gkq, Vicedo:2019dej, Delduc:2019whp} for additional references on this variant of Chern-Simons theory and its relation to integrable systems). Note that, in the terminology of~\cite{Costello:2019tri}, we restrict our attention here to 4d Chern-Simons theory with disorder defects. It was shown in~\cite{Costello:2019tri} that the PCM with Wess-Zumino term and the integrable $\sigma$-model coupling $N$ of its copies can be obtained from this approach. It was subsequently shown in~\cite{Vicedo:2019dej} that the integrable 2d field theories obtained from 4d Chern-Simons theory with disorder defects are realisations of AGM. Moreover, it was explained in~\cite{Delduc:2019whp} how the Yang-Baxter model and the $\lambda$-model can also be derived following this approach. It is thus natural to search for a generalisation of these results for the AGM coupling together $N_1$ copies of the Yang-Baxter model and $N_2$ copies of the $\lambda$-model, which is the integrable field theory constructed in the present article.

\subsection{4d semi-holomorphic Chern-Simons theory and integrable field theories}
\label{SubSec:CSInt}

In this subsection, we will briefly sketch the method developed in~\cite{Costello:2019tri} to generate integrable 2d field theories from 4d semi-holomorphic Chern-Simons theory. We will not explain this method in details here and mainly focus on the aspects that will be concretely relevant for the purpose of this article (we then refer to~\cite{Costello:2019tri,Delduc:2019whp} for details). We will follow here the conventions of~\cite{Delduc:2019whp}, which are in agreement with the ones used in the rest of this article.

\paragraph{4d Chern-Simons theory.} The semi-holomorphic Chern-Simons theory is defined on the 4d manifold $\mathbb{R}\times\mathbb{D} \times \mathbb{P}^{1}$: the $\mathbb{R}\times\mathbb{D}$ part of this manifold corresponds to the 2d space-time with coordinates $(t,x)$ of the resulting integrable field theory (here the spatial manifold $\mathbb{D}$ can be either the real line $\mathbb{R}$ or the circle $S^1$, as in the rest of this article), while the Riemann sphere $\mathbb{P}^1$ gives rise to the spectral parameter $z$ of this integrable model. The 4d Chern-Simons theory is partly characterised by the choice of a meromorphic $1$-form $\omega = \vp(z)\text{d} z$ on $\mathbb{P}^1$: as shown in~\cite{Vicedo:2019dej}, the corresponding rational function $\vp(z)$ is the twist function of the resulting integrable model. The dynamical fields of the four-dimensional theory are the components $A_+$, $A_-$ and $A_{\bar{z}}$ of a $\g$-valued gauge field $A$ along the light-cone directions $x^\pm$ of $\mathbb{R}\times\mathbb{D}$ and the anti-holomorphic direction $\bar z$ of $\mathbb{P}^1$ (note that the component of $A$ in the $z$-direction decouples from the theory and is not a physical degree of freedom). In addition to the choice of $\omega$ made above, the theory is then fully determined by specifying appropriate boundary conditions on the field $A$ at the poles $\mathcal{Z}\subset\mathbb{P}^1$ of $\omega$, \textit{i.e.} at the poles of the twist function (see~\cite{Costello:2019tri,Delduc:2019whp} and the next subsection for details). The action of the semi-holomorphic Chern-Simons theory is defined as~\cite{Costello:2013zra}
\begin{equation}\label{Eq:CS}
S_{\text{CS}}[A] = \frac{i}{4\pi} \int_{\mathbb{R}\times\mathbb{D} \times \mathbb{P}^{1}} \omega \wedge \text{CS}(A),
\end{equation}
where $\text{CS}(A)$ is the standard Chern-Simons 3-form of $A$. 

\paragraph{Parametrisation of the gauge field.} In order to relate the 4d Chern-Simons theory to an integrable 2d model, one parametrises the gauge field components in the following form
\begin{equation}\label{Eq:A}
A_{\bar z} = \gh\, \p_{\bar z} \gh\,^{-1}, \;\;\;\;\; A_\pm = \gh\, \p_\pm \gh\,^{-1} + \gh \,\Lc_\pm \,\gh\,^{-1},
\end{equation}
where $\gh$ and $\Lc_\pm$ are fields respectively valued in the group $G$ and the Lie algebra $\g$. The equation of motion obtained by varying the action \eqref{Eq:CS} with respect to $A_{\bar z}$ then ensures that the fields $\Lc_\pm$ depend meromorphicaly on $z$. Moreover, the equations of motion obtained by varying $A_\pm$ show that they also satisfy a zero curvature equation $\p_+\Lc_- - \p_-\Lc_+ + \bigl[\Lc_+,\Lc_-\bigr] = 0$ on $\mathbb{R}\times\mathbb{D}$. These two properties make the field $\Lc_\pm$ a good candidate for being the Lax pair of a 2d integrable model on $\mathbb{R}\times\mathbb{D}$.

\paragraph{The fields of the 2d theory.} Let us now explain how this integrable 2d field theory is constructed. For $z$ in the Riemann sphere $\mathbb{P}^1$ and a field $\phi$ on $\mathbb{R}\times\mathbb{D} \times \mathbb{P}^{1}$, we will denote by $\phi|_z$ the field on $\mathbb{R}\times\mathbb{D}$ obtained by evaluating $\phi$ at the point $z$ on the Riemann sphere. It is explained in~\cite{Costello:2019tri,Delduc:2019whp} that for a point $z\in\mathbb{P}^1\setminus\mathcal{Z}$ which is not a pole of $\omega$, the 2d field $\gh|_z$ can be set to a constant field equal to the identity of $G$ by an appropriate gauge transformation on the gauge field $A$. The fact that we restrict here to points $z$ which are not poles of $\omega$ is due to the fact that this gauge transformation on $A$ should preserve the boundary conditions imposed on $A$ at these poles and mentioned above (see~\cite{Costello:2019tri,Delduc:2019whp} for details). Thus, the 2d fields $\gh|_z$, $z\in\mathbb{P}^1\setminus\mathcal{Z}$, are not physical degrees of freedom of the model. The dynamical fields of the 2d model we aim to construct are then defined to be the remaining degrees of freedom contained in $\gh$, \textit{i.e.} its evaluations $\{\gh|_{z_0}\}_{z_0\in\mathcal{Z}}$ at the poles of $\omega$. Let us mention that in general, one should also consider the fields $\p_z^p\,\gh|_{z_0}$ obtained by evaluating derivatives of $\gh$ at the points $z_0\in\mathcal{Z}$: however, as explained in~\cite{Costello:2019tri,Delduc:2019whp}, for the boundary conditions considered in these references and that we shall consider in this article, these degrees of freedom can also be eliminated by gauge transformations.

So far, we have considered only the degrees of freedom contained in the field $\gh$, which, as we see from Equation \eqref{Eq:A}, encodes the component $A_{\bar z}$ of the gauge field. Let us now consider the component $A_\pm$ and thus the field $\Lc_\pm$. As explained above, the equation of motion of $A_{\bar z}$ ensures that $\Lc_\pm$ is meromorphic in $z$. In fact, it also implies that $\Lc_\pm$ can have poles in $\mathbb{P}^1$ only at the zeroes of $\omega$. This constrains quite strongly the dependence of $\Lc_\pm$ in terms of the variable $z\in\mathbb{P}^1$. Let us be more precise. As $\omega$ will ultimately be given by the twist function of the resulting 2d theory, let us denote its zeroes $\{\zeta_i\}_{i\in\{1,\cdots,M\}}$, in agreement with what was done in the rest of this article. These zeroes can be separated into two sets $\lbrace \zeta_i \rbrace_{i\in\mathcal{I}_\pm}$, labelled by subsets $\mathcal{I}_+$ and $\mathcal{I}_-$ of $\lbrace 1,\cdots,M\rbrace$, depending on which of the component $\Lc_+$ or $\Lc_-$ has a pole at $\zeta_i$ (see~\cite{Delduc:2019whp} for details). This fixes the $z$-dependence of the fields $\Lc_\pm$: more precisely, they are of the form
\begin{equation}\label{Eq:LaxCS}
\Lc_\pm(z) = \sum_{i\in\mathcal{I}_\pm} \frac{U_i}{z-\zeta_i} + U_\pm^\infty,
\end{equation}
for some 2d $\g$-valued fields $U_i$, $U_+^\infty$ and $U_-^\infty$ on $\mathbb{R}\times\mathbb{D}$. In this equation, we have written the Lax pair as $\Lc_\pm(z)$ to stress its dependence on the spectral parameter $z$: note however that it also depends on the coordinates $(t,x)\in\mathbb{R}\times\mathbb{D}$, through the 2d fields $U_i$ and $U_\pm^\infty$.

Recall that the gauge field $A$ obeys some specific boundary conditions at the poles $z_0 \in \mathcal{Z}$ of $\omega$, which translate into conditions on the evaluations $\{\Lc_\pm|_{z_0}\}_{z_0\in\mathcal{Z}}$ and $\{\gh|_{z_0}\}_{z_0\in\mathcal{Z}}$. As observed in~\cite{Costello:2019tri,Delduc:2019whp} and as we shall see in this article, these boundary conditions, combined with the $z$-dependence \eqref{Eq:LaxCS} of $\Lc_\pm$, specify completely $\Lc_\pm$ in terms of the 2d fields $\{\gh|_{z_0}\}_{z_0\in\mathcal{Z}}$. The field $\Lc_\pm$ then does not contain any additional degrees of freedom and is interpreted as the Lax pair of the resulting 2d field theory on $\{\gh|_{z_0}\}_{z_0\in\mathcal{Z}}$ (indeed, recall also from the previous paragraph that, on-shell, it satisfies a zero curvature equation on $\mathbb{R}\times\mathbb{D}$).

Let us end this paragraph by the following remark. As argued above, the fields $\{\gh|_{z_0}\}_{z_0\in\mathcal{Z}}$ describe all the degrees of freedom of the resulting 2d model. However, in general, these degrees of freedom are not all physical: there are some residual gauge symmetries acting on these fields, which depend on the type of boundary conditions considered. Moreover, there always exists an additional redundancy on these fields which consists on multiplying all of them on the right by an arbitrary $G$-valued field $h$ on $\mathbb{R}\times\mathbb{D}$ (see~\cite{Costello:2019tri,Delduc:2019whp}). This redundancy can be used to fix one of the fields $\{\gh|_{z_0}\}_{z_0\in\mathcal{Z}}$ to the identity.

\paragraph{The effective 2d action.} To complete the description of the 2d field theory obtained through this method, one has to describe its action. This is done by performing the integration over $\mathbb{P}^1$ in the 4d action \eqref{Eq:CS}, resulting on an effective 2d action on $\mathbb{R}\times\mathbb{D}$ depending on the 2d fields $\{\gh|_{z_0}\}_{z_0\in\mathcal{Z}}$. However, we will not need the details of this procedure in the following and thus refer to~\cite{Costello:2019tri,Delduc:2019whp} for details. In particular, it was shown in~\cite{Delduc:2019whp} that, for the type of boundary conditions that we shall consider in this article, this 2d action simply reads:
\begin{eqnarray}\label{Eq:ActionCS}
S\bigl[ \{\gh|_{z_0}\}_{z_0\in\mathcal{Z}} \bigr] &=& \frac{1}{4} \sum_{z_0\in\mathcal{Z}} \iint_{\mathbb{R}\times\mathbb{D}} \text{d} t \, \text{d} x\; \left( \kappa\Bigl( \res_{z=z_0} \vp(z)\Lc_+(z)\text{d} z, j_-^{\{z_0\}} \Bigr)
- \kappa\Bigl(j_+^{\{z_0\}} ,  \res_{z=z_0} \vp(z)\Lc_-(z)\text{d} z \Bigr) \right) \notag \\
&& \hspace{30pt} - \frac{1}{2} \sum_{z_0\in\mathcal{Z}}  \left( \res_{z=z_0} \vp(z)\text{d}z \right) \Ww{ \gh|_{z_0} }, 
\end{eqnarray}
where $\Ww{ \gh|_{z_0} }$ is the Wess-Zumino term of $\gh|_{z_0}$ and $j_\pm^{\{z_0\}}$ is defined as the Maurer-Cartan current
\begin{equation*}
j_\pm^{\{z_0\}} = \gh|_{z_0}^{-1} \p_\pm \gh|_{z_0}.
\end{equation*}

\subsection{The models}

Our aim in this section is to show explicitly that a certain class of 2d integrable field theories obtained using the Chern-Simons approach described in the previous subsection can be identified with the affine Gaudin models coupling together an arbitrary number of copies of inhomogeneous Yang-Baxter realisations and $\lambda$-realisations, as considered in the rest of this article. Let us then start by defining the particular class of 4d Chern-Simons theories that we shall consider here.

As explained in~\cite{Costello:2019tri,Delduc:2019whp} and recalled in the previous subsection, the 4d semi-holomorphic Chern-Simons theory is characterised by the choice of the meromorphic 1-form $\omega$ and of the boundary conditions on $A$ at the poles $\mathcal{Z}$ of $\omega$. Let us then define the 1-form and boundary conditions that we shall consider here.

\subsubsection[1-form $\omega$]{1-form $\bm\omega$} Following~\cite{Vicedo:2019dej} (see also the summary in the previous subsection), the meromorphic 1-form $\omega$ characterising the models obtained from the 4d Chern-Simons approach should coincide with $\vp(z)\text{d}z$, where $\vp(z)$ is the twist function of these models when seen as realisations of AGM. As we aim to recover the models constructed in this article, we will then choose $\omega$ to be given by the twist function~\eqref{Eq:TwistZeroes} considered in the previous sections, \textit{i.e.}
\begin{equation}\label{Eq:Omega}
\omega = -\ell^{\infty}\frac{\prod_{i=1}^{2N}{(z-\zeta_i)}}{\prod_{s=1}^N{(z-z_r^+)(z-z_r^-)}} \; \text{d} z.
\end{equation}
This 1-form has $2N$ simple poles at the points $z_r^\pm$ and a double pole at $\infty$. In the language of the previous subsection, one then has $\mathcal{Z}=\lbrace z_1^+,z_1^-,\cdots,z_N^+,z_N^-,\infty\rbrace$. Following the notations of this article, let us define $\ell_r^\pm$ as the residues of $\omega$ at the poles $z_r^\pm$, which coincide with the levels of the model when seen as a realisation of AGM.

\subsubsection{Boundary conditions}
\label{SubSubSec:BC}

\paragraph{Boundary condition at the double pole at infinity.} Let us consider the double pole $\infty$ of $\omega$. Following~\cite{Costello:2019tri} (see also~\cite{Delduc:2019whp}), we will impose at this pole the following simple boundary condition on the Chern-Simons gauge field $A$:
\begin{equation}\label{Eq:BCInfinity}
A_\pm |_{\infty} = 0.
\end{equation}

\paragraph{Boundary conditions at the simple poles $\bm{z_r^\pm}$.} Let us now consider a pair of simple poles $z_r^\pm$ and the corresponding evaluations $A_\pm|_{z^+_r}$ and $A_\pm|_{z^-_r}$ of the gauge field at these points. A systematic study of the consistent boundary conditions that can be imposed on these evaluations has been presented in~\cite{Delduc:2019whp} (see also~\cite{Costello:2017dso,Costello:2019tri}). We will consider here two of them.

\paragraph{Yang-Baxter boundary condition.} The first one, that we shall call Yang-Baxter boundary condition, is characterised by the choice of a skew-symmetric $R$-matrix $R_r$ satisfying the mCYBE \eqref{Eq:mCYBE}, with $c_r=1$ if the poles $z_r^\pm$ are real and $c_r=i$ if they are complex conjugate, and satisfying $R^3_r = c^2_r R_r$. Consider the residues $\ell^\pm_r$ of $\omega$ at $z_r^\pm$, as defined above. Let us define from them the following parameters:
\begin{equation*}
\kay_r = - \frac{\ell^+_r + \ell^-_r}{2}, \;\;\;\;\;\;\; \gamma_r = \frac{1}{c_r(\ell^+_r-\ell^-_r)}  \;\;\;\;\;\; \text{ and } \;\;\;\;\;\;\; \delta_r = \frac{1-\sqrt{1-4c_r^2\kay_r^2\gamma_r^2}}{2c_r\kay_r\gamma_r}.
\end{equation*}
These coincide with the coefficients $\kay_r$, $\gamma_r$ and $\delta_r$ considered in the rest of this article for an inhomogeneous Yang-Baxter realisation with Wess-Zumino term (see Subsection \ref{Examples}). Let us note that the coefficient $\delta_r$ satisfy $(\delta_r+1)^2\ell^+_r+(\delta_r-1)^2\ell^-_r =0$ so that $-c_r\delta_r$ coincides with the parameter $\theta$ considered in~\cite[Subsection 5.6]{Delduc:2019whp}. The Yang-Baxter boundary condition can then be seen as requiring that the evaluations $A_\pm|_{z^+_r}$ and $A_\pm|_{z^-_r}$ satisfy
\begin{equation}\label{Eq:BCYB}
(R_r-c_r\delta_r\Pi_r+c_r)A_\pm|_{z^+_r}=(R_r-c_r\delta_r\Pi_r-c_r)A_\pm|_{z^-_r},
\end{equation}
with $\Pi_r=\text{Id}-R_r^2/c_r^2$ as in Subsection \ref{Examples}.

\paragraph{$\bm\lambda$-boundary condition.} Let us describe the second type of boundary condition at the pair of simple poles $z_r^\pm$ that we shall consider, which we call the $\lambda$-boundary condition. It can be imposed only if the poles $z_r^\pm$ and the residues $\ell^\pm_r$ are real and satisfy the additional condition $\ell^+_r+\ell^-_r=0$ (note that this is identical to the condition \eqref{Eq:ConstraintLevel} that one should impose to consider a $\lambda$-realisation in an affine Gaudin model). The $\lambda$-boundary condition is then simply given by
\begin{equation}\label{Eq:BCLambda}
A_\pm|_{z^+_r}=A_\pm|_{z^-_r}.
\end{equation}
For a $\lambda$-boundary condition, we define the parameter $\kay_r=\ell^-_r/2=-\ell_r^+/2$, which is equal to the Wess-Zumino coefficent $\kay_r$ defined for a $\lambda$-realisation (see Subsection \ref{Examples}).

\subsubsection{Fields of the model}
\label{SubSubSec:FieldsCS}

Let us consider a 4d Chern-Simons theory with $\omega$ as in Equation \eqref{Eq:Omega} and with $N_1$ Yang-Baxter boundary conditions and $N_2$ $\lambda$-boundary conditions. Let us describe what are the dynamical fields of this model. As recalled in the previous subsection, these fields are given by the evaluations $\gh|_{z_0}$ of the field $\gh$ at the poles $z_0\in\mathcal{Z}$ of $\omega$ and thus by the $2N+1$ fields $\gh|_\infty$, $\gh|_{z_r^+}$ and $\gh|_{z_r^-}$.

However, as mentionned in the previous subsection and explained in~\cite{Costello:2019tri,Delduc:2019whp}, we can eliminate some of these degrees of freedom. In particular, recall from the previous subsection that we can fix one of the fields $\gh|_{z_0}$ to the identity: here, we will choose to fix the field at infinity $\gh|_\infty$. Moreover, as explained in~\cite{Delduc:2019whp}, if one considers a Yang-Baxter boundary condition or a $\lambda$-boundary condition at the pair of poles $z_r^\pm$, there exists a residual gauge symmetry on the fields $\gh|_{z_r^+}$ and $\gh|_{z_r^-}$. In the case of a Yang-Baxter boundary condition, this gauge symmetry can be fixed by imposing $\gh|_{z_r^+}=\gh|_{z_r^-}$: we then define $g^{(r)}$ as their common value. For the $\lambda$-boundary condition, it can be fixed instead by imposing $\gh|_{z_r^-}=\text{Id}$: we then define $g^{(r)}=\gh|_{z_r^+}$. To summarise, the fields of the model are the $N$ group-valued fields $g^{(1)},\cdots,g^{(N)}$ and we have
\begin{equation*}
\gh|_\infty = \text{Id}, \;\;\;\;\;\;\;\;\;\;\; \text{YB-BC: } \;\; \gh|_{z^+_r}=\gh|_{z^-_r}=g^{(r)}, \;\;\;\;\;\;\;\;\;\;\; \lambda\text{-BC: } \;\; \gh|_{z^+_r}=g^{(r)}, \;\; \gh|_{z^-_r}=\text{Id}.
\end{equation*}

\subsection{Identification of the two approaches}

Let us consider the 2d integrable field theory defined in the previous subsection with $N_1$ Yang-Baxter boundary conditions and $N_2$ $\lambda$-boundary conditions. We will prove in this subsection that it can be identified with the AGM with $N_1$ Yang-Baxter realisations and $N_2$ $\lambda$-realisations studied in the previous sections. In order to do so, we shall show that the two approaches lead to the same Lax pair as well as the same action.

\subsubsection{Identification of the Lax pairs}

Let us consider the Lax pair of the model coming from 4d Chern-Simons theory as given by Equation \eqref{Eq:LaxCS}. Let us now express it in terms of the fields $g^{(r)}$ of the model, using the boundary conditions that are imposed on the gauge field $A$ at the poles $z_0\in\mathcal{Z}$ of $\omega$.

\paragraph{Pole at infinity.} Let us start with the pole at $z_0=\infty$, for which the boundary condition is simply defined by Equation \eqref{Eq:BCInfinity}. From the fact that $\gh|_\infty=\text{Id}$ (see Subsection \ref{SubSubSec:FieldsCS}) and the expression \eqref{Eq:LaxCS} of the Lax pair $\Lc_\pm(z)$, it is clear that the evaluation of the gauge field \eqref{Eq:A} at $z=\infty$ gives
\begin{equation*}
A_\pm|_\infty = \Lc_\pm(\infty) = U_\pm^\infty.
\end{equation*}
Combining this with the boundary condition \eqref{Eq:BCInfinity}, we then get that $U_\pm^\infty=0$.

\paragraph{Pair of poles with Yang-Baxter boundary condition.} Let us know consider a pair of simple poles $z_r^\pm$ and let us suppose that we imposed on this pair a Yang-Baxter boundary condition \eqref{Eq:BCYB}. As explained in Subsection \ref{SubSubSec:FieldsCS}, in this case, we have $\gh|_{z_r^+}=\gh|_{z_r^-}=g^{(r)}$. Thus, the evaluation of the gauge field \eqref{Eq:A} at $z_r^\varepsilon$, for $\varepsilon\in\lbrace +1,-1\rbrace$, is given by
\begin{equation*}
A_\pm|_{z_r^\varepsilon} = g^{(r)} \p_\pm g^{(r)\,-1} +  g^{(r)} \Lc_\pm(z_r^\varepsilon) g^{(r)\,-1} = \Ad_{g^{(r)}} \Bigl( \Lc_\pm(z^\varepsilon_r) - j_\pm^{(r)} \Bigr),
\end{equation*}
where $j_\pm^{(r)}=g^{(r)\,-1} \p_\pm g^{(r)}$ are the Maurer-Cartan currents of the field $g^{(r)}$. After a few manipulations, the Yang-Baxter boundary condition \eqref{Eq:BCYB} then becomes
\begin{equation*}
j_\pm^{(r)} = \frac{1}{2} \Ad_{g^{(r)}}^{-1}\left( \text{Id} + \frac{R_r}{c_r} - \delta_r \Pi_r \right) \Ad_{g^{(r)}} \, \Lc_\pm(z_r^+) + \frac{1}{2} \Ad_{g^{(r)}}^{-1}\left( \text{Id} - \frac{R_r}{c_r} + \delta_r \Pi_r \right) \Ad_{g^{(r)}} \, \Lc_\pm(z_r^-).
\end{equation*}
Noting that $R_r$ is skew-symmetric and $\Pi_r$ is symmetric, this can be rewritten as
\begin{equation*}
j^{(r)}_\pm = \null^{t}\A^{(r)}_+ \Lc_\pm(z_r^+) + \null^{t}\A^{(r)}_- \Lc_\pm(z_r^-), \;\;\;\;\;\; \text{ with } \;\;\;\;\; \A^{(r)}_\pm = \frac{1}{2} \left( \text{Id} \mp \frac{R^{(r)}}{c_r} \mp \delta_r \Pi^{(r)} \right),
\end{equation*}
where $R^{(r)}=\Ad_{g^{(r)}}^{-1}\circ R_r \circ \Ad_{g^{(r)}}$ and $\Pi^{(r)}=\Ad_{g^{(r)}}^{-1}\circ \Pi_r \circ \Ad_{g^{(r)}}$. The operators $\A_\pm^{(r)}$ found here coincide exactly with the operators, denoted in the same way in the rest of this article, coming from a Yang-Baxter realisation (see Subsection \ref{Examples}). The above equation is then equivalent to the equation \eqref{Eq:jL} obtained in the context of affine Gaudin models.

\paragraph{Pair of poles with $\bm{\lambda}$-boundary condition.} Let us now consider a pair of simple poles $z_r^\pm$ with the $\lambda$-boundary condition \eqref{Eq:BCLambda}. In this case, we have $\gh|_{z^+_r}=g^{(r)}$ and $\gh|_{z^-_r}=\text{Id}$ (see Subsection \ref{SubSubSec:FieldsCS}). Thus, the evaluations of the gauge field \eqref{Eq:A} at $z_r^+$ and $z_r^-$ read
\begin{equation*}
A_\pm|_{z_r^+} = \Ad_{g^{(r)}} \Bigl( \Lc_\pm(z_r^+) - j_\pm^{(r)} \Bigr) \;\;\;\;\; \text{ and } \;\;\;\;\; A_\pm|_{z_r^-} = \Lc_\pm(z_r^-).
\end{equation*}
Similarly to what was done in the previous paragraph for the Yang-Baxter boundary condition, the $\lambda$-boundary condition \eqref{Eq:BCLambda} can then be rewritten
\begin{equation*}
j^{(r)}_\pm = \null^{t}\A^{(r)}_+ \Lc_\pm(z_r^+) + \null^{t}\A^{(r)}_- \Lc_\pm(z_r^-), \;\;\;\;\;\; \text{ with } \;\;\;\;\; \A^{(r)}_+ = \text{Id} \;\;\; \text{ and } \;\;\; \A^{(r)}_- = \Ad_{g^{(r)}}.
\end{equation*}
The operators $\A_\pm^{(r)}$ coincide with the ones introduced in the previous sections for a $\lambda$-realisation (see Subsection \ref{Examples}). As for the Yang-Baxter boundary condition, we then recover the equation \eqref{Eq:jL} obtained through the affine Gaudin model approach.

\paragraph{Summary.} Let us summarise the results of this subsection. We have proved from the boundary condition at $z=\infty$ that the fields $U_\pm^\infty$ vanish. The component $\Lc_\pm(z)$ of the Lax pair \eqref{Eq:LaxCS} has then no constant term and has simple poles at the zeroes $\{\zeta_i\}_{i\in\mathcal{I}_\pm}$. Thus, it has the same meromorphic $z$-dependence as the Lax pair \eqref{Laxlight} of the corresponding affine Gaudin model. Moreover, we showed that the boundary conditions imposed at the pairs of simple poles $z_r^\pm$ in the Chern-Simons approach coincide exactly with the Equation \eqref{Eq:jL} obtained in the affine Gaudin model approach. Recall from Subsection \ref{Lax} that this equation, combined with the meromorphic $z$-dependence mentioned above, allowed us to express the Lax pair $\Lc_\pm(z)$ in terms of the Maurer-Cartan currents $j^{(r)}_\pm$ by means of interpolation techniques. This ensures that the Lax pairs obtained from the two approaches can be identified.

\subsubsection{Identification of the actions}

Let us end this section by showing that the action obtained by the Chern-Simons approach for the model with $N_1$ Yang-Baxter and $N_2$ $\lambda$-boundary conditions coincides with the one of the AGM with $N_1$ Yang-Baxter and $N_2$ $\lambda$-realisations, computed in Section \ref{Sec:Lag}. As recalled in Subsection \ref{SubSec:CSInt}, the former is given by Equation \eqref{Eq:ActionCS}. Since we proved in the previous subsection that the Lax pair $\Lc_\pm(z)$ of the two models coincide, one can then re-insert in this equation the expression \eqref{Eq:Interpolation} of $\Lc_\pm(z)$ obtained in the AGM approach using interpolation techniques. As the twist function has simple poles at $z_r^\pm$ with residues $\ell^\pm_r$, we then get
\begin{equation*}
\res_{z=z_r^\pm} \vp(z)\Lc_\pm(z) \, \text{d}z = \ell^\pm_r J^{(r)}_\pm \;\;\;\;\;\; \text{ and } \;\;\;\;\;\; \res_{z=z_r^\mp} \vp(z)\Lc_\pm(z) \, \text{d}z = \ell^\mp_r \sum_{s=1}^N \frac{\vp_{\pm,s}(z_s^\pm)}{\vp_{\pm,s}(z_r^\mp)} J^{(s)}_\pm.
\end{equation*}
Moreover, recall that the field $\gh|_\infty$ has been set to the identity. The action \eqref{Eq:ActionCS} then becomes
\begin{equation}\label{Eq:SCS}
S = \sum_{r=1}^N \iint \text{d}t\,\text{d}x\;\Upsilon_r \;  -  \sum_{r=1}^N \left(  \frac{\ell^+_r}{2} \Ww{ \gh|_{z_r^+} } + \frac{\ell^-_r}{2} \Ww{ \gh|_{z_r^-} } \right),
\end{equation}
where
\begin{equation*}
\Upsilon_r =  \frac{\ell^+_r}{4} \kappa\bigl( J^{(r)}_+, j_-^{\{z_r^+\}} \bigr) - \frac{\ell^-_r}{4} \kappa\bigl( j_+^{\{z_r^-\}}, J^{(r)}_- \bigr) + \sum_{s=1}^N \left( \frac{\ell^-_r}{4} \frac{\vp_{+,s}(z_s^+)}{\vp_{+,s}(z_r^-)} \kappa \bigl( J^{(s)}_+, j_-^{\{z_r^-\}} \bigr) -  \frac{\ell^+_r}{4} \frac{\vp_{-,s}(z_s^-)}{\vp_{-,s}(z_r^+)} \kappa \bigl( j_+^{\{z_r^+\}}, J^{(s)}_- \bigr) \right).
\end{equation*}
Recall from Subsection \ref{SubSubSec:FieldsCS} that the fields $\gh|_{z_r^\pm}$ are related to the fundamental fields $g^{(r)}$ of the model, in a way which depends on the type of boundary conditions considered at the poles $z_r^\pm$. Equation \eqref{Eq:SCS} then expresses the action of the model in terms of the Maurer-Cartan currents $j_\pm^{(r)}$, the currents $J_\pm^{(r)}$ and the Wess-Zumino terms of the fields $g^{(r)}$. In the AGM approach, we obtained a similar expression for the action in Equation \eqref{Eq:ActionJj}. In the rest of this subsection, we shall show that these two expressions coincide, thus proving that the models obtained from the 4d Chern-Simons and the AGM approaches are identical. For that, we will prove that for every $r\in\lbrace 1,\cdots,N \rbrace$, we have
\begin{equation}\label{Eq:CSWZ}
\frac{\ell^+_r}{2} \Ww{ \gh|_{z_r^+} } + \frac{\ell^-_r}{2} \Ww{ \gh|_{z_r^-} } = - \kay_r \Ww{ g^{(r)} }
\end{equation} 
and
\begin{equation}\label{Eq:CSV}
\Upsilon_r = \frac{1}{2} \sum_{s=1}^N \left(\kappa\left(  \Lop^+_{rs} J_+^{(s)},j_-^{(r)}\right) + \kappa\left(j_+^{(r)}, \Lop^-_{rs} J_-^{(s)} \right) \right).
\end{equation}
In order to show these identities, one needs to distinguish the cases where the pair of poles $z_r^\pm$ is associated with a Yang-Baxter boundary condition or a $\lambda$-boundary condition in the Chern-Simons approach and, correspondingly, with a Yang-Baxter realisation or a $\lambda$-realisation in the AGM approach.

\paragraph{Yang-Baxter boundary condition.} Let us start with the  Yang-Baxter boundary condition. In this case, recall that $\gh|_{z_r^+}=\gh|_{z_r^-}=g^{(r)}$ and that we defined the Wess-Zumino coefficient to be $\kay_r=-(\ell_r^++\ell_r^-)/2$. The Wess-Zumino terms in Equation \eqref{Eq:SCS} corresponding to these poles thus satisfy Equation \eqref{Eq:CSWZ}. Let us now focus on the term $\Upsilon_r$. Note first that $j^{\{z_r^+\}}_\pm=j^{\{z_r^-\}}_\pm=j^{(r)}_\pm$. One can then rewrite $\Upsilon_r$ as
\begin{equation*}
\Upsilon_r = \frac{1}{2} \sum_{s=1}^N \left( \rho^+_{rs} \; \kappa\left( J_+^{(s)},j_-^{(r)}\right) + \rho^-_{rs}\; \kappa\left(j_+^{(r)}, J_-^{(s)} \right) \right),
\end{equation*}
with $\rho^\pm_{rs}$ given by Equation~\eqref{Eq:Rho2}. For a Yang-Baxter realisation, the operators $\A^{(r)}_\pm$ and $\B^{(r)}_\pm$ are related by $\B_\pm^{(r)} = \ell_r^\pm \,\text{Id} + \A_\pm^{(r)}$. This implies that the operators $\Kop^\pm_{rs}$ and $\Lop^\pm_{rs}$ defined in Equations~\eqref{Eq:DefK} and~\eqref{Eq:DefL} satisfy
\begin{equation*}
\Lop^\pm_{rs} = \rho^\pm_{rs}\,\text{Id} \pm \frac{\kay_r}{2} \Kop_{rs}^\pm.
\end{equation*}
Using the expression \eqref{Eq:j2J} of $J^{(r)}_\pm$, we then get
\begin{equation*}
\sum_{s=1}^N \rho^\pm_{rs} J_\pm^{(s)} = \sum_{s=1}^N \Lop^\pm_{rs} J_\pm^{(s)} \mp \frac{\kay_r}{2} \sum_{s=1}^N \Kop^\pm_{rs} J^{(s)}_\pm = \sum_{s=1}^N \Lop^\pm_{rs} J_\pm^{(s)} \mp \frac{\kay_r}{2} j^{(r)}_\pm.
\end{equation*}
Re-inserting this identity in the above expression for $\Upsilon_r$ then shows that it satisfies Equation \eqref{Eq:CSV}, as required.

\paragraph{$\bm\lambda$-boundary condition.} Let us consider now a pair of poles $z_r^\pm$ associated with a $\lambda$-boundary condition. One then has $\gh|_{z_r^+}=g^{(r)}$ and $\gh|_{z_r^-}=\text{Id}$ (see Subsection \ref{SubSubSec:FieldsCS}). Recall moreover from Subsection \ref{SubSubSec:BC} that the Wess-Zumino coefficient is defined for $\lambda$-boundary conditions as $\kay_r=-\ell_r^+/2$. Thus, the Wess-Zumino terms corresponding to the poles $z_r^\pm$ in the action \eqref{Eq:SCS} are given by Equation \eqref{Eq:CSWZ}. Let us now compute $\Upsilon_r$. For a $\lambda$-boundary condition, one has $j^{\{z_r^+\}}_\pm=j^{(r)}_\pm$ and $j^{\{z_r^-\}}_\pm=0$. Thus, $\Upsilon_r$ is given by:
\begin{equation}\label{Eq:LambdaCS1}
\Upsilon_r = -\frac{\kay_r}{2} \kappa\bigl( J^{(r)}_+, j_-^{(r)} \bigr) + \frac{\kay_r}{2} \sum_{s=1}^N \frac{\vp_{-,s}(z_s^-)}{\vp_{-,s}(z_r^+)} \kappa \bigl( j_+^{(r)}, J^{(s)}_- \bigr).
\end{equation}
Let us note that for a $\lambda$-realisation, one has $\B_+^{(r)}=-\kay_r \A_+^{(r)}$ and $\B_-^{(r)}=\kay_r \A_-^{(r)}$ (see Subsection \ref{Examples}). In terms of the operators $\Kop^+_{rs}$ and $\Lop^+_{rs}$ defined in Equations \eqref{Eq:DefK} and \eqref{Eq:DefL}, this implies
\begin{equation*}
\Lop^+_{rs} = \frac{\kay_r}{2} \Kop_{rs}^+ - \kay_{r}\delta_{rs}\text{Id}.
\end{equation*}
Using the expression \eqref{Eq:j2J} of the currents $J^{(s)}_\pm$, we then obtain
\begin{equation}\label{Eq:LambdaCS2}
-\kay_r J^{(r)}_+ = \sum_{s=1}^N \Lop^+_{rs} J^{(s)}_+  - \frac{\kay_r}{2} \sum_{s=1}^N \Kop^+_{rs} J^{(s)}_+ = \sum_{s=1}^N \Lop^+_{rs} J^{(s)}_+  - \frac{\kay_r}{2} j^{(r)}_+.
\end{equation}
Moreover, using $\B_-^{(r)}=\kay_r \A_-^{(r)}$, $\A_+^{(r)}=\text{Id}$ and $\B_+^{(r)}=-\kay_r\text{Id}$ (cf. Subsection \ref{Examples}), one gets
\begin{equation}\label{Eq:LambdaCS3}
\kay_r \sum_{s=1}^N \frac{\vp_{-,s}(z_s^-)}{\vp_{-,s}(z_r^+)} J_-^{(s)} = \sum_{s=1}^N \Lop^-_{rs} J_-^{(s)} + \frac{\kay_r}{2}  \sum_{s=1}^N \Kop^-_{rs} J_-^{(s)} = \sum_{s=1}^N \Lop^-_{rs} J_-^{(s)} + \frac{\kay_r}{2}  j_-^{(r)}.
\end{equation}
Reinserting Equations \eqref{Eq:LambdaCS2} and \eqref{Eq:LambdaCS3} in the expression \eqref{Eq:LambdaCS1} of $\Upsilon_r$, one sees that $\Upsilon_r$ satisfies Equation \eqref{Eq:CSV}, as required.

\section{Conclusion and perspectives}

In this article, we constructed integrable deformations of the coupled $\s$-model introduced in~\cite{Delduc:2018hty}, using the formalism of affine Gaudin models. In particular, we obtained explicit expressions of the action and Lax pair of the deformed models corresponding to arbitrary combinations of Yang-Baxter and $\lambda$-deformations. Moreover, we showed that the integrable coupled $\lambda$-models introduced recently in~\cite{Georgiou:2016urf,Georgiou:2017jfi,Georgiou:2018hpd,Georgiou:2018gpe} can be seen as particular limits of the models constructed here. Let us now conclude by discussing some perspectives of the present work.\\

As explained in Subsection \ref{SubSec:Undef}, the deformed models constructed in this article which involve a Yang-Baxter realisation without Wess-Zumino term possess a corresponding $q$-deformed Poisson-Lie symmetry, which replaces the left translation symmetry of the undeformed model. It is well known that the Yang-Baxter model (with one copy and without Wess-Zumino term) in fact possesses a larger (infinite) symmetry algebra, satisfying the relations of an affine $q$-deformed Poisson algebra~\cite{Delduc:2017brb} (see also~\cite{Kawaguchi:2012ve,Kawaguchi:2012gp,Kawaguchi:2013gma}), which replaces the Yangian symmetry of the undeformed Principal Chiral Model~\cite{MacKay:1992he}. It would be interesting to understand whether such infinite extensions of the $q$-deformed symmetries also exist for the deformed coupled models and what would be their underlying algebraic structure.

The integrable deformed models constructed in this article still possess an undeformed symmetry, corresponding to the diagonal symmetry of the underlying affine Gaudin model, which acts on the fields $g^{(r)}$ by right multiplication ($g^{(r)}\mapsto g^{(r)}h$) or conjugacy ($g^{(r)}\mapsto h^{-1}g^{(r)}h$), depending on whether the realisations at sites $(r,\pm)$ are Yang-Baxter realisations or $\lambda$-realisations. It was explained in~\cite{Lacroix:2019xeh} that for a general realisation of affine Gaudin model of the type considered in~\cite{Delduc:2019bcl}, one can construct an integrable Yang-Baxter deformation which breaks its diagonal symmetry. Thus, one can introduce a further integrable deformation of the deformed coupled $\s$-models constructed in this article. As explained in~\cite{Lacroix:2019xeh}, this deformation procedure involves gauging the model and thus requires treating Hamiltonian first-class constraints. For brevity, we chose not to treat these deformations in the present article: however, we expect that they can be studied using similar methods to the ones developed here. For the case with one copy only, it was conjectured in~\cite{Lacroix:2019xeh} that these further deformed theories coincide with already known models, namely the bi-Yang-Baxter model (see~\cite{Klimcik:2008eq,Klimcik:2014bta,Delduc:2015xdm} for the case without Wess-Zumino term and~\cite{Delduc:2017fib} for the case with Wess-Zumino term) and the generalised $\lambda$-model~\cite{Sfetsos:2015nya}.

It is known that the Yang-Baxter and $\lambda$-models are Poisson-Lie T-dual~\cite{Klimcik:1995ux,Klimcik:1995jn,Klimcik:1995dy} to one another~\cite{Vicedo:2015pna, Hoare:2015gda, Sfetsos:2015nya, Klimcik:2015gba}, while the Yang-Baxter model with Wess-Zumino term is Poisson-Lie T-dual to itself with different parameters~\cite{Demulder:2017zhz}. It would be interesting to investigate the various possible dualities between the coupled models constructed in this article and how they would manifest themselves in the underlying geometry of their target space $G_0^N$. The study of Poisson-Lie T-dualities between deformed $\s$-models led to their reformulation as $\mathcal{E}$-models~\cite{Klimcik:2015gba,Klimcik:2017ken,Klimcik:2019kkf}, making their duality properties manifest. A natural direction to explore is thus to search for a similar reformulation of the coupled models constructed here. 

The results of Section \ref{Sec:CS} illustrate once again the deep relation existing between the approaches to two-dimensional integrable field theories from affine Gaudin models~\cite{Vicedo:2017cge} and from four-dimensional semi-holomorphic Chern-Simons theory~\cite{Costello:2019tri}, first established in~\cite{Vicedo:2019dej} and further supported in~\cite{Delduc:2019whp}. In particular, the analysis conducted in this section strengthens the apparent correspondence between the choice of realisations in the first approach and the choice of boundary conditions in the second one. It would be interesting to understand in more details this correspondence. Let us also note that the construction of the Yang-Baxter and $\lambda$-boundary conditions introduced in~\cite{Delduc:2019whp}, which uses isotropic subalgebras of the complex or real double of $\g_0$, is reminiscent of the structure underlying Poisson-Lie T-duality and $\mathcal{E}$-models and could thus provide interesting directions for investigating the questions raised in the previous paragraph.

It would also be interesting to explore the quantum properties of these classically integrable deformed $\s$-models. For example, a natural question is whether these models are one-loop renormalisable and if there exist conformal fixed points in this space of models. The results obtained in~\cite{Georgiou:2017jfi,Georgiou:2018hpd,Georgiou:2018gpe} about the renormalisation of the coupled $\lambda$-models introduced in these references (which are limits of the models considered here), already show a rich structure in their renormalisation group flow. As a further possible step, it would be interesting to investigate the higher-loops renormalisability of these models and, if needed, the corresponding quantum corrections of their underlying geometry, as recently studied in~\cite{Hoare:2019ark,Hoare:2019mcc,Georgiou:2019nbz} for non-coupled models.

\paragraph{Acknowledgments.} We would like to thank G. Arutyunov, F. Delduc, M. Magro and B. Vicedo for useful discussions and comments on the draft. This work is funded by the Deutsche Forschungsgemeinschaft (DFG, German Research Foundation) under Germany’s Excellence Strategy -- EXC 2121 ``Quantum Universe'' -- 390833306.

\begin{appendices}

\section{Proof of the identities \eqref{Identities}}
\label{Identities1}

In this appendix we will present the calculation of the non-ultralocal terms (\textit{i.e.} terms containing derivatives of the delta distribution) in the bracket \eqref{PoissonJ}, using the ansatz \eqref{Currents} for the currents $\mathcal{J}_\pm$ in terms of the operators $\A_\pm$ and $\B_\pm$. In particular, we will show that this computation implies that these operators satisfy the identities \eqref{Identities}. Let us start by noting that in order to perform this computation, we need the Poisson brackets between the following objects: $\A_\pm$, $Y$, $\B_\pm$ and $j$. However, let us recall that we have assumed the operators $\A_\pm$ and $\B_\pm$ to depend only on the field $g$ (and not on its derivative $\p_xg$). Thus, the non-ultralocal terms in the brackets of $\mathcal{J}_\pm$ can only come from the brackets between the fields $Y$ and $j$. More precisely, for $\epsilon, \sigma \in \{\pm\}$, we have
\begin{align*}
\{\mathcal{J}_{\epsilon \underline{\mathbf{1}}}(x),\mathcal{J}_{\sigma \underline{\mathbf{2}}}(y)\} &= \{\A_\epsilon Y_{\underline{\mathbf{1}}}(x), \B_\sigma j_{\underline{\mathbf{2}}}(y)\} + \{\B_\epsilon j_{\underline{\mathbf{1}}}(x), \A_\sigma Y_{\underline{\mathbf{2}}}(y)\} + \text{[ultralocal terms]} \\
&= \A_{\epsilon \underline{\mathbf{1}}}(x) \B_{\sigma \underline{\mathbf{2}}}(y) \{Y_{\underline{\mathbf{1}}}(x), j_{\underline{\mathbf{2}}}(y)\} + \B_{\epsilon \underline{\mathbf{1}}}(x) \A_{\sigma \underline{\mathbf{2}}}(y) \{j_{\underline{\mathbf{1}}}(x), Y_{\underline{\mathbf{2}}}(y)\} + \text{[u.l.]}.
\end{align*}
From here, using the form of the Poisson bracket between $Y$ and $j$, which can be simply found from \eqref{Poissonj} and \eqref{PoissonW1}, we find that
\begin{align*}
\{\mathcal{J}_{\epsilon \underline{\mathbf{1}}}(x),\mathcal{J}_{\sigma \underline{\mathbf{2}}}(y)\} &= -\A_{\epsilon \underline{\mathbf{1}}}(x) \B_{\sigma \underline{\mathbf{2}}}(y) C_{\underline{\mathbf{12}}}\delta'_{xy} - \B_{\epsilon \underline{\mathbf{1}}}(x) \A_{\sigma \underline{\mathbf{2}}}(y) C_{\underline{\mathbf{12}}}\delta'_{xy}  + \text{[u.l.]} \\
&= -(\A_{\epsilon \underline{\mathbf{1}}}(x) \B_{\sigma \underline{\mathbf{2}}}(x) + \B_{\epsilon \underline{\mathbf{1}}}(x) \A_{\sigma \underline{\mathbf{2}}}(x)) C_{\underline{\mathbf{12}}}\delta'_{xy}  + \text{[u.l.]},
\end{align*}
where we have used the fact that for any function $f$, $f(y)\delta'_{xy}=f(x)\delta'_{xy}+f'(x)\delta_{xy}=f(x)\delta'_{xy}+\text{[u.l.]}$. Finally, using the fact that for an operator $\mathcal{O}$ on the Lie algebra $\mathfrak{g}$, $\mathcal{O}_{\underline{\mathbf{1}}}C_{\underline{\mathbf{12}}} = {^t}\mathcal{O}_{\underline{\mathbf{2}}}C_{\underline{\mathbf{12}}}$, we get
\begin{equation}\label{Eq:PBJ}
\{\mathcal{J}_{\epsilon \underline{\mathbf{1}}}(x),\mathcal{J}_{\sigma \underline{\mathbf{2}}}(y)\} = -(\A_\epsilon {^t}\B_\sigma + \B_\epsilon {^t}\A_\sigma)_{\underline{\mathbf{1}}}(x) C_{\underline{\mathbf{12}}}\delta'_{xy}  + \text{[u.l.]}.
\end{equation}
As we want $\mathcal{J}_\pm$ to be Poisson commuting Kac-Moody currents of levels $\ell^\pm$, one should have
\begin{eqnarray*}
\lbrace \mathcal{J}_{\pm\,\underline{\mathbf{1}}}(x), \mathcal{J}_{\pm\,\underline{\mathbf{2}}}(y) \rbrace &=& -\ell^\pm C_{\underline{\mathbf{12}}} \delta'_{xy} + \text{[u.l.]}, \\
\lbrace \mathcal{J}_{\pm\,\underline{\mathbf{1}}}(x), \mathcal{J}_{\mp\,\underline{\mathbf{2}}}(y) \rbrace &=& 0.
\end{eqnarray*}
Comparing with Equation \eqref{Eq:PBJ}, we then see that the operators $\A_\pm$ and $\B_\pm$ should satisfy the identities \eqref{Identities}.

\section{Simplification of the action \eqref{Eq:ActionNonLorentz}}
\label{Lorentz}

In this appendix, we show that the non-Lorentz invariant terms appearing in the second line of the action \eqref{Eq:ActionNonLorentz} cancel with the term in the first line containing the Hamiltonian. For that, let us start by computing the expression of the Hamiltonian in terms of Lagrangian fields.

\paragraph{Hamiltonian in terms of Lagrangian fields.} We will proceed here in a similar fashion to what has been done in \cite{Delduc:2019bcl}. Let us start by noting that, combining the equations \eqref{Charges} and \eqref{Hamiltonian}, the Hamiltonian can be rewritten as
\begin{equation}\label{Eq:Ham}
\mathcal{H} = \int_{\mathbb{D}} \text{d}x \ \left(\sum_{i\in\mathcal{I}_-}\frac{1}{2\varphi'(\zeta_i)} \kappa(\Gamma(\zeta_i),\Gamma(\zeta_i)) - \sum_{i\in\mathcal{I}_+}\frac{1}{2\varphi'(\zeta_i)} \kappa(\Gamma(\zeta_i),\Gamma(\zeta_i))\right),
\end{equation}
where we have used the fact that $\epsilon_i=\pm 1$ for $i\in\mathcal{I}_\pm$. We then need to look for the Lagrangian expression of the quantities $\Gamma(\zeta_i)$. This is done by relating them to residues of the Lax pair. More precisely, let us fix $i\in\mathcal{I}_\pm$: from \eqref{Laxlight}, we have
\begin{equation*}
\Gamma(\zeta_i) = \pm \frac{1}{2}\varphi'(\zeta_i)\underset{z = \zeta_i}{\operatorname{res}} \mathcal{L}_\pm(z).
\end{equation*}
Using the Lagrangian expression \eqref{Eq:Interpolation} of $\Lc_\pm(z)$ and the fact that $\vp'(\zeta_i)=\vp_\pm'(\zeta_i)\vp_\mp(\zeta_i)$, we then get
\begin{equation*}
\Gamma(\zeta_i) = \pm \frac{\ell^\infty}{2}\sum_{r=1}^N \frac{\varphi_{\pm,r}(z_r^\pm)\varphi_\mp(\zeta_i)}{z_r^\pm - \zeta_i}J_\pm^{(r)}.
\end{equation*}
Substituting back into equation \eqref{Eq:Ham}, we arrive at the following expression for $\mathcal{H}$:
\begin{equation}\label{Eq:HamLag}
\mathcal{H} = \int_{\mathbb{D}} \text{d}x \ \left(\sum_{r,s = 1}^Nc_{rs}^+\,\kappa\bigl(J_+^{(r)},J_+^{(s)}\bigr) + \sum_{r,s = 1}^Nc_{rs}^-\,\kappa\bigl(J_-^{(r)},J_-^{(s)}\bigr)\right),
\end{equation}
with
\begin{equation*}
c_{rs}^\pm = \pm \frac{\ell^\infty}{8}\varphi_{\pm,r}(z_r^\pm)\varphi_{\pm,s}(z_s^\pm)\sum_{i\in\mathcal{I}_\pm}\frac{1}{z_r^\pm-\zeta_i}\frac{1}{z_s^\pm-\zeta_i}\frac{\varphi_\mp(\zeta_i)}{\varphi_\pm'(\zeta_i)}.
\end{equation*}
Before proceeding, we note that one can prove the above coefficients to be equal to
\begin{equation} \label{crs}
c_{rs}^\pm = \pm \delta_{rs} \frac{\ell_r^\pm}{8} \pm \frac{1}{2} \sum_{k=1}^N \ell_k^\mp \frac{\varphi_{\pm,r}(z_r^\pm)}{\varphi_{\pm,r}(z_k^\mp)} \frac{\varphi_{\pm,s}(z_s^\pm)}{\varphi_{\pm,s}(z_k^\mp)}.
\end{equation}

\paragraph{Simplification of non Lorentz invariant terms in the action.} Let us consider the terms in the second line of \eqref{Eq:ActionNonLorentz}. Using the expression \eqref{Eq:J2j} of $j^{(r)}_\pm$, they can be rewritten in the following way:
\begin{align*}
&\frac{1}{2} \sum_{r,s}^N \iint\text{d}t \,\text{d}x \ \left[\kappa\left(\Lop_{rs}^+ J_+^{(s)},j_+^{(r)}\right) + \kappa\left(\Lop_{rs}^- J_-^{(s)},j_-^{(r)}\right)\right] \\
&= \frac{1}{2} \sum_{r,s,t}^N \iint\text{d}t \,\text{d}x \ \left[\kappa\left(J_+^{(s)},{^t}\Lop^+_{rs} \;\Kop^+_{rt} J_+^{(t)}\right) + \kappa\left(J_-^{(s)}, {^t}\Lop^-_{rs} \;\Kop^-_{rt} J_-^{(t)}\right)\right] \\
&= \frac{1}{4} \sum_{r,s,t}^N \iint\text{d}t \,\text{d}x \ \left[\kappa\left(J_+^{(s)}, \bigl( {^t}\Lop^+_{rs} \;\Kop^+_{rt} + {^t}\Kop^+_{rs} \,\Lop^+_{rt} \bigr) J_+^{(t)}\right) + \kappa\left(J_-^{(s)}, \bigl( {^t}\Lop^-_{rs}\; \Kop^-_{rt} + {^t}\Kop^-_{rs} \,\Lop^-_{rt} \bigr) J_-^{(t)}\right)\right]
\end{align*}
We want to prove that these terms are cancelled by the term in \eqref{Eq:ActionNonLorentz} containing the Hamiltonian. From the expression \eqref{Eq:HamLag} of the Hamiltonian, one sees that this is the case upon using the following identity:
\begin{equation*}
\frac{1}{4}\sum_{r=1}^N\left({^t}\Lop^\pm_{rs} \;\Kop^\pm_{rt} + {^t}\Kop^\pm_{rs} \,\Lop^\pm_{rt}\right) = c_{st}^\pm\,\Id,
\end{equation*}
which can be proved using the identities \eqref{Identities} and the form \eqref{crs} of the coefficients $c_{st}^\pm$.

\end{appendices}

\bibliographystyle{JHEP}

\providecommand{\href}[2]{#2}\begingroup\raggedright\endgroup

\end{document}